\documentclass[11pt,a4paper]{article}
\usepackage{jheppub}
\usepackage{graphicx}
\usepackage{slashed}
\usepackage{physics}
\usepackage{amsmath,amsbsy,amsfonts,amsmath,bbm,latexsym,amssymb,bm}
\usepackage{epsfig}
\usepackage{wasysym}
\usepackage{array}
\usepackage{mathrsfs}
\usepackage[super]{nth}
\newcolumntype{C}[1]{>{\centering\arraybackslash}m{#1}}

\newenvironment{Eqnarray}%
         {\arraycolsep 0.14em\begin{eqnarray}}{\end{eqnarray}}
\newcommand{\be}{\begin{equation}}
\newcommand{\ee}{\end{equation}}
\newcommand{\ba}{\begin{Eqnarray}}
\newcommand{\ea}{\end{Eqnarray}}
\newcommand{\bs}{\begin{subequations}}
\newcommand{\es}{\end{subequations}}
\newcommand{\no}{\nonumber\\}

\newcommand{\grts}{\raise.3ex\hbox{$>$\kern-.75em\lower1ex\hbox{$\sim$}}}
\newcommand{\lets}{\raise.3ex\hbox{$<$\kern-.75em\lower1ex\hbox{$\sim$}}}

\def\eq#1{eq.~(\ref{#1})}
\def\eqst#1#2{eqs.~(\ref{#1})--(\ref{#2})}
\def\eqs#1#2{eqs.~(\ref{#1}) and (\ref{#2})}
\def\Eq#1{Eq.~(\ref{#1})}
\def\Eqs#1#2{Eqs.~(\ref{#1}) and (\ref{#2})}
\def\eqss#1#2#3{eqs.~(\ref{#1}), (\ref{#2}) and (\ref{#3})}

\def\half{\tfrac12}
\def\phaa{\phantom{AA}}
\def\idtN{ \bm{1}_{2N \times 2N}}

\def\ifmath#1{\relax\ifmmode #1\else $#1$\fi}
\def\lsub#1{\ifmath{_{\lower1.5pt\hbox{$\scriptstyle #1$}}}}
\def\lsup#1{^{\lower 6pt\hbox{$\scriptstyle#1$}}}
\def\phm{\phantom{-}}
\def\newcdot{\kern.06em{\cdot}\kern.06em}
\def\iso{\mathchoice{\cong}{\cong}{\isoS}{\cong}}
\def\isoS{\vbox{\baselineskip 0pt  \lineskip 0.5pt
    \ialign{$ \mathsurround=0pt  \scriptstyle \hfil ## \hfil $\crcr
        \sim \crcr = \crcr}}}

\def\vb#1{\vbox to #1 pt{}}

\notoc

\allowdisplaybreaks

\begin{document}
\begin{flushright}
CFTP-17-005\\[1mm]
SCIPP-17/10
\end{flushright}

\title{Multi-Higgs doublet models: physical parametrization,
sum rules and unitarity bounds}

\author[a]{Miguel\ P.\ Bento}
\author[b]{Howard E.~Haber}
\author[a]{J.\ C.\ Rom\~{a}o}
\author[a]{Jo\~{a}o P.\ Silva}

\affiliation[a]{Centro de F\'{\i}sica Te\'{o}rica de Part\'{\i}culas (CFTP) and
    Departamento de F\'{\i}sica,
    Instituto Superior T\'{e}cnico, Technical University of Lisbon,
    1049-001 Lisboa, Portugal}

\affiliation[b]{Santa Cruz Institute for Particle Physics,
University of California,
Santa Cruz, California 95064, USA}
\emailAdd{miguelfilipebento@gmail.com}
\emailAdd{haber@scipp.ucsc.edu}
\emailAdd{jorge.romao@tecnico.ulisboa.pt}
\emailAdd{jpsilva@cftp.tecnico.ulisboa.pt}

\date{\today}

\abstract{If the scalar sector of the Standard Model is non-minimal,
one might expect multiple generations of the hypercharge-$1/2$
scalar doublet analogous to the generational structure of the fermions.
In this work, we examine the structure of a Higgs sector consisting of
$N$ Higgs doublets (where $N\geq 2$).
It is particularly convenient to work in the so-called
\textit{charged Higgs basis}, in which the neutral Higgs vacuum
expectation value resides entirely in the first Higgs doublet,
and the charged components of remaining $N-1$ Higgs doublets are
mass-eigenstate fields.
We elucidate the interactions of the gauge bosons with the
physical Higgs scalars and the Goldstone bosons and show that
they are determined by an $N \times 2N$ matrix.
This matrix depends on $(N-1)(2N-1)$ real parameters
that are associated with the mixing of the neutral Higgs
fields in the charged Higgs basis.
Among these parameters, $N-1$ are unphysical (and can be
removed by rephasing the physical charged Higgs fields),
and the remaining $2(N-1)^2$ parameters are physical.
We also demonstrate a particularly simple form for the cubic
interaction and some of the quartic interactions of the Goldstone
bosons with the physical Higgs scalars.
These results are applied in the derivation of Higgs coupling
sum rules and tree-level unitarity bounds that restrict the
size of the quartic scalar couplings.
In particular, new applications to three Higgs doublet models
with an order-4 CP symmetry and with a $\mathbb{Z}_3$ symmetry,
respectively, are presented.
}




\maketitle
\flushbottom

\section{Introduction}

The discovery of the Higgs boson at the Large Hadron Collider (LHC)
appears to complete the story of the Standard Model (SM) of
particle physics~\cite{Aad:2012tfa,Chatrchyan:2012xdj}.
In particular,  the subsequent experimental measurements of the
properties of the observed scalar with mass 125 GeV are so far
consistent with those of the SM Higgs boson~\cite{Khachatryan:2016vau}.
The current Higgs data set is still statistically limited, so a more
precise statement is that the observed scalar behaves as a SM-like Higgs boson,
to within an accuracy of about $20\%$.
Future experimental studies of the Higgs boson at the LHC  will
continue to search for deviations from SM behavior, as well as
evidence for additional scalar states that might comprise an extended Higgs sector.

Nevertheless, there are numerous reasons to suspect that there must
exist new physical phenomena beyond the SM.
For example, the SM cannot accommodate dark matter~\cite{darkmatter},
massive neutrinos~\cite{numass}, baryogenesis~\cite{White:2016nbo} and the gravitational interaction.
Indeed, there is no fundamental understanding of how the electroweak
scale arises, and why this scale is many orders of magnitude smaller
than the Planck scale~\cite{hierarchy}.
However, the casual observer examining the structure of the SM in
its present form might also be puzzled that the set of fundamental scalar
fields consists of a single neutral CP-even Higgs boson.
After all, the SM employs a direct product of three separate
gauge groups and the elementary fermionic matter comes in three generations.
Why should one expect a scalar sector to consist only of a single
physical state?

In light of the non-minimal structure of the fermionic and gauge
bosonic sectors of the SM, one is tempted to suppose that the scalar
sector is likewise non-minimal.
It is a simple matter to construct an extension of the SM
that incorporates an enlarged Higgs sector.
In order to preserve the tree-level relation,
$\rho_0\equiv m_W/m_Z\cos\theta_W=1$ (which is confirmed by the
electroweak data after accounting for electroweak radiative corrections~\cite{rhoparm}),
the electroweak quantum numbers of the Higgs scalar multiplets are constrained~\cite{HHG,Ivanov:2017dad}.
For example,
a Higgs sector employing hypercharge-$1/2$ scalar doublets and hypercharge-zero scalar singlets yields $\rho_0=1$,
independently of the vacuum expectation values (vevs) of the
neutral scalar fields.
Following the generational pattern of the fermions,
we shall simply replicate the SM Higgs doublet and consider
an extended Higgs sector consisting of $N$ hypercharge-$1/2$ scalar doublets.
For $N\geq 2$, one must further require that at the minimum
of the scalar potential, only the neutral components of the
$N$ scalar fields acquire vevs~\cite{NHiggs,Nishi:2007nh}.\footnote{The requirement
that the electric charge preserving vacuum is a global minimum
of the scalar potential imposes some constraints on the scalar
potential parameters.  Henceforth, we assume that these constraints are respected.}

In this paper, we wish to explore various relations among Higgs
couplings and bounds on scalar masses that arise in an $N$ Higgs
doublet extension of the SM.
Electroweak gauge invariance plays a central role in determining
the structure of the Higgs couplings.
Our primary focus here will be the couplings of Higgs bosons
to gauge bosons and the cubic and quartic scalar self-couplings.
The couplings of the Higgs bosons to fermions is governed by the
Yukawa couplings, which must be highly constrained in order
to avoid tree-level flavor-changing neutral currents mediated by neutral scalars~\cite{Glashow:1976nt,Paschos:1976ay}.
In this paper, we shall simply postpone the consideration of the
scalar-fermion interactions~\cite{postpone}.

Our main tool for obtaining relations among Higgs couplings and
constraints on Higgs masses is tree-level unitarity.
If one computes the scattering amplitudes for $2\to 2$ scattering
of gauge and Higgs bosons, under the assumption that all Higgs
couplings are independent of one another, then one finds that
some of the scattering amplitudes grow with the center of mass energy.  Such
behavior is not consistent with unitarity.
Of course, there is no paradox here since the assumption of
independent Higgs couplings is incorrect.
Electroweak gauge invariance imposes relations among the
couplings that guarantee that the bad high energy behavior of
any scattering amplitude must exactly cancel.
One can turn this argument around and derive relations
among the Higgs couplings that are required to cancel the bad
high energy behavior of all scattering amplitudes~\cite{LlewellynSmith:1973yud,Cornwall:1974km}.
This procedure allows one to deduce a variety of sum rules that
relate various Higgs couplings~\cite{Weldon:1984wt, Gunion:1990kf,Grinstein:2013fia}.

Having canceled the bad high energy behavior, one finds that
scattering amplitudes in the high energy limit either approach
a constant value or vanish in the limit of large center of mass energy.
In the former case, the condition of tree-level unitarity imposes an
upper limit on the value of this constant.
Ultimately, one can show that this constant is a function of
dimensionless quartic couplings that appear in the scalar potential.
Thus, the imposition of tree-level unitarity yields an upper bound
on the values of various combinations of quartic scalar couplings.
This in turn can provide upper bounds on some combinations of scalar masses~\cite{Lee:1977yc,Lee:1977eg}.

In section~\ref{sec:model} we consider the most general $N$
Higgs doublet model (NHDM).
We explicitly write out the couplings of the Higgs bosons to
the gauge bosons, which arise from the scalar field kinetic energy
terms after replacing the ordinary derivative with the gauge
covariant derivatives of the electroweak theory.
Remarkably, these couplings are controlled by an
$N \times 2N$ matrix $B$, whose physical significance is
explained below.  We also note the appearance of the
$2N \times 2N$ matrix $A = \Im(B^\dagger B)$.
The matrix $A$ is an orthogonal antisymmetric matrix that is
governed by $N(N-1)$ parameters.
These parameters are independent of the basis of scalar fields used
to define the model, and thus are related to physical
observables.\footnote{Given a set of $N$ scalar doublet fields
$\{\Phi_k\}$, one is always free to consider another basis of
scalar fields, $\{\Phi^\prime_\ell\}$ that is related to the
original set of scalar fields by a unitary transformation.
Any physical observable must be independent of the choice of basis.}
The matrix $B$ is governed by $2(N-1)^2$ physical parameters, which
include the $N(N-1)$ parameters that already appear in the matrix $A$.
In addition, the matrix $B$ depends on $N-1$ (unphysical) phases
that can be eliminated by appropriately rephasing the
physical charged Higgs fields of the model.
We also examine some aspects of the scalar self-couplings.
We find that some specific cubic and quartic couplings
involving Goldstone boson fields also depend
exclusively on the matrices $A$ and $B$.
This is a consequence of the fact that these interactions
terms are related by gauge-fixing to terms appearing in the
gauge-covariant scalar kinetic energy terms.
The structure of the other cubic and quartic couplings are
not as simple, and involve more complicated invariant expressions
involving the coefficients of the scalar potential.

In section~\ref{sec:sumrules}, we derive a variety a sum
rules involving the couplings of gauge bosons and Higgs
bosons and the couplings of Goldstone bosons and Higgs bosons.
In section~\ref{sec:unitaritybounds}, we present an efficient technique
to impose the condition
of tree-level unitarity, leading to upper bounds on various
combinations of couplings and scalar masses.
We apply this to known results and present new results
for the three Higgs doublet models (3HDMs) with $\mathbb{Z}_3$ symmetry and
with order-4 CP symmetry, respectively.
Conclusions are given in section~\ref{sec:conclusions}.

In our analysis of the NHDM,
one can define a new basis of scalar fields such that
the neutral scalar vev resides entirely in
one of the Higgs doublet fields, denoted by $\Phi^H_1$.
This is the well-known Higgs basis~\cite{Georgi:1978ri,LS,BLS,DH}, in which the charged
component of $\Phi^H_1$ is identified as the charged Goldstone
boson field and the imaginary part of the neutral component
of $\Phi^H_1$ is the neutral Goldstone boson field.
The Higgs basis is not unique, since one is free
to make an arbitrary U($N-1$) transformation among the remaining $N-1$ doublet fields.
In particular, one can employ this transformation to
diagonalize the physical charged Higgs squared-mass matrix.
This procedure yields the \textit{charged Higgs basis}, as discussed
in appendix~\ref{app:HiggsBasis}.
Note that the resulting basis is unique up to an arbitrary
separate rephasing of the $N-1$ scalar doublets that contain the physical
charged Higgs boson fields.

In appendix~\ref{app:C2HDM}, we apply the analysis of the
NHDM given in section~\ref{sec:model}
to the two-Higgs doublet model (2HDM).  We first discuss the
complex two-Higgs doublet model (C2HDM), in which
a $\mathbb{Z}_2$-symmetric scalar potential is softly broken by a complex
squared-mass parameter.   We then generalize to the
most general 2HDM, which is treated using the
basis-independent formalism of ref.~\cite{Haber:2006ue}.
In both cases, we display the explicit 2HDM forms for
the matrices $A$ and $B$
and exhibit the unphysical parameters in the matrix $B$ that can be
eliminated by an appropriate rephasing of the physical
charged Higgs fields.
In appendix~\ref{app:count}, we demonstrate how to count
the number of parameters that govern the matrices $A$
and $B$ of the NHDM and identify which
of these parameters are physical.

In our derivation of coupling sum rules and unitarity relations,
the coupling of the Goldstone bosons fields to
the physical Higgs fields play an important role.
Although the initial forms of these expressions are quite complicated,
the corresponding cubic couplings end up reducing to remarkably simple forms.
As an example, we provide in appendix~\ref{app:cubic} the details of the derivation
and simplification of the coupling of two neutral Goldstone fields and a physical Higgs scalar.
Finally, in appendix~\ref{app:SR}, we rederive the sum
rules obtained directly from the NHDM interaction Lagrangian
using an alternative method,
which imposes the cancellation of bad high energy behavior in the
$2\to 2$ scattering amplitudes of processes involving the gauge and Higgs bosons.  The relation between sum rules
involving the neutral Higgs boson couplings to $W^+W^-$ and $ZZ$ is clarified in
appendix~\ref{app:ZZS=WWS}.

\section{\label{sec:model} $N$ Higgs doublet models}

In this section we discuss the
bosonic Lagrangian of the
most general
NHDM.
The field content consists of
the $SU(2)_L \times U(1)_Y$ gauge bosons and
$N$ hypercharge-$1/2$ Higgs doublet fields,
parameterized as,
\be
\Phi_k =
\left(
\begin{array}{c}
\varphi_k^+ \\
\tfrac{1}{\sqrt{2}} ( v_k + \varphi_k^0)
\end{array}
\right),\qquad
\text{for $k=1,\ldots,N$.}
\label{Phi-1}
\ee
\enlargethispage{1.5\baselineskip}%
The Higgs-fermion Yukawa interactions will be examined in a subsequent work~\cite{postpone}.

\subsection{\label{subsec:pot}The scalar potential}

For the scalar potential, we follow the notation
of \cite{BS,BLS}:
\be
V_H
=
\mu_{ij}  ( \Phi_i^\dagger \Phi_j )
+
\lambda_{ij,kl} ( \Phi_i^\dagger \Phi_j ) ( \Phi_k^\dagger \Phi_l )
= - {\cal L}_{\rm Higgs},
\label{VH}
\ee
where, by hermiticity,
\be
\mu_{ij} = \mu_{ji}^\ast,
\hspace{3ex}
\lambda_{ij,kl} \equiv \lambda_{kl,ij} = \lambda_{ji,lk}^\ast.
\label{hermiticity}
\ee
Using eq.~\eqref{Phi-1},
the Higgs potential becomes
\be
V_H = V_0 + V_1 + V_2 + V_3 + V_4,
\label{VHiggs}
\ee
where,\footnote{It is useful to note that,
because $(M^2_\pm)_{ij}$ and $\lambda_{ik,lj} v_k v_l^\ast$ are hermitian in $(i,j)$,
the second term in eq.~\eqref{V2} may be written as
$$
\left[ (M^2_\pm)_{ij} + \lambda_{ik,lj} v_k v_l^\ast \right]
\varphi_i^{0 \ast} \varphi_j^0
=
\textrm{Re} \left\{ \left[ (M^2_\pm)_{ij} + \lambda_{ik,lj} v_k v_l^\ast \right]
\varphi_i^{0 \ast} \varphi_j^0 \right\}.
$$
\label{fn3}}
\ba
2 V_0
&=&
\mu_{ij}\, (v_i^\ast v_j)
+
\tfrac{1}{2} \lambda_{ij,kl}  \, (v_i^\ast v_j) (v_k^\ast v_l),
\label{V0}\\*[2mm]
2 V_1
&=&
v_i^\ast \left[ \mu_{ij} + \lambda_{ij,kl} v_k^\ast v_l \right] \varphi_j^0
+
\varphi_i^{0 \ast} \left[ \mu_{ij} + \lambda_{ij,kl} v_k^\ast v_l \right] v_j ,
\label{V1}\\*[2mm]
V_2
&=&
(M^2_\pm)_{ij}\, \varphi^-_i \varphi^+_j
+
\tfrac{1}{2}
\left[ (M^2_\pm)_{ij} + \lambda_{ik,lj} v_k v_l^\ast \right]
\varphi_i^{0 \ast} \varphi_j^0
+
\tfrac{1}{2} \textrm{Re}
\left\{ \lambda_{ik,jl} v_k v_l\,  \varphi_i^{0 \ast} \varphi_j^{0 \ast} \right\},
\label{V2}\\*[2mm]
V_3
&=&
\lambda_{ij,kl}\, \varphi_i^- \varphi_j^+
\left[ \varphi_k^{0 \ast} v_l + v_k^\ast \varphi_l^0 \right]
+
\tfrac{1}{2} \lambda_{ij,kl}\, \varphi_i^{0 \ast} \varphi_j^0
\left[ \varphi_k^{0 \ast} v_l + v_k^\ast \varphi_l^0 \right],
\label{V3}\\*[2mm]
V_4
&=&
\lambda_{ij,kl} (\varphi_i^- \varphi_j^+)(\varphi_k^- \varphi_l^+)
+
\lambda_{ij,kl} (\varphi_i^- \varphi_j^+)(\varphi_k^{0 \ast} \varphi_l^0)
+
\tfrac{1}{4} \lambda_{ij,kl} (\varphi_i^{0 \ast} \varphi_j^0)(\varphi_k^{0 \ast} \varphi_l^0),
\label{V4}
\ea
and
\be
(M^2_\pm)_{ij} = \mu_{ij} + \lambda_{ij,kl} v_k^\ast v_l
\label{Mpm}
\ee
is the mass matrix for the charged scalar fields.
Requiring the absence of linear terms yields the
stationarity condition,
\be
V_1 = 0 \ \ \Longrightarrow \ \
\left[ \mu_{ij} + \lambda_{ij,kl} v_k^\ast v_l \right] v_j
= (M^2_\pm)_{ij}\, v_j
= 0.
\label{stationarity}
\ee
The vacuum of these models has been studied in ref.~\cite{NHiggs};
here we assume only that the electromagnetic $U(1)_\textrm{em}$ remains
unbroken.\footnote{Some features of the vacuum of $N$ Higgs doublet models
have also been studied using the bilinear formalism in
refs.~\cite{Nishi:2006tg, Nishi:2007nh, Ivanov:2010ww, Maniatis:2015gma}.
}
Expanding the neutral fields in terms of their real and imaginary
components, the second and third terms of $V_2$ may be written as
\be
\frac{1}{2}\,
\left(
\begin{array}{cccccc}
\textrm{Re}(\varphi_1^0), &
\dots &
, \textrm{Re}(\varphi_N^0), &
\textrm{Im}(\varphi_1^0), &
\dots &
, \textrm{Im}(\varphi_N^0)
\end{array}
\right)
\left(
\begin{array}{cc}
\hspace{1ex}\\
M_R^2 & \quad M_{RI}^2\\
\hspace{2ex}
\\
(M_{RI}^2)^T & \quad M_I^2\\
\hspace{1ex}
\end{array}
\right)
\left(
\begin{array}{c}
\textrm{Re}(\varphi_1^0)\\
\vdots\\
\textrm{Re}(\varphi_N^0)\\*[2mm]
\textrm{Im}(\varphi_1^0)\\
\vdots\\
\textrm{Im}(\varphi_N^0)
\end{array}
\right),
\label{M_neutrals}
\ee
where [cf.~footnote~\ref{fn3}],
\ba
(M_R^2)_{ij} &=&
\textrm{Re} \left\{
(M^2_\pm)_{ij} + \lambda_{ik,lj} v_k v_l^\ast + \lambda_{ik,jl} v_k v_l
\right\},
\label{MR}\\
(M_I^2)_{ij} &=&
\textrm{Re} \left\{
(M^2_\pm)_{ij} + \lambda_{ik,lj} v_k v_l^\ast - \lambda_{ik,jl} v_k v_l
\right\},
\label{MI}\\
(M_{RI}^2)_{ij} &=&
- \textrm{Im} \left\{
(M^2_\pm)_{ij} + \lambda_{ik,lj} v_k v_l^\ast - \lambda_{ik,jl} v_k v_l
\right\}.
\label{MRI}
\ea
Using eq.~\eqref{hermiticity},
we conclude that $M_\pm^2$ is hermitian,
the matrices $M_R^2$ and $M_I^2$ are real and symmetric,
and $M_{RI}^2$ is a general real matrix.
Thus,
the mass matrix in eq.~\eqref{M_neutrals} is real and symmetric,
as expected.
Our mass matrices agree with those in ref.~\cite{NHiggs},
after noting that their $\nu^d_k = v_k/\sqrt{2}$.
Our eq.~\eqref{MRI} differs by a minus sign in the last term with respect
to a similar eq.~(A17) of ref.~\cite{GL1}.  However, this sign error is the result of a misprint,
in light of the agreement in signs between our results and eqs.~(A18)--(A22) of
ref.~\cite{GL1}.

For later use,
we note that
\be
(M_{RI}^2)^T_{ij} =
\textrm{Im} \left\{
(M^2_\pm)_{ij} + \lambda_{ik,lj} v_k v_l^\ast + \lambda_{ik,jl} v_k v_l
\right\},
\label{MRI_transp}
\ee
arises from eqs.~\eqref{MRI} and \eqref{hermiticity}.
From eqs.~\eqref{MR}--\eqref{MRI_transp},
we find
\ba
2\, \lambda_{ik,lj} v_k v_l^\ast &=&
- 2 (M^2_\pm)_{ij}
+
\left[ (M_R^2)_{ij} +  (M_I^2)_{ij} \right]
+ i
\left[
  (M_{RI}^2)^T_{ij} - (M_{RI}^2)_{ij}
\right],
\label{Kprime_1}
\\
2\, \lambda_{ik,jl} v_k v_l &=&
\left[
 (M_R^2)_{ij} -  (M_I^2)_{ij}
\right]
+ i
\left[
 (M_{RI}^2)^T_{ij} +  (M_{RI}^2)_{ij}
\right].
\label{K_1}
\ea

Eqs.~\eqref{V0}--\eqref{V4} are written in with respect to a generic basis of the scalar
doublet fields.
One can define a new set of charged and neutral scalar fields denoted,
respectively, by $S^\pm_a\ \ (a=1,\ldots,N)$ and $S^0_\beta\ \ (\beta=1\,\ldots,2N)$,
via
\ba
\varphi_k^+ &=&
\sum_{a=1}^N U_{k a} S_a^+,
\label{S+}\\
\varphi_k^0 &=&
\sum_{\beta=1}^{2N} V_{k \beta} S_\beta^0,
\label{S0}
\ea
where $U$ is an $N \times N$ unitary matrix,
and $V$ is a complex $N \times 2N$ matrix.
It is convenient to define the real $2N\times 2N$ matrix,
\be
\tilde{V}
=
\left(
\begin{array}{c}
\textrm{Re}\ V\\
\textrm{Im}\ V
\end{array}
\right)
\label{tildeV}
\ee
Note that the transformation given in \eq{S0} results in a real orthogonal similarity transformation of
the $2N \times 2N$ symmetric squared-mass matrix given in eq.~\eqref{M_neutrals}.  That is $\tilde{V}$ is
a $2N \times 2N$ real orthogonal matrix.
As a result,
we find
\be
\bm{1}_{2N \times 2N}
=
\tilde{V}^T \tilde{V}
=
\textrm{Re} V^T\ \textrm{Re} V + \textrm{Im} V^T\ \textrm{Im} V
=
\textrm{Re}  \left( V^\dagger V \right),
\label{VTV}
\ee
where $\bm{1}_{2N \times 2N}$ is the $2N\times 2N$ identity matrix.
Similarly,
from $\bm{1}_{2N \times 2N} = \tilde{V} \tilde{V}^T $,
we obtain,
\ba
\textrm{Re} V\ \textrm{Re} V^T
&= \bm{1}_{N \times N} =&
\textrm{Im} V\ \textrm{Im} V^T,
\nonumber\\
\textrm{Re} V\ \textrm{Im} V^T
&= \bm{0}_{N \times N} =&
\textrm{Im} V\ \textrm{Re} V^T.
\label{VVT_parts}
\ea
\clearpage

\noindent
Hence, it follows that,
\be
V V^T = \textrm{Re}(V V^T ) + i \textrm{Im}(V V^T ) = 0.
\label{VVT}
\ee
Note that we have used the convenient notation of refs.~\cite{GL2,GL3},
which in turn
was inspired by refs.~\cite{GL1,Grimus:1989pu}.
In addition,
\begin{equation}
V V^\dagger = \Re(V V^\dagger) + i \Im(VV^\dagger) = 2 \newcdot \bm{1}_{N \times N}.
\end{equation}

Finally,
one can show that the matrix $\textrm{Im}  \left( V^\dagger V \right)$
is antisymmetric.  Moreover, 
using eqs.~\eqref{VVT_parts}, this matrix satisfies,
\be
\left[
\textrm{Im}  \left( V^\dagger V \right)
\right]^2
=
-
\bm{1}_{2N \times 2N}.
\label{Asquare=-1}
\ee
Thus,
$\textrm{Im}  \left( V^\dagger V \right)$
is the only nontrivial piece of $V V^\dagger$ and $V^\dagger V$.
As we shall see below,
it has the crucial role of controlling scalar couplings involving the
$Z$ boson or its corresponding Goldstone boson.
Combining the antisymmetry with eq.~\eqref{Asquare=-1},
we conclude that $\textrm{Im}  \left( V^\dagger V \right)$ is orthogonal
\be
\left[ \textrm{Im}  \left( V^\dagger V \right)\right]^T
\left[ \textrm{Im}  \left( V^\dagger V \right)\right]
=
\bm{1}_{2N \times 2N}.
\ee
In particular,
all of its entries satisfy
\be
\left| \textrm{Im}  \left( V^\dagger V \right)_{\alpha \beta}\right| \leq 1.
\ee
This will be of interest later.

Let us consider matrices such that
\ba
U_{k1} &=& \hat{v}_k,
\label{U_k1}
\\
V_{k1} &=& i \hat{v}_k,
\label{V_k1}
\ea
where we have defined
\be
\hat{v}_k \equiv \frac{v_k}{v}.
\label{omk}
\ee
With these choices,
the unitarity of $U$ implies
\be
(U^\dagger \hat{v})_a = \delta_{1 a} = (\hat{v}^\dagger U)_a
\ \ \ \ (a=1, \dots, N),
\label{omU}
\ee
while eq.~\eqref{VTV} implies
\ba
- \textrm{Im} (V^\dagger \hat{v})_\beta =
\textrm{Im} (\hat{v}^\dagger V)_\beta &=& \delta_{1 \beta}
\ \ \ \ (\beta=1, \dots, 2N),
\label{ImVDomega}
\\
\textrm{Re} (V^\dagger \hat{v})_\beta =
\textrm{Re} (\hat{v}^\dagger V)_\beta &=& - \textrm{Im} (V^\dagger V)_{1 \beta}
\ \ \ \ (\beta=2, \dots, 2N),
\label{ReVDomega}
\\
\textrm{Re} (V^\dagger \hat{v})_1 =
\textrm{Re} (\hat{v}^\dagger V)_1 &=& - \textrm{Im} (V^\dagger V)_{1 1} = 0,
\label{omV}
\ea
where the last equality holds because $\textrm{Im} (V^\dagger V)_{\alpha \beta}$
is antisymmetric in $\alpha \beta$.
We wish to study the squared-mass matrix of the charged scalars
with respect to the charged scalar fields $S_a^+$,
\be
M_C^2 = U^\dagger M^2_\pm U.
\ee
Using eqs.~\eqref{Mpm}, \eqref{U_k1}, and \eqref{stationarity},
it is easy to show that
\be
\left(U^\dagger M^2_\pm U \right)_{1b} = \left(U^\dagger M^2_\pm U \right)_{a1} = 0.
\ee
This means that with respect to the  charged scalar fields $S_a^+$ [reached by transformations with
\eqref{U_k1}],
the first row and first column of the transformed squared-mass matrix of the charged scalars vanishes.
This identifies $S_1^\pm$ with the charged would-be Goldstone boson~$G^\pm$,
\be
S_1^\pm = G^\pm.
\label{S1+_is_G+}
\ee

Next, we turn to the squared-mass matrix of the neutral scalar fields.
\ba
M^2_N
&=&
\tilde{V}^T\
\left(
\begin{array}{cc}
\hspace{1ex}\\
M_R^2 & M_{RI}^2\\
\hspace{2ex}
\\
(M_{RI}^2)^T & M_I^2\\
\hspace{1ex}
\end{array}
\right)
\ \tilde{V}
\label{MN2}\\
&=&
\textrm{Re} V^T\, M_R^2\, \textrm{Re} V
+
\textrm{Im} V^T\, \left(M_{RI}^2\right)^T\, \textrm{Re} V
+
\textrm{Re} V^T\, M_{RI}^2\, \textrm{Im} V
+
\textrm{Im} V^T\, M_I^2\, \textrm{Im} V.
\nonumber
\ea
Using eq.~\eqref{V_k1},
we start by looking at
\ba
\left( M_R^2 \textrm{Re} V \right)_{i1} +
\left( M_{RI}^2 \textrm{Im} V \right)_{i1}
&=&
- \textrm{Re} \left\{
(M^2_\pm)_{ij} + \lambda_{ik,lj} v_k v_l^\ast + \lambda_{ik,jl} v_k v_l
\right\}\
\textrm{Im}(\hat{v}_j)
\nonumber\\
&&
- \textrm{Im} \left\{
(M^2_\pm)_{ij} + \lambda_{ik,lj} v_k v_l^\ast - \lambda_{ik,jl} v_k v_l
\right\}\
\textrm{Re}(\hat{v}_j)
\nonumber\\
&=&
- \textrm{Im} \left\{ (M^2_\pm)_{ij} \hat{v}_j
+ \lambda_{ik,lj} v_k v_l^\ast \hat{v}_j
\right\}
+
\textrm{Im} \left\{ \lambda_{ik,jl} v_k v_l \hat{v}_j^\ast
\right\}
\nonumber\\
&=&
0.
\label{aux_MN_1}
\ea
For the first equality,
we have used eqs.~\eqref{MR} and \eqref{MRI}.
To reach the last line of \eq{aux_MN_1},
we have used eq.~\eqref{omk} and the
stationarity condition given in eq.~\eqref{stationarity}.
Similarly,
\ba
\left( (M_{RI}^2)^T \textrm{Re} V \right)_{i1} +
\left( M_I^2 \textrm{Im} V \right)_{i1}
&=&
- \textrm{Im} \left\{
(M^2_\pm)_{ij} + \lambda_{ik,lj} v_k v_l^\ast + \lambda_{ik,jl} v_k v_l
\right\}\
\textrm{Im}(\hat{v}_j)
\nonumber\\
&&
+ \textrm{Re} \left\{
(M^2_\pm)_{ij} + \lambda_{ik,lj} v_k v_l^\ast - \lambda_{ik,jl} v_k v_l
\right\}\
\textrm{Re}(\hat{v}_j)
\nonumber\\
&=&
\textrm{Re} \left\{ (M^2_\pm)_{ij} \hat{v}_j
+ \lambda_{ik,lj} v_k v_l^\ast \hat{v}_j
\right\}
-
\textrm{Re} \left\{ \lambda_{ik,jl} v_k v_l \hat{v}_j^\ast
\right\}
\nonumber\\
&=&
0.
\label{aux_MN_2}
\ea
Multiplying eq.~\eqref{aux_MN_1} by $(\textrm{Re} V^T)_{\alpha i}$,
multiplying eq.~\eqref{aux_MN_2} by $(\textrm{Im} V^T)_{\alpha i}$,
and summing over~$i$,
we conclude from eq.~\eqref{MN2} that
\be
(M_N^2)_{1 \beta} = (M_N^2)_{\alpha 1} = 0,
\ee
where the first equality holds since $M_N^2$ is a real symmetric matrix.
This means that with respect to the scalar fields $S_\beta^0$,
[reached by transformations with
\eqref{V_k1}],
the first row and first column of the transformed squared-mass matrix of the neutral scalars vanishes.
This identifies $S_1^0$ with the neutral would-be Goldstone boson
$G^0$,
\be
S_1^0 = G^0.
\label{S10_is_G0}
\ee

One can choose matrices $U$ and $V$ in such a way that the transformations in
eqs.~\eqref{S+} and \eqref{S0} yield the
charged and neutral scalar mass eigenstate fields, respectively,
\ba
U^\dagger\, M_\pm^2\, U
&=&
D_\pm^2 =
\textrm{diag} \left( m_{\pm, 1}^2=0,  m_{\pm, 2}^2, \dots,  m_{\pm, N}^2 \right)
\label{Dpm2}
\\*[2mm]
\tilde{V}^T\,
\left(
\begin{array}{cc}
\hspace{1ex}\\
M_R^2 & M_{RI}^2\\
\hspace{2ex}\\
(M_{RI}^2)^T & M_I^2\\
\hspace{1ex}
\end{array}
\right)
\, \tilde{V}
&=&
D_0^2 =
\textrm{diag} \left( m_1^2=0,  m_2^2, \dots,  m_{2N}^2 \right).
\label{D02}
\ea
Since we have identified $S_1^\pm=G^\pm$ and $S_1^0=G^0$, it follows from our above analysis that
the matrices $U$ and $V$ must satisfy eqs.~\eqref{U_k1} and \eqref{V_k1}, respectively.
In this case,
$S^\pm_a\ \ (a=2,\ldots,N)$ and $S^0_\beta\ \ (\beta=2,\ldots,2N)$
denote the fields of the physical charged and neutral scalar
particles, respectively.
Their corresponding masses are $m_{\pm, k}^2\ \ (k=1,2, \ldots, N)$
and $m_\beta^2\ \ (\beta=1,2, \ldots, 2N)$.%
\footnote{In appendix~\ref{app:HiggsBasis} we discuss two other ways to
identify the neutral and charged scalar mass eigenstate fields,
involving intermediate steps which simplify some of the analysis.}

Using eqs.~\eqref{Dpm2}--\eqref{D02} in
eqs.~\eqref{Kprime_1}--\eqref{K_1},
we end up with
\ba
2\, \lambda_{ik,lj} v_k v_l^\ast &=&
- 2 \left(U\, D_\pm^2\, U^\dagger \right)_{ij}
+
\left( V\, D_0^2\, V^\dagger \right)_{ij},
\label{Kprime_2}
\\
2\, \lambda_{ik,jl} v_k v_l &=&
\left( V\, D_0^2\, V^T \right)_{ij}.
\label{K_2}
\ea
These are the only combinations of quartic couplings
(and vevs) that one can obtain from
the diagonalization of the scalar squared-mass matrices.
Thus,
only those cubic and quartic terms of the scalar potential involving
these combinations will be related to scalar masses.
Eqs.~\eqref{Kprime_2}--\eqref{K_2} constitute a crucial result of our paper,
since, they will enable us to relate the gauge-Higgs couplings
with the scalar-scalar couplings.\footnote{This is, of course, consistent with
gauge-fixing and is needed, in particular,
for the equivalence theorem.}

\subsection{\label{subsec:Gauge-higgs}Gauge-Higgs couplings}

When expressed in terms of the physical gauge fields,
the gauge covariant derivative may be written as
\be
i D_\mu
=
i \partial_\mu
- \frac{g}{2} (\tau_+ W^+_\mu + \tau_- W^-_\mu )
- e Q A_\mu
- \frac{g}{c_W}
\left( \frac{\tau_3}{2} - Q s_W^2 \right) Z_\mu,
\label{D_mu}
\ee
where $g$ is the $SU(2)$ coupling constant,
$c_W = \cos{\theta_W}$,
$s_W = \sin{\theta_W}$,
$e$ is the electric charge of the positron,
$Q$ is the charge operator,
and\footnote{In this paper, we normalize the hypercharge $\mathcal{Y}$ such that $Q=T_3+\mathcal{Y}$, where $\boldsymbol{\vec T}\equiv\half\boldsymbol{\vec{\tau}}$.}
\be
\tau_+ =
\left(
\begin{array}{cc}
0 & \quad\sqrt{2}\\
0 & \quad 0
\end{array}
\right),
\ \ \
\tau_- =
\left(
\begin{array}{cc}
0 & \quad 0 \\
\sqrt{2} & \quad0
\end{array}
\right),
\ \ \
\tau_3 =
\left(
\begin{array}{cc}
1 & \quad \phm0 \\
0 & \quad -1
\end{array}
\right),
\ee
when acting on SU(2) doublets.
Note that the signs of the
coupling constants and gauge fields above correspond to
choosing all the $\eta_k$ equal to $+1$ in the notation of
ref.~\cite{Romao:2012pq}.
This coincides with the conventions of ref.~\cite{HHG},
but differs in the signs in $g$ from refs.~\cite{djouadi1, djouadi2}.\footnote{The signs in refs.~\cite{BLS, GL2} correspond to yet another choice,
which yields an unexpected sign in the relation $\tan{\theta_W} = -g^\prime/g$.
}
This also has an impact on any Feynman rules proportional to
$M_W$ or $M_Z$.

The kinetic term for the scalar fields is
\be
\mathcal{L}_{K \Phi}
=
\sum_{k=1}^N \left(D^\mu \Phi_k \right)^\dagger \left( D_\mu \Phi_k\right).
\label{LkH}
\ee
We substitute eq.~\eqref{D_mu} and parameterize the $\Phi_k$ as in eq.~\eqref{Phi-1}.
Next, we employ eqs.~\eqref{S+}--\eqref{S0} to express the charged and neutral fields in terms of the
mass eigenstate fields,
and we use the properties in eqs.~\eqref{VTV}--\eqref{omV}.
We end up with,
\ba
{\cal L}_{K \Phi}
&=&
\sum_{a=1}^N \left(\partial^\mu S^-_a \right)^\dagger \left( \partial_\mu S^+_a \right)
+
\frac{1}{2} \sum_{\beta=1}^{2N}
\left(\partial^\mu S^0_\beta \right) \left( \partial_\mu S^0_\beta \right)
+M_W^2 W^{+\, \mu} W^-_\mu + \frac{1}{2} M_Z^2 Z^\mu Z_\mu
\no
&+&
i M_W
\left[
W_\mu^+ (\partial^\mu G^-)
-
W_\mu^- (\partial^\mu G^+)
\right]
-
M_Z Z_\mu (\partial^\mu G^0)
+\mathcal{L}_{VVS}+\mathcal{L}_{VSS}+\mathcal{L}_{VVSS}\,,\nonumber \\[0.01in]
\phantom{line}
\label{kinetic}
\ea
where,\footnote{We employ the notation where $S_a^+\overset{\leftrightarrow}{\partial}\lsup{\mu} S_a^-\equiv
S_a^+(\partial^\mu S_a^-) -(\partial^\mu S_a^+) S_a^-$.}
\ba
\mathcal{L}_{VVS}&=&
\left[
e M_W A^\mu - g M_Z s_W^2 Z^\mu
\right]\,
\left[
W_\mu^- G^+
+
W_\mu^+ G^-
\right] \no
&-&g
\left[
M_W W^{+\, \mu} W^-_\mu + \frac{M_Z}{2 c_W} Z^\mu Z_\mu
\right]
\sum_{\beta=2}^{2N}
S^0_\beta\,
A_{1 \beta}\,,
\label{VVS_lag} \\*[2mm]
\mathcal{L}_{VSS}&=&
i
\left[
e A_\mu + \frac{g\, c_{2W}}{2 c_W} Z_\mu
\right]
\sum_{a=1}^N
S_a^+\overset{\leftrightarrow}{\partial}\lsup{\mu} S_a^- +\frac{g}{4 c_W} Z_\mu
\sum_{\gamma \neq \beta = 1}^{2N} (S_\gamma^0\overset{\leftrightarrow}{\partial}\lsup{\mu} S_\beta^0) A_{\beta\gamma}
\no
&+&
i \frac{g}{2}
\sum_{a=1}^N
\sum_{\beta=1}^{2N}
\left[
B_{a \beta}
W_\mu^+
(S_\beta^0 \overset{\leftrightarrow}{\partial}\lsup{\mu} S_a^-)
+
(B^\dagger)_{\beta a}
W_\mu^-
(S_a^+ \overset{\leftrightarrow}{\partial}\lsup{\mu} S_\beta^0)
\right],
\label{Z_NHNH}
\\*[2mm]
\mathcal{L}_{VVSS}&=&
\left[
\frac{g^2}{4} W^{+\, \mu} W^-_\mu
+
\frac{g^2}{8 c_W^2} Z^\mu Z_\mu
\right]
\sum_{\beta=1}^{2N} (S_\beta^0)^2
\no
&+&
\left[
\frac{g^2}{2} W^{+\, \mu} W^-_\mu
+
e^2 A^\mu A_\mu
+
\frac{e g\, c_{2W}}{c_W} A^\mu Z_\mu
+
\frac{g^2\, c_{2W}^2}{4 c_W^2} Z^\mu Z_\mu
\right]
\sum_{a=1}^{N} S_a^- S_a^+
\no
&+&
\left[
\frac{e g}{2} A^\mu
-\frac{g^2 s_W^2}{2 c_W} Z^\mu
\right]
\sum_{a=1}^N
\sum_{\beta=1}^{2N}
S^0_\beta
\left[
(B^\dagger)_{\beta a} W^-_\mu S_a^+
+
B_{a \beta} W^+_\mu S_a^-
\right],
\label{GG_CHCH}
\ea
with $c_{2W}=\cos{(2 \theta_W)}$ and
\ba
B &\equiv&
U^\dagger V\hspace{38mm} \text{\footnotesize $[N \times 2N]$},
\label{def_B}
\\
A &\equiv&
\textrm{Im} (V^\dagger V) = \textrm{Im} (B^\dagger B)
\ \ \ \ \ \ \ \text{\footnotesize $[2N \times 2N]$},
\label{A_definition}
\ea
are matrices of dimension $N\times 2N$ and $2N\times 2N$, respectively.
\clearpage

The matrix $A$ is basis-independent and hence physical,
whereas the matrix $B$ is basis-independent up to unphysical phases
that can be absorbed into the definition of the physical charged Higgs
fields, $S^\pm_a$ (for $a=2,3,\ldots,N$).
Further details will be provided in the next section.
Eqs.~\eqref{kinetic}--\eqref{GG_CHCH} agree with
eqs.~(29a)--(29p) of ref.~\cite{GL2},
if we notice that,
because of the different sign in the coupling $g$,
we have
$g=-g^{\textrm{GLOO}}$ , $M_W=-M_W^{\textrm{GLOO}}$ ,
and $M_Z=-M_Z^{\textrm{GLOO}}$,
where the subscript ``GLOO'' stands for ref.~\cite{GL2}.
The only exception is in the $ZS_\gamma S_\beta$
coupling given in \eq{Z_NHNH} which,
after the difference in notation is properly accounted for,
disagrees with the sign of eq.~(29h) of ref.~\cite{GL2}.
We have checked in both notations that their incorrect sign can be attributed to a misprint.

\subsection{\label{subsec:matrix_B}The $A$ and $B$ matrices}

In section~\ref{subsec:Gauge-higgs} we showed that the couplings
arising from the kinetic Lagrangian depend exclusively on the
matrix $B$ in the charged scalar sector
and on
$A = \textrm{Im}(B^\dagger B)$
in the neutral scalar sector.   $A$ and $B$ are defined in terms of
the matrices $U$ and $V$, which relate the charged and neutral fields in a generic basis of scalar fields, $\{\Phi_k\}$, to the corresponding mass-eigenstate scalar fields, respectively.   The choice of basis is of course arbitrary.  For example,
another set of scalar fields $\{\Phi^\prime_\ell\}$, with
\be
\Phi_\ell = X_{k\ell} \Phi^\prime_\ell,
\label{basis_change}
\ee
where $X$ is some $N \times N$ unitary matrix, could have been employed.
The scalar field kinetic energy terms are invariant with
respect to \eq{basis_change}, since
\be
\mathcal{L}_{K \Phi}
=
\sum_{k=1}^N \left(D^\mu \Phi_k \right)^\dagger \left( D_\mu \Phi_k\right)=
\label{LkHp}\sum_{k=1}^N \left(D^\mu \Phi^\prime_k \right)^\dagger \left( D_\mu \Phi^\prime_k\right)\,,
\ee
which follows from $X^\dagger X= \bm{1}_{N \times N}$.
Consequently, the interactions of the scalars with the gauge bosons given by \eqst{kinetic}{GG_CHCH} are basis-independent.
Indeed, any physical observable cannot depend on the choice of basis.
We would like to use these observations to address the behavior of the
matrices $A$ and $B$ under an arbitrary change of basis.
The interaction Lagrangian given by \eqst{kinetic}{GG_CHCH} is
written in terms of the scalar mass eigenstates $S_a^\pm$ and $S_\beta^0$,
which are related via \eqs{S+}{S0} to the scalar fields in a generic basis.
The diagonalization of the neutral scalar squared-mass matrix is
given by \eq{MN2} and yields real neutral scalar mass-eigenstate fields.
The overall sign of the neutral scalar mass-eigenstate fields are not physical.
However, the standard practice is to fix this sign by appropriately
restricting the range of the angles that parameterize the diagonalization matrix.
Having adopted this convention where the sign of the neutral
scalar mass-eigenstate fields are fixed, it follows that the matrix
$A$ that appears in \eqs{VVS_lag}{Z_NHNH} is basis-independent and
hence physical.

In contrast, the diagonalization of the charged scalar squared mass
matrix yields complex mass-eigenstate charged scalar fields.
By convention, the phase of the charged Goldstone field is fixed.
In particular, it is convenient to choose $X=U$ in \eq{basis_change},
which yields the scalar field basis,
\be
\Phi^C_1 =
\left(
\begin{array}{c}
G^+\\*[2mm]
\tfrac{1}{\sqrt{2}}
\left( v + H^0 + i G^0 \right)
\end{array}
\right),
\ \ \
\Phi^C_2 =
\left(
\begin{array}{c}
S^+_2\\*[2mm]
\tfrac{1}{\sqrt{2}} \varphi^{C0}_2
\end{array}
\right),
\ \ \
\ldots\,,
\ \ \
\Phi^C_N =
\left(
\begin{array}{c}
S^+_N\\*[2mm]
\tfrac{1}{\sqrt{2}} \varphi^{C0}_N
\end{array}
\right),
\ee
where $S_2^+,\ldots,S^+_N$ are the physical (mass-eigenstate) charged Higgs fields
with corresponding masses $m^2_{\pm, i}$.
This is called the \textit{charged Higgs basis} and has two defining properties:
\begin{enumerate}
\item $S_2^+$, $S_3^+$, $\ldots$, $S_N^+$,
are the charged scalar mass-eigenstate fields;
\item the first doublet field, $\Phi_1^C$, has the massless would-be
Goldstone boson $G^+$ as its charged (upper) component, and its phase has been
chosen such that the (real and positive) vev,
$v\simeq 246$~GeV, is contained in the
real part of its neutral (lower) component.
\end{enumerate}
The Higgs basis (which is defined by property 2 alone) and
the charged Higgs basis are discussed in detail
in appendix~\ref{app:HiggsBasis}.
Notice that the fields $H^0$ and $ \varphi^{C0}_2,\ldots,\varphi^{C0}_N$
are not the neutral scalar mass-eigenstate fields.
$B$ is the matrix that transforms these fields,
written in the charged Higgs basis, into the neutral scalar mass-eigenstates.

The charged Higgs basis is unique up to a possible rephasing of
the charged Higgs fields,
$S^+_a\to e^{i\chi_a}S_a^+$ [for $a=2,3,\ldots,N$].
That is, the charged Higgs basis is a family of scalar bases
that is characterized by $N-1$ phases, $\chi_a$, as discussed at the end of appendix~\ref{app:HiggsBasis}.
In \eqs{Z_NHNH}{GG_CHCH},
the invariant combinations $B_{a\beta}S_a^-$ and its charged
conjugate appear.
Hence, it follows that the matrix $B$ is not quite basis-independent,
since $B_{a\beta}\to e^{i\chi_a}B_{a\beta}$ [for $a=2,3,\ldots,N$]
under the rephasing of the charged Higgs fields to preserve the
invariance of \eqs{Z_NHNH}{GG_CHCH}.

In contrast, the mixing matrices $U$ and $V$ that relate the charged and
neutral fields in a generic basis of scalar fields, $\{\Phi_k\}$,
to the corresponding mass-eigenstate scalar fields, respectively,
are basis dependent.
It is instructive to examine the question of basis dependence and
work out the explicit forms of the matrices $A$ and $B$ in the more
familiar 2HDM.
In this case, the matrix $U$ contains the angle $\beta = \tan^{-1}{(v_2/v_1)}$.
As discussed at length in ref.~\cite{Haber:2006ue},
this means that the angle $\beta$ does not in general have
a physical meaning, since the vevs $v_1$ and $v_2$
(and hence $\tan\beta$) transform under the basis transformation
given by \eq{basis_change}.
Similarly, the mixing angle parameters appearing in the
matrix $V$ that transforms the neutral scalar fields of the
generic basis into the neutral scalar mass-eigenstate fields are
also basis dependent.
To find basis-independent quantities,
one must consider the neutral mixing angle parameters
\textit{relative} to the angle $\beta$.
That is, under a basis transformation, both the neutral mixing
angle parameters and the angle $\beta$ shift by the same amount so
that their difference is invariant.
Thus, one way to determine the invariant neutral mixing angle
parameters is to work in the Higgs basis, in which $\beta=0$.
Further details can be found in appendix~\ref{app:C2HDM}.
The matrix $B$ is the NHDM generalization of the neutral mixing
angle parameters relative to the angle $\beta$.
As noted above, the matrix $B$ is almost basis-invariant,
since unphysical phases remain that reflect the possible
rephasing of the charged Higgs fields.

From section~\ref{subsec:pot}, we can deduce the following
properties of the matrices $A$ and $B$.
First, it is convenient to re-express $A$ in terms of the matrix
$\tilde{V}$ defined in \eq{tildeV}.
We introduce the $2N\times 2N$ orthogonal antisymmetric matrix
$\tilde{J}$, which in block form is defined by
\be
\tilde{J}\equiv \begin{pmatrix} 0 &\quad \bm{1}_{N \times N} \\
-\bm{1}_{N \times N} & \quad 0\end{pmatrix}\,.
\ee
Then $A=\Im(V^\dagger V)$ can be rewritten as,
\be \label{VJV}
A=\tilde{V}^T \tilde{J}\tilde{V}\,.
\ee
It immediately follows that $A$ is a real orthogonal
\textit{and} antisymmetric matrix; i.e.,
\be \label{roa}
A A^T = \bm{1}_{2N \times 2N},
\qquad\quad
A^T = - A.
\ee
In particular, the orthogonality of $A$ implies that
$|A_{\alpha \beta}| \leq 1$.

Given an $N\times N$ unitary matrix $U$, it is always possible to represent this matrix by the following $2N\times 2N$ real orthogonal matrix,
\be \label{Ureal}
\tilde{U}_R\equiv \begin{pmatrix} \Re U & \quad -\Im U\\ \Im U & \quad \phm \Re U\end{pmatrix}\,.
\ee
Henceforth, we shall always identify the real orthogonal representation of a unitary matrix with a subscript $R$ and a tilde (to indicate that the dimensionality of the matrix has been doubled).

Using this notation, it is convenient to construct the $2N\times 2N$ matrix $\tilde B$,
\be \label{tildeB}
\tilde{B}
\equiv
\left(
\begin{array}{c}
\Re B\\*[1mm]
\Im B
\end{array}
\right)
=\tilde{U}_R^T\tilde{V}\,.
\ee
Since $\tilde{U}_R$ and $\tilde{V}$ are real orthogonal
$2N \times 2N$ matrices,  it follows that
$\tilde B$ is also a real orthogonal $2N \times 2N$ matrix.
Moreover,
\be
\tfrac12 B B^\dagger
= \bm{1}_{N \times N},
\qquad\quad
\Re (B^\dagger B)
=
\bm{1}_{2N \times 2N}.\label{bbid}
\ee
Noting that $B=U^\dagger V$ and $A=\Im(B^\dagger B)$, one can write
\be \label{BIA}
B^\dagger B=V^\dagger V=\bm{1}_{2N \times 2N}+iA\,.
\ee
Finally, we note that
\ba
B_{a1} &=& i \delta_{a1},
\nonumber\\
B_{1 \beta} &=& - A_{1 \beta} + i \delta_{1 \beta}.
\label{properties_B}
\ea

The central point of this section is the following.
Unless there is an (imposed symmetry) reason to single out some
specific basis, the best way to count parameters and to set up
a numerical simulation is to write the potential in the
charged Higgs basis ab initio.
As seen from appendix~\ref{app:HiggsBasis},
the charged Higgs basis is (almost) unique,
up to the separate rephasing of the
$N-1$ doublets with zero vev.
As a result, all the parameters of the scalar potential, when written
in the charged Higgs basis,
are either invariant or pseudo-invariant quantities with respect
to arbitrary scalar basis transformations.
Here, pseudo-invariant means invariant up to an overall phase that
arises from the rephasing of the
$N-1$ doublets that contain the physical charged Higgs fields.
It is straightforward to construct invariants from appropriate
products of pseudo-invariants in which the overall phase ambiguity cancels.
All such observable quantities are potential experimental observables.
This provides the generalization of the parameters
$ Y_1, Y_2, Y_3$ and $Z_1, Z_2,  \ldots, Z_7 $
championed for the 2HDM in ref.~\cite{Zs}.

\subsection{\label{subsec:counting}Parameter counting}

We begin by asking the following question.
How many parameters govern the matrices $A$ and $B$ and how many
of these parameters are physical?
To address this question, we first examine the matrices $V$ and $U$,
which transform the neutral and charged scalar fields of a
generic basis to the corresponding scalar-mass eigenstates,
respectively.

Starting from a scalar basis, $\{\Phi_k\}$,
one can compute the neutral scalar squared-mass matrix as
shown in \eq{D02}.
The real orthogonal $2N\times 2N$ diagonalizing matrix $\tilde V$
is related to the matrix $V$ defined in \eq{S0} via \eq{tildeV}.
With respect to a new scalar basis $\{\Phi^\prime_\ell\}$,
one obtains a new matrix $V$ given by
\be \label{vprime}
V^\prime=X^\dagger V\,,
\ee
after using \eqs{S0}{basis_change}.
Employing \eqs{tildeV}{vprime}, one obtains a new real
orthogonal diagonalizing matrix
$\tilde{V}^\prime$ given by
\be \label{tildeVp}
\tilde{V}^\prime=\tilde{X}_R^T\tilde{V}\,,
\ee
where [following \eq{Ureal}] the $2N\times 2N$ real orthogonal matrix $\tilde{X}_R$ is defined by
\be \label{VV}
\tilde{X}_R\equiv\begin{pmatrix} \Re X & \quad -\Im X \\ \Im X
& \quad \phm\Re X\end{pmatrix}\,.
\ee
Using \eq{rcurp}, one can decompose the real orthogonal $2N\times 2N$
matrix $\tilde V=\widetilde{W}_R \tilde{R}_c$,
where $\widetilde{W}_R$ and $\tilde{R}_c$ are the real orthogonal
$2N\times 2N$ matrices given in \eq{rpdef}.
Inserting this result back into \eq{VV} yields,
\be \label{Vnewbasis}
\tilde{V}^\prime=X_R^T \widetilde{W}_R \tilde{R}_c\,.
\ee

Using the definition of the matrix $A$ given in \eq{A_definition},
one can determine $A$ with respect to a new scalar basis $\{\Phi^\prime_\ell\}$,
\be
A^\prime= \Im(V^{\prime\,\dagger}V^\prime)=\Im(V^\dagger V)=A\,.
\ee
Note that the same result can be obtained by employing \eqs{VJV}{tildeVp},
since
\be \label{Aprime}
A^\prime=\tilde{V}^{\prime\,T}\tilde{J}\tilde{V}^\prime=\tilde{V}^T\tilde{J}\tilde{V}=A\,,
\ee
after noting that $\tilde{X}_R\tilde{J}\tilde{X}_R^T=\tilde{J}$
(the latter makes use of the fact that $X$ is unitary).
That is, $A$ is basis-independent.
In particular, if we choose $X=W$ in \eq{Vnewbasis},
then $\tilde{V}^\prime=\tilde{R}_c$, which can be expressed in terms of $N(N-1)$
parameters.\footnote{Since $\tilde{V}^\prime=\tilde{R}_c$,
we can use the results of appendix~\ref{embed} to show that
$\tilde{R}_c$, which depends on two $N\times N$ real antisymmetric matrices,
is governed by $N(N-1)$ parameters.}
Consequently, \eq{Aprime} implies that $A=\tilde{R}_c^T\tilde{J}\tilde{R}_c$,
which depends on $N(N-1)$ physical
parameters.\footnote{This result is consistent with
\eq{roa}, since the most general real orthogonal
antisymmetric $2N\times 2N$ matrix can be expressed in
terms of $N(N-1)$ independent parameters, as shown in
appendix~\ref{app:Acount}.}

Likewise, starting from a scalar basis, $\{\Phi_k\}$,
one can compute the charged scalar squared-mass matrix
as shown in \eq{Dpm2}.
The $N\times N$ unitary diagonalizing matrix $U$ defined in \eq{S+}
depends on the basis.
One also must take into account that the charged Higgs basis is not uniquely defined, as discussed at the end of appendix~\ref{app:HiggsBasis}.  Hence, the physical charged Higgs fields may acquire phases under the basis change.
If we perform a basis transformation given by \eq{basis_change}, we must allow for the possibility
that  $S^+_a\to e^{i\chi_a}S_a^+$ [for $a=2,3,\ldots,N$].   We can write the latter as
\be \label{schange}
S_a^+=\sum_{b} K_{ab}S^{\prime\,+}_b\,,
\ee
where the primes indicate quantities associated with the transformed basis and
\be \label{diagph}
K\equiv{\rm diag}(1\,,\,e^{-i\chi_2}\,,\,e^{-i\chi_3}\,,\,\ldots\,,\,e^{-i\chi_N})\,.
\ee
Combining \eqss{S+}{basis_change}{schange} yields
\be
\varphi^{\prime\,+}_k=\sum_{a=1}^N U^\prime_{ka}S_a^{\prime\,+}\,,
\ee
where
\be \label{xuk}
U'=X^\dagger UK\,.
\ee
That is,
one obtains a new unitary diagonalizing matrix $U^\prime$ given by
\be \label{uprime}
U_{j1}^\prime=(X^\dagger)_{jk} U_{k1}\,,\qquad\quad U_{ja}^\prime=e^{-i\chi_a}(X^\dagger)_{jk} U_{ka}\,,
\ee
where there is an implicit sum over the repeated index $k$.

We can now compute the transformation of $B$ under a change of scalar basis.
With respect to the basis $\{\Phi^\prime_\ell\}$,
we have
\ba
B^\prime_{1\beta}
&=&
(U^{\prime\,\dagger}V^\prime)_{1\beta}
=
(U^\dagger V)_{1\beta}
=
B_{1\beta}\,,\\
B^\prime_{a\beta}
&=&
(U^{\prime\,\dagger}V^\prime)_{a\beta}
=
e^{i\chi_a}(U^\dagger V)_{a\beta}
=
e^{i\chi_a}B_{a\beta}\,.
\ea
after making use of \eqs{vprime}{uprime}.

If we consider the charged Higgs basis where $X=U$, then \eq{xuk} yields $U'=K$
and
\be
\tilde{B}^\prime
=
\tilde{U}^{\prime\,T}_R\tilde{V}^\prime
=
\tilde{U}^{\prime\,T}_R(\tilde{U}_R^T \widetilde{W}_R)\tilde{R}_c\,,
\ee
after making use of \eqs{tildeB}{Vnewbasis} with $X=U$.
The $N-1$ phases $\chi_a$ (for $a=2,3,\ldots,N$) that appear $U'=K$
(and similarly are contained in
$\tilde U_R^{\prime\,T}$) are unphysical and can be
absorbed into the definition of the charged Higgs fields $S_a^+$.
Since $W\neq U$ (unless the neutral Higgs fields are mass
eigenstates in the charged Higgs basis),
it follows that $\tilde{B}$ contains additional parameters
beyond the $N(N-1)$ physical parameters that determine the matrix $\tilde{R}_c$.
As shown in appendix~\ref{app:Bcount},
the matrix $B$ (and $\tilde{B}$) depends on an additional
$(N-1)(N-2)$ physical parameters.
That is, after absorbing the $N-1$ phases into a redefinition
of the charged Higgs fields,
there are $2(N-1)^2$  physical parameters remaining
in the matrix $B$ (and $\tilde B$).

The C2HDM provides an interesting example,
discussed in detail in  appendix~\ref{app:C2HDM}.
One sees in eq.~\eqref{beemat}
that $\tilde B$ (or equivalently $B$) depends on $3$ angles,
where one of the angles is unphysical and corresponds to the
freedom to rephase the second scalar doublet with zero vev.
In contrast,
$A$ in eq.~\eqref{amat} depends only on the two invariant angles
contained in $B$.
The case of $N=2$ is special in that the parameters that
define the matrix $A$ corresponding precisely to the physical
parameters that appear in the matrix $B$.

The $2(N-1)^2$ physical parameters contained in the matrix $B$ are
sufficient to parameterize all the gauge boson--Higgs boson interactions.
But, the three-scalar and four-scalar interactions derived from the
scalar potential necessarily involve additional parameters.
We have already emphasized that the charged Higgs basis is especially
useful in identifying the invariant (and pseudo-invariant) scalar
self-coupling coefficients.
In particular, in the charged Higgs basis, the scalar potential is given by
\be
V_H
=
Y_{ij}  ( \Phi_i^{C \dagger} \Phi^C_j )
+
Z_{ij,kl} ( \Phi_i^{C \dagger} \Phi^C_j ) ( \Phi_k^{C \dagger} \Phi^C_l ).
\label{VH_chHbasis}
\ee
The number of relevant parameters is shown in table~\ref{parameters}.
%
\begin{table}[tbp]
\centering
\begin{tabular}{|C{2cm}|C{2cm}|C{2cm}|C{2cm}|@{}m{0pt}@{}}
\hline
& parameters & magnitudes  &  phases & \\*[2mm]
\hline
$Y$ & $N^2$ & $\frac{N(N+1)}{2}$ & $\frac{N(N-1)}{2}$ & \\*[2mm]
\hline
$Z$ &  $\frac{N^2(N^2+1)}{2}$
             & $\frac{N^2(N^2+3)}{4}$ & $\frac{N^2(N^2-1)}{4}$ & \\*[2mm]
\hline
$Y$ and $Z$ & $\frac{N^2(N^2+3)}{2}$
             & $\frac{N^4+5N^2+2N}{4}$ & $\frac{N^4+N^2-2N}{4}$ & \\*[2mm]
\hline
\end{tabular}
\caption{Number of parameters in the $Y$ and $Z$ coefficients
of the Higgs potential.}
\label{parameters}
\end{table}

The number of parameters gets reduced for two reasons.
First, our definition of the charged Higgs basis allows
for a rephasing of the $N-1$ scalar doublets with zero vevs.
This reduces the number of phases by $N-1$.
In addition,
the stationarity conditions written in the charged Higgs basis
relate some $Y$ and $Z$ parameters.
More generally, eq.~\eqref{Mpm_CH} can be used to obtain
a relation of the $Y$ and $Z$ parameters with the charged scalar masses,
\be
Y_{ij} + v^2 Z_{ij,11}
= \delta_{ij}\, m^2_{\pm, i},
\ee
which can be used
to trade in the $Y_{ij}$ for the $Z$ parameters and the charged scalar masses.
Hence, using the charged Higgs basis as parameters,
we need only the $\tfrac14 N^2(N^2+3)$ magnitudes and the
$\tfrac14 N^2(N^2-1)$ phases in $Z$,
of which only $\tfrac14 N^2(N^2-1) - (N-1)$ phases are physical.

For example, in the 2HDM and in the notation of
ref.~\cite{Zs},
we find
\ba
Y_1 &=& -\tfrac{1}{2} Z_1 v^2,
\nonumber\\
Y_3 &=& -\tfrac{1}{2} Z_6 v^2,
\nonumber\\
Y_2 &=& -\tfrac{1}{2} Z_3 v^2 + m_\pm^2.
\ea
The seven magnitudes correspond to
$|Z_1|, |Z_2|,\ldots,|Z_7|$
and the two independent phases are $\Im(Z_5^\ast Z_6^2)$
and  $\Im(Z_5^\ast Z_7^2)$,
first identified in
ref.~\cite{LS} as basis invariant measures of CP violation.
Although not independent in the case
$Z_5, Z_6, Z_7 \neq 0$,
the possibility of $Z_5=0$ is only covered by considering
in addition the third invariant phase, $\Im(Z_6 Z_7^\ast)$ \cite{LS,DH}.

Many different parameter choices exist in the literature.
For example, one can employ: (i)~$m_{\pm a}$ and the quartic coefficients $Z$ of the scalar potential in the charged Higgs basis; or (ii) the quadratic coefficients $Y$ not fixed by the scalar potential minimum conditions and all of the quartic coefficients $Z$;
or (iii) the physical parameters in $B$, the physical charged and neutral
scalar masses,
and (if needed) some invariant combination of scalar self-couplings.
In extended Higgs sectors with additional symmetries (where the basis in which the symmetries are manifest becomes physical), one can also employ the various ratios of scalar vevs in the list of parameters.
For example,
in the CP-conserving 2HDM with a softly-broken $\mathbb{Z}_2$-symmetric scalar potential,
some authors choose parameters consisting of  $\beta= \tan^{-1}{(v_2/v_1)}$,
$\beta-\alpha$ [which appears in the matrix $B$ in our notation],
the charged scalar mass $m_\pm$, the neutral scalar masses
($m_h$, $m_H$, and $m_A$),
and the soft $\mathbb{Z}_2$ breaking squared-mass term $m_{12}^2$.

\subsection{\label{subsec:scalar-scalar}Scalar self-couplings}

In this section, we demonstrate that some of the scalar self-couplings are related to
the kinetic terms obtained in section~\ref{subsec:Gauge-higgs}.
In particular, it is convenient to enquire which scalar couplings
may be written exclusively in terms of the matrix $B$ (and $A$),
that appear in the kinetic terms, and the scalar masses $m_{\pm a}$ and $m_\beta$.

We first consider the cubic scalar self-couplings.
The scalar potential in eqs.~\eqref{VHiggs}--\eqref{V4}
can be expressed in terms of the physical fields using the mixing matrices
$U$ and $V$ in eqs.~\eqref{S+}--\eqref{S0}.
The cubic vertex interactions can therefore be written as
\ba
V_3  &=& \lambda_{ij,kl} (U^\dagger)_{ai} U_{jb} \left[
(V^\dagger)_{\beta k} v_l + v^*_k V_{l \beta} \right]
S^0_\beta S^-_a S^+_b
\nonumber\\
&&
 +\, \tfrac{1}{2} \lambda_{ij,kl} (V^\dagger)_{\delta i}
 V_{j \gamma} \left[ (V^\dagger)_{\beta k} v_l
 + v^*_k V_{l \beta} \right] S^0_\beta
 S^0_\gamma S^0_\delta .
\label{V3_mass}
\ea
In contrast to the terms of the interaction Lagrangian that couple the scalar and vector bosons given in \eqst{kinetic}{GG_CHCH},
additional structures appear beyond those combinations of $U$ and $V$ that define the matrices $A$ and $B$.
However, if we focus on the cubic couplings that involve at
least one Goldstone field, the form of the
cubic interaction terms simplify significantly and can be expressed in
terms of the $A$ and $B$ matrices and the squared masses of the physical
scalars.\footnote{In particular, we find that the couplings
$G^0 S^- S^+$, $G^0 G^\pm S^\mp$, $G^0 G^0 G^0$,
and $G^0 G^- G^+$ vanish.  The vanishing of the $G^0 G^0 G^0$,
and $G^0 G^- G^+$ couplings is a well-known result of the SM.}
Indeed, this is to be expected, as these interaction terms are
related by gauge-fixing to the
pure gauge boson terms arising from the kinetic Lagrangian,
which can be expressed in terms of the matrices $A$ and $B$,
as shown in \eqst{kinetic}{GG_CHCH}.

In order to simplify the non-vanishing cubic potential terms
we employ eqs.~\eqref{VTV}--\eqref{omV},
from which we obtain the further useful relations
\ba
\left[
U^\dagger V
\right]
\left[
U^\dagger V
\right]^\dagger
=
2 \newcdot \bm{1}_{N \times N}
& \ \ \Longrightarrow\ \  &
\sum_{\beta=1}^{2N} (U^\dagger V)_{a \beta} (V^\dagger U)_{\beta b}
= 2 \delta_{a b}\, ,
\label{eq:rel3}
\\
\left[
U^\dagger V
\right]^\dagger
\left[
U^\dagger V
\right]
=
\bm{1}_{N \times N} + i\, \textrm{Im} (V^\dagger V)
&\ \  \Longrightarrow\ \  &
\sum_{a=1}^{N}  (V^\dagger U)_{\alpha a} (U^\dagger V)_{a \beta}
= \delta_{\alpha \beta} + i \Im(V^\dagger V)_{\alpha \beta}\, , \nonumber \\
\phantom{line}
\ea
and
\begin{eqnarray}
(U^\dagger V)_{a 1}
&=& i \delta_{a1}\, ,
\\
(U^\dagger V)_{1 \beta}
&=&
- \Im(V^\dagger V)_{1 \beta} + i \delta_{1 \beta}\, .
\end{eqnarray}
Moreover, we must use the crucial relations given in
eqs.~\eqref{Kprime_2}--\eqref{K_2},
relating some combinations of quartic couplings with the
scalar masses. This simplifies considerably the final expressions.
The end result is,\footnote{The expressions for the cubic couplings of the Goldstone bosons and physical Higgs scalars in terms of scalar squared-masses was first obtained in the CP-conserving 2HDM in Ref.~\cite{Gunion:2002zf}.}
\begin{itemize}
\item For $S^0 G^- S^+ + S^0 S^- G^+$ :
\begin{equation} \label{eq:feyn1}
V_3 \supset \frac{1}{v}
\left[
B_{a \beta}
\left( m_\beta^2 - m_{\pm a}^2 \right) S^0_\beta S_a^- G^+
+
(B^\dagger)_{\beta a}
\left(m_\beta^2 - m_{\pm a}^2\right) S^0_\beta S^+_a G^- \right]\, .
\end{equation}
\item For $S^0 G^+ G^-$ :
\begin{equation} \label{eq:feyn2}
V_3 \supset - \frac{1}{v} G^+ G^-
m_\beta^2
A_{1 \beta}
\, S^0_\beta
\ \ \ (\beta \geq 2)\, .
\end{equation}
\item For $G^0 S^0 S^0$ :
\begin{equation} \label{eq:feyn3}
V_3 \supset \frac{1}{v} G^0
m_\beta^2
A_{\beta \gamma}
S_\beta^0 S_\gamma^0
\ \ \ (\gamma \neq \beta \geq 2)\, .
\end{equation}
\item For $G^0 G^0 S^0$ :
\begin{equation} \label{eq:feyn4}
V_3 \supset - \frac{1}{2 v} G^0 G^0
m_\beta^2
A_{1 \beta}
S_\beta^0
\ \ \ (\beta \geq 2)\,,
\end{equation}
\end{itemize}
where there is an implicit sum over repeated indices.
Achieving the simplified forms of the couplings above is rather laborious.
For example, the explicit derivation of \eq{eq:feyn4} is given in appendix~\ref{app:cubic}.

The cubic couplings that are present in the kinetic
lagrangian are also present in parallel with the scalar
potential.
A comparison is shown in table~\ref{table:FD_cubic},
where we have collected the relevant Feynman rules.
%
\begin{table}[tbp]
\centering
		\begin{tabular}{|c|c|c|c|}
		\hline
\multicolumn{2}{|c|}{Kinetic lagrangian} &
\multicolumn{2}{|c|}{Scalar potential} \\
\hline
Coupling
& Feynman rule
& Coupling
&  Feynman rule  \\*[5pt]
\hline
\rule{0pt}{3ex}
$[W^\pm G^\mp Z]$
& $ - i g M_Z s_W^2 g_{\mu \nu} $
& $[G^\pm G^\mp G^0]$
& Null \\*[5pt]
\hline
\rule{0pt}{3ex}
$[Z S^+_a S^-_a]$
& $ - \frac{ i g\, c_{2W}}{2 c_W}
(p_{a}^+ - p_{a}^-)^\mu $ , \, $\forall a$
& $[G^0 S^+_a S^-_a]$
&  Null , \, $\forall a$ \\*[5pt]
\hline
\rule{0pt}{3ex}
$[Z S^0_\beta S^0_\gamma]$
& $ \frac{g}{2 c_W}
(p_\beta^0 - p_\gamma^0)^\mu
A_{\beta \gamma} $
, \, $\forall \beta \neq \gamma$
& $[G^0 S^0_\beta S^0_\gamma]$
& $ - \frac{i}{v} (m^2_\beta  - m^2_\gamma )
A_{\beta \gamma} $
, \, $\forall \beta \neq \gamma \geq 2$ \\*[5pt]
\hline
\rule{0pt}{3ex}
$[Z G^0 S^0_\beta]$
& $ \frac{g}{2 c_W}
(p_G^0 - p_\beta^0)^\mu
A_{1 \beta} $
, \, $\forall \beta \neq 1$
& $[G^0 G^0 S^0_\beta]$
& $ \frac{i}{v} m^2_\beta
A_{1 \beta} $
, \, $\forall \beta \neq 1$ \\*[5pt]
\hline
\rule{0pt}{3ex}
$[W^+ S^-_a S^0_\beta]$
& $ \frac{i g}{2} (p_a^- - p_\beta^0)^\mu
B_{a \beta} $
, \, $\forall a, \forall \beta \neq 1$
& $[G^+ S^-_a S^0_\beta]$
& $ - \frac{i}{v} [ m_\beta^2 - m_{\pm a}^2 ]
B_{a \beta} $
, \, $\forall a, \forall \beta \neq 1$  \\*[5pt]
\hline
\rule{0pt}{3ex}
$[W^+ G^- S^0_\beta]$
& $ - \frac{i g}{2} (p_G^- - p_\beta^0)^\mu
A_{1 \beta} $
, \, $ \forall \beta \neq 1$
& $[G^+ G^- S^0_\beta]$
& $ \frac{i}{v} m^2_\beta
A_{1 \beta} $
, \, $\forall \beta \neq 1$  \\*[5pt]
\hline
\rule{0pt}{3ex}
$[W^+ G^- G^0]$
& $ - \frac{ g}{2} (p_G^- - p_G^0)^\mu $
& $[G^+ G^- G^0]$
& Null \\*[5pt]
\hline
\rule{0pt}{3ex}
$[Z Z S^0_\beta]$
& $ - \frac{i g M_Z}{c_W}
A_{1 \beta}
\, g_{\mu \nu} $
, \, $\forall \beta \neq 1$
& $[G^0 G^0 S^0_\beta]$
& $ \frac{i}{v} m^2_\beta
A_{1 \beta} $
, \, $\forall \beta \neq 1$  \\*[5pt]
\hline
\rule{0pt}{3ex}
$[W^+ W^- S^0_\beta]$
& $ - i g M_W
A_{1 \beta}
\, g_{\mu \nu} $
, \, $\forall \beta \neq 1$
& $[G^+ G^- S^0_\beta]$
& $ \frac{i}{v} m^2_\beta
A_{1 \beta} $
, \, $\forall \beta \neq 1$  \\*[5pt]
\hline
		\end{tabular}
		\caption{\label{table:FD_cubic}All cubic couplings from the kinetic
        lagrangian (except photon) and their scalar potential counterparts
        obtained by substituting \emph{all} gauge bosons by the
        corresponding Goldstone bosons.}
\end{table}
%
We notice that both sets of couplings
depend on the same parameters:
\be
A_{1 \beta} = \Im(V^\dagger V)_{1 \beta},
\ \ \
A_{\alpha \beta} = \Im(V^\dagger V)_{\alpha \beta},
\ \ \
\textrm{and}\ \ \  B_{a \beta} = (U^\dagger V)_{a \beta}.
\label{FD_couplings}
\ee
This is not surprising,
since gauge boson couplings and Goldstone boson couplings
are related by the gauge-fixing, as previously noted.
Although the equivalence theorem \cite{Veltman:1989ud}
is not a requirement on the couplings but rather
on full processes \cite{miguel},
the relation between these couplings insures
that the equivalence theorem is satisfied.

For the calculation of the quartic couplings we use the
general expression
\begin{eqnarray}
V_4
&=&
\lambda_{ij,kl} \, (U^\dagger)_{ai} \, U_{jb} \, (U^\dagger)_{ck}
\, U_{ld} \, (S^-_a S^+_b) (S^-_c S^+_d)
\nonumber\\
& &
+\, \lambda_{ij,kl} \, (U^\dagger)_{ai} \, U_{jb} \,
(V^\dagger)_{\alpha k} \,
V_{l \beta} \, (S^-_a S^+_b) (S^0_\alpha S^0_\beta)
\nonumber\\
& &
+\, \tfrac{1}{4} \lambda_{ij,kl} \, (V^\dagger)_{\alpha i} \,
V_{j \beta} \, (V^\dagger)_{\gamma k} \, V_{l \delta} \,
(S^0_\alpha S^0_\beta) (S^0_\gamma S^0_\delta) .
\end{eqnarray}
It is instructive to focus on the quartic couplings that involve the Goldstone boson fields.
After some very long simplifications,
we find the non-vanishing Goldstone-scalar quartic couplings that involve an even number of Goldstone boson fields listed below.
\begin{itemize}
\item For $G^- G^+ G^- G^+$:
\begin{equation}
\label{eq:qfeyn41}
V_4 \supset \frac{1}{2v^2} \, (G^- G^+)^2
\sum_{\beta = 2}^{2N}
m_\beta^2 \,
[A_{1 \beta}]^2
\end{equation}
\item For $G^- G^+ S^- S^+$:
\begin{equation}
\label{eq:qfeyn42}
V_4 \supset \frac{1}{v^2} \, G^- G^+ S_a^- S_b^+
\left[ \sum_{\beta=2}^{2N} m_\beta^2
B_{a\beta} (B^\dagger)_{\beta b}
- 2 Y_{a b} \right]
\end{equation}
\item For $G^- G^- S^+ S^+$ + h.c.:
\begin{equation}
V_4 \supset \frac{1}{2v^2} G^- G^- S^+_a S^+_b
(B^* D_0^2 B^\dagger)_{a b} + {\rm h.c.}
\end{equation}
\item For $G^- G^+ G^0 G^0$:
\begin{equation} 
V_4 \supset \frac{1}{2v^2} \, G^- G^+ G^0 G^0
\sum_{\beta = 2}^{2N}
m_\beta^2 \,
[A_{1 \beta}]^2
\end{equation}
\item For $G^0 G^0 G^0 G^0$:
\begin{equation}
\label{eq:qfeyn42b}
V_4 \supset \frac{1}{8v^2} G^0 G^0 G^0 G^0
\sum_{\beta=2}^{2N} \, m_\beta^2
[ A_{1 \beta} ]^2
\end{equation}
\item For $G^0 G^0 S^0 S^0$:
\begin{equation} 
V_4 \supset - \frac{1}{2v^2} G^0 G^0 S^0_\alpha S^0_\beta
\left[ \left( B^\dagger Y B \right)_{\alpha \beta} -
\left( A D_0^2 A \right)_{\alpha \beta} \right]
\end{equation}
\item For $G^- G^+ S^0 S^0$:
\begin{equation}
\label{eq:qfeyn43}
V_4 \supset \frac{1}{v^2} \, G^- G^+ S_\alpha^0 S_\beta^0
\left[
(B^\dagger\,  D_{\pm}^2\, B)_{\alpha \beta}
- (B^\dagger Y\, B)_{\alpha \beta} \right]
\end{equation}
\item For $G^0 G^0 S^- S^+$:
\begin{equation}
\label{eq:qfeyn44}
V_4 \supset \frac{1}{v^2} \, G^0 G^0 S_a^- S_b^+
\left[ (D_{\pm}^2)_{a b}
 - Y_{a b} \right]
\end{equation}
\item For $G^- G^0 S^0 S^+$ + h.c.:
\begin{equation}
\label{eq:qfeyn45}
V_4 \supset - \frac{1}{ v^2} G^- S^+_b G^0 S^0_\beta\,
\left[
 i (B^\dagger D_\pm^2)_{\beta b}  + (A D_0^2 B^\dagger)_{\beta b} \right]
 + {\rm h.c.}
\end{equation}
\end{itemize}
The corresponding expressions involving an odd number of Goldstone
fields are more complicated and will
not be given here.

In contrast to the cubic couplings given in \eqst{eq:feyn1}{eq:feyn4},
additional structures beyond
the matrices $A$, $B$ and the squared masses of the physical scalars
appear in the expressions above.  For example, the quadratic coefficient
of the scalar potential in the charged Higgs basis, $Y$, appears
on the right hand side of eq.~\eqref{eq:qfeyn42}.
Note that there is no simple relation between quartic couplings in the
kinetic Lagrangian and quartic terms from the scalar potential.
For example,
as in the SM,
there is a ${(G^0)}^4$ coupling,
but no $Z^4$ coupling.
This does not violate the equivalence theorem since the processes involving
quartic couplings typically involve other diagrams.  In particular, only
the sum of all contributing diagrams must obey the equivalence
principle.

\section{Sum rules}
\label{sec:sumrules}

\subsection{Coupling relations and sum rules from the Lagrangian}

One can obtain a large number of
sum rules from the kinetic part of the Lagrangian and
 the scalar potential of the NHDM.  In the case of the 2HDM,
two of the sum rules that are usually exhibited
(see, e.g.,
refs.~\cite{Gunion:1990kf, Grinstein:2013fia, abdss, Arhrib:2015gra})
are
\begin{eqnarray}
\sum_k [S^0_k V V]^2
&=&
1,
\label{S0k V V}
\nonumber\\
{}
[S^0_k V V]^2 + \left| [S^0_k  W^\mp H^\pm] \right|^2
&=&
1\ \ (\textrm{any } k),
\end{eqnarray}
where $k$ identifies some specific neutral scalar
physical field.
It is easy to find the corresponding sum rules
in the general NHDM:
\begin{eqnarray}
\label{eq:1stsum}
\sum_\beta [S^0_\beta V V]^2 &=& 1,
\\
{}
[S^0_\beta V V]^2 + \sum_{b=2}^N
\left| [S^0_\beta  W^\pm S^\mp_b] \right|^2
&=& 1\ \ (\textrm{for } \beta>1)  \, ,
\label{S0betaVV}
\end{eqnarray}
where the indices follow the same notation as above.
In this section we use a simplified notation for the couplings,
strictly related to the matrices $A$ and $B$,
in which some coupling $[X_a Y_b Z_c]$ is identified as the term in the Lagrangian
that depends explicitly on family type indices.
For example,
in the Lagrangian term
\be
{\cal L} \supset C_1\, f(a,b,c) X_a Y_b Z_c,
\ee
involving the fields $X_a$, $Y_b$, $Z_c$
and the constant $C_1$, we identify
$[X_a Y_b Z_c]=f(a,b,c)$.
In cases where the corresponding coupling also exists in the SM,
this procedure means simply that we have divided out by the SM
coupling $C_1$.

In addition to these,
we have found many sum rules for
an arbitrarily extended scalar doublet sector.
For example,
\begin{equation}
\left| [S_\beta^0 W^\pm G^\mp] \right|^2 = [S_\beta^0 VV]^2
\ \ (\textrm{for } \beta>1)  \, .
\label{couprel1}
\end{equation}
Using this result in eq.~\eqref{S0betaVV},
and recalling that $S_1^\pm = G^\pm$,
we find
\begin{eqnarray}
\label{eq:2ndsum}
 \sum_{b=1}^N \left| [S^0_\beta  W^\pm S^\mp_b ] \right|^2
&=& 1\ \ (\textrm{any } \beta)  \, ,
\\
\sum_{\beta=1}^{2N} [W_\mu^+ S_a^- S_\beta^0]
[Z_\mu S^0_\beta S^0_\gamma]
&=&
-i [W_\mu^+ S_a^- S_\gamma^0]\ \ (\textrm{any } a, \gamma)  \, .
\end{eqnarray}
Further equations arising from the kinetic
Lagrangian are
\begin{eqnarray}
\left| [S^0_\beta W^\pm S_b^\mp] \right|^2
&=&
\left| [S^0_\beta W^\pm S_b^\mp Z] \right|^2
\ \ (\textrm{any } b, \beta)\, ,
\label{couprel2}
\\
\label{eq:surul3}
\sum_{\beta=1}^{2N} [Z S^0_\alpha S^0_\beta]^2 &=& 1,
\\
\sum_{\alpha \neq \beta=1}^{2N} [Z S^0_\alpha S^0_\beta]^2
&=& 2N .
\end{eqnarray}
The relations among different couplings in the kinetic
Lagrangian imply that one can write eq.~\eqref{eq:surul3}
in terms of massive fields (i.e., without the Goldstone bosons) as
\begin{equation} \label{eq:surul4}
[S^0_\alpha VV]^2 + \sum_{\beta=2}^{2N}
[Z S^0_\alpha S^0_\beta]^2 = 1 .
\end{equation}
We have checked that all our relations are verified
when using the couplings in the special case of the C2HDM.
Notice that eqs.~\eqref{couprel1} and \eqref{couprel2} are not proper sum rules,
but rather relations among couplings valid in our particular
model.

These sum rules have been found by employing the NHDM interaction Lagrangian.
But we know that sum rules can also be found for generic
Lagrangians by using unitarity arguments,
as in refs.~\cite{Gunion:1990kf,Grinstein:2013fia}.
The sum rules we have derived in eqs.~\eqref{eq:1stsum} and
\eqref{eq:2ndsum}--\eqref{couprel2} can also be found in that fashion.
We will revisit this question in section~\ref{subsec:compare} and in
appendix~\ref{app:SR}.

In the quartic couplings sector,
we have found further interesting sum rules.
For example,
using the couplings in eq.~\eqref{eq:qfeyn42},
we find\footnote{Note that $\trace[\mu]=\trace[Y]$ is a basis-invariant quantity.}
\begin{equation} \label{eq:rule42}
\sum_{a=1}^N \, [G^- G^+ S^-_a S^+_a] =
\trace[D_0^2] - 2 \trace[Y] \, ,
\end{equation}
which has the peculiar feature that a sum of couplings
involving solely charged particles
yields a contribution that depends on the masses of the neutral fields.
Similarly,
using the couplings in eq.~\eqref{eq:qfeyn43},
we find
\begin{equation} \label{eq:rule43}
\sum_{\beta=1}^{2N} \, [G^- G^+ S^0_\beta S^0_\beta] =
2 \trace[D_\pm^2] - 2 \trace[Y] \, .
\end{equation}
We can recombine a number of these sum rules as
\begin{equation} \label{eq:rule44}
\sum_{a=1}^N \, [G^- G^+ S^-_a S^+_a]
- 2 \sum_{b=1}^N \, [G^0 G^0 S^-_b S^+_b] =
\trace[D_0^2] - 2 \trace[D_\pm^2] \, ,
\end{equation}
or
\begin{equation} \label{eq:rule45}
\sum_{a=1}^N \, [G^- G^+ S^-_a S^+_a]
- \sum_{\alpha=1}^{2N} \, [G^- G^+ S^0_\alpha S^0_\alpha] =
\trace[D_0^2] - 2 \trace[D_\pm^2] \, ,
\end{equation}
because the following equation is satisfied,
\begin{equation} \label{eq:rule46}
2 \sum_{a=1}^N \, [G^0 G^0 S^-_a S^+_a]
= \sum_{\alpha=1}^{2N} \, [G^- G^+ S^0_\alpha S^0_\alpha] .
\end{equation}
The sum rules involving quartic couplings
presented in eqs.~\eqref{eq:rule42}--\eqref{eq:rule46}
are new.

\subsection{\label{subsec:compare}Sum rules from unitarity}

The constraints from unitarity have had an historical impact
on the development of particle physics.
The idea is that the scattering among vector bosons and/or
scalars cannot grow with energy and must
obey the optical theorem.
Imagine that one writes the most general effective couplings
between these states, up to dimension four.
Forcing the sum of all terms growing like the fourth power of the
center of mass energy ($E$) to vanish immediately restricts the
vector bosons to originate from a gauge theory
\cite{Cornwall:1974km, LlewellynSmith:1973yud}.
Requiring that all $E^2$ terms vanish forces the scalar-gauge
couplings to originate from a gauge theory and further constrains
the couplings \cite{Weldon:1984wt, Gunion:1990kf}.
But unitarity also limits the value of the constant
in the $E^0$ terms.
This has been used in the past to place limits on
(combinations) of the scalar masses and couplings
\cite{Lee:1977yc, Lee:1977eg, Weldon:1984wt,Grinstein:2013fia}.

In the previous section we derived the coupling relations
\eqref{couprel1} and \eqref{couprel2} and many sum rules
directly from the Lagrangian of the NHDM.
We have checked that most sum rules involving triple couplings
can be obtained from those presented in ref.~\cite{Gunion:1990kf},
by applying them to the NHDM and cycling through all possible indices.
The exception arises in applying eqs.~\eqref{eq:1stsum} and \eqref{S0betaVV}
for the case of $VV=ZZ$,
which are presented in ref.~\cite{Gunion:1990kf} under the assumption of
CP conservation.
For completeness, we include in appendix~\ref{app:SR} a
careful and detailed derivation of the sum rules given in
eqs.~(2.4)--(2.6) of ref.~\cite{Gunion:1990kf}.
The derivation of these sum rules does not make use of
CP conservation.
Nevertheless,
when applied to models with scalars in their section IV,
the authors of ref.~\cite{Gunion:1990kf} focus on
models that conserve CP.
In particular,
they point out that requiring
CP conservation, $m_W^2 = m_Z^2 c_W^2$,
and $[W^+ Z S^-_a] = 0$ leads,
through their eq.~(4.5),
to $[Z Z S_\beta^0] = [W^+ W^- S_\beta^0]$,
thus turning sum rules that involve
$[W^+ W^- S_\beta^0]$ into sum rules involving
$[Z Z S_\beta^0]$ in the case of a CP conserving Higgs sector.
In contrast,
the general NHDM may (or not) violate CP; nevertheless,
the same sum rules apply.
We include
in appendix~\ref{app:ZZS=WWS} a proof that
$[Z Z S_\beta^0] = [W^+ W^- S_\beta^0]$,
which makes no assumption concerning CP conservation, and yet is
valid for scalars in any representation of $SU(2)_L$,
provided that the theory satisfies  $m_W^2 = m_Z^2 c_W^2$ without fine tuning
of the various vevs.

\section{Tree-level unitarity bounds}
\label{sec:unitaritybounds}

In this section,
we present an algorithm for the determination of the tree-level unitarity
bounds.
We check that it reproduces results available in the literature,
and present its application to two new cases with three Higgs doublets.
Finally,
the techniques presented here can be used for faster numerical
implementation of unitarity bounds in more complicated cases,
not amenable to closed-form solutions.

The problem of finding the constraints imposed by tree-level unitarity
has been addressed in the case of the 2HDM, both with $\mathbb{Z}_2$
symmetry~\cite{Ginzburg:2003fe} and also for most general
case~\cite{Ginzburg:2005dt, Kanemura:2015ska}.
For the 2HDM with a $\mathbb{Z}_2$ symmetric scalar potential, the results are
simple, but for the most general case one has to compute the eigenvalues
of $4\times 4$ and $3\times 3$ matrices~\cite{Ginzburg:2005dt, Kanemura:2015ska}.
For the case of the 3HDM a
solution is known for the case of $S_3$ symmetry~\cite{Das:2014fea}
where because of the symmetry the solutions are again simple, although
the method to obtain them is already quite complicated.

Clearly, for the NHDM one needs an algorithm that can be easily
implemented numerically.
As explained in the references above,
since one is interested in the high energy limit,
one just needs to evaluate the scattering
$S$-matrix for the two body scalar bosons, and these
arise exclusively from the
quartic part of the potential, $V_4$.
Then, the first part of the problem consists in finding the
set of two body states that can contribute.
For the cases of low $N$ and high symmetry we can choose conveniently
the sets to take advantage of
this~\cite{Ginzburg:2003fe,Ginzburg:2005dt}, but if we are going to
solve the problem numerically it is better to have a simple
algorithm of general applicability.
Since the electric charge and the hypercharge are conserved
in this high energy scattering, we can separate the states according
to these quantum numbers. For this it is better not to separate the
real and imaginary parts of the neutral components, using the following
notation for Higgs doublets
\begin{equation}
  \label{equnit:3}
  \Phi=
  \begin{bmatrix}
    w^+_i\\
    n_i
  \end{bmatrix}\, ,
\quad
  \Phi^\dagger=
  \begin{bmatrix}
    w^-_i\\
    n^*_i
  \end{bmatrix}^T\, .
\end{equation}
The relevant two body states are given in the
entries of table~\ref{tab:unitarity1},
and their complex conjugates.
\begin{table}[h!]
  \centering
  \begin{tabular}{|c|c|l|l|}\hline
    $Q$&$2\mathcal{Y}$&State&Number of states\\[+1mm]\hline\hline
2 &2 &$S_{\alpha}^{++}=w_i^+ w_j^+$, ($i=1,N,j=i,N$) &
$\alpha=1,\ldots,\frac{1}{2} N(N+1)$  \\[+1mm]\hline\hline
1 &2 & $S_{\alpha}^{+}=w_i^+ n_j$, ($i=1,N,j=1,N$)
&$\alpha=1,\ldots,N^2$ \\[+1mm]\hline
1 &0 & $T_{\alpha}^{+}=w_i^+ n^*_j$, ($i=1,N,j=1,N$)
&$\alpha=1,\ldots,N^2$ \\[+1mm]\hline\hline
0 &2 & $S_{\alpha}^{0}=n_i n_j$, ($i=1,N,j=i,N$)
& $\alpha=1,\ldots,\frac{1}{2} N(N+1)$ \\[+1mm]\hline
0 &0 &$T_{\alpha}^{0}=\{w_i^- w_j^+,n_i n^*_j\}$, ($i=1,N,j=1,N$)
&$\alpha=1,\ldots,2\,N^2$  \\[+1mm]\hline
  \end{tabular}
  \caption{List of two body scalar states separated by $(Q,\mathcal{Y})$.}
  \label{tab:unitarity1}
\end{table}

\noindent
As an example,
for $N=2$ we have,
\begin{align}
S_{\alpha}^{++}=&\{w_1^+ w_1^+,w_1^+ w_2^+,w_2^+ w_2^+\}, \label{equnit:8a}\\
S_{\alpha}^{+}=&\{w_1^+ n_1,w_1^+ n_2,w_2^+ n_1,w_2^+ n_2\}, \label{equnit:8b}\\
T_{\alpha}^{+}=&\{w_1^+ n^*_1,w_1^+ n^*_2,w_2^+ n^*_1,w_2^+ n^*_2\}, \label{equnit:8c}\\
S_{\alpha}^{0}=&\{n_1 n_1,n_1 n_2,n_2 n_2\}, \label{equnit:8d}\\
T_{\alpha}^{0}=&\{w_1^- w_1^+,w_1^- w_2^+,w_2^- w_1^+,w_2^- w_2^+,n_1
n^*_1,n_1 n^*_2,n_2 n^*_1,n_2 n^*_2\}, \label{equnit:8e}
\end{align}
plus their complex conjugates.
It is important to note that the index
$\alpha$ is a compound index;
it refers to a set of $\{i,j\}$ indices.
Also note that in eqs.~\eqref{equnit:8a} and \eqref{equnit:8d} the
two body states with equal particles have a normalization of
$1/\sqrt{2}$ that we have not written here (see below).

Now we separate the different partial wave amplitudes for the
different charges and hypercharges. We have
\begin{align}
  \label{equnit:5}
16\pi \left(a_{0,\mathcal{Y}=1}^{++}\right)_{\alpha\beta}
\equiv& \left(M^{++}_2\right)_{\alpha\beta}
\equiv \frac{\partial^2 (V_4)}{\partial S_{\alpha}^{--}\, \partial S_{\beta}^{++}}
&& \left[\frac{N(N+1)}{2}\times\frac{N(N+1)}{2}\right],
\\
16\pi \left(a_{0,\mathcal{Y}=1}^{+}\right)_{\alpha\beta}
\equiv& \left(M^{+}_2\right)_{\alpha\beta}
\equiv \frac{\partial^2 (V_4)}{\partial S_{\alpha}^{-}\, \partial   S_{\beta}^{+}}
&& \left[N^2\times N^2\right],
\\
16\pi \left(a_{0,\mathcal{Y}=0}^{+}\right)_{\alpha\beta}
\equiv& \left(M^{+}_0\right)_{\alpha\beta}
\equiv \frac{\partial^2 (V_4)}{\partial T_{\alpha}^{-}\, \partial T_{\beta}^{+}}
&& \left[N^2\times N^2\right],
\\
16\pi \left(a_{0,\mathcal{Y}=1}^{0}\right)_{\alpha\beta}
\equiv& \left(M^{0}_2\right)_{\alpha\beta}
\equiv \frac{\partial^2 (V_4)}{\partial S_{\alpha}^{0}\, \partial S_{\beta}^{0}}
&& \left[\frac{N(N+1)}{2}\times\frac{N(N+1)}{2}\right],
\\
16\pi \left(a_{0,\mathcal{Y}=0}^{0}\right)_{\alpha\beta}
\equiv& \left(M^{0}_0\right)_{\alpha\beta}
\equiv \frac{\partial^2 (V_4)}{\partial T_{\alpha}^{0}\, \partial  T_{\beta}^{0}}
&& \left[2N^2\times 2N^2\right],
\end{align}
where we have indicated on the right-hand side the dimensionality of
the resulting matrices.
The compound nature of the index $\alpha$ should be taken in account.
For instance
\begin{equation}
  \label{equnit:6}
  \frac{\partial V_4}{\partial S_{\alpha}^{--}} = N_{ij} \frac{\partial^2
    V_4}{\partial w_i^-\ \partial w_j^-}
\end{equation}
where the set $\{i,j\}$ corresponds to $\alpha$ and the normalization
$N_{ij}$ is $1/\sqrt{2}$ for the 2 body states with equal particles
and 1 in all the other cases. This factor can be understood in the
following way. When we take the derivative with respect to the 2 body
state with equal particles we should divide by the normalization
$1/\sqrt{2}$. But on the right-hand side we
are taking derivatives with respect to the individual fields. To avoid
double counting we should divide by two in the case of identical
particles, so $\frac{1}{2}(\frac{1}{\sqrt{2}})^{-1} =
\frac{1}{\sqrt{2}}$.

This procedure can be
easily implemented in an algebraic program like \texttt{Mathematica}
and we can easily obtain the five matrices
$M^{++}_2,M^{+}_2,M^{+}_0,M^{0}_2,M^{0}_0$.
For simple cases we can obtain the eigenvalues;
for the more complicated cases we can obtain the characteristic equation
and solve it numerically for the eigenvalues.

For illustration, consider the case of the 2HDM. The quartic part of the potential contains
\begin{align}
  \label{equnit:4}
  V_4 \supset &
\frac{1}{2} \lambda_1 (w_1^-)^2 (w_1^+)^2
+\frac{1}{2} \lambda_2  (w_2^-)^2 (w_2^+)^2
+\lambda_3 w_2^- w_1^- w_1^+  w_2^+
+\lambda_4 w_2^- w_1^- w_1^+  w_2^+\nonumber\\
&+\frac{1}{2} \lambda_5 (w_1^-)^2 (w_2^+)^2
+\frac{1}{2} \lambda_5^* (w_2^-)^2 (w_1^+)^2
+\lambda_6  (w_1^-)^2 w_1^+ w_2^+
+\lambda_6^* w_2^- w_1^-  (w_1^+)^2 \nonumber\\
&+\lambda_7 w_2^- w_1^-  (w_2^+)^2
+\lambda_7^* (w_2^-)^2 w_1^+ w_2^+
\end{align}
Using the procedure described above one can easily get for $M_2^{++}$,
\begin{equation}
  \label{equnit:8}
  M_2^{++}=
  \begin{bmatrix}
    \lambda_1 & \quad \sqrt{2}\lambda_6&\quad \lambda_5\\
    \sqrt{2}\lambda_6^* &\quad  \lambda_3+\lambda_4&\quad \sqrt{2} \lambda_7 \\
    \lambda_5^* &\quad \sqrt{2}\lambda_7^* &\quad \lambda_2
  \end{bmatrix},
\end{equation}
which coincides with the result of ref.~\cite{Kanemura:2015ska},
up to an interchange of rows and columns.

We have applied this procedure for the known cases in the literature
and, to illustrate the power of the method, we also present two new cases.

\subsection{An (almost) trivial case: the Standard Model}

Consider the Standard Model with $n_H=1$.
With the conventions of ref.~\cite{Romao:2012pq}, we have
\begin{equation}
  \label{eq:SM9}
  V_4= \lambda \left(\Phi^\dagger \Phi\right)^2,\quad \text{where}\quad
 \lambda = \frac{g^2}{8 m_W^2} m_H^2 = \frac{m_H^2}{2 v^2}\ .
\end{equation}
In this case the $M$ matrices reduce to
\begin{equation}
  \label{eq:SM11}
  M^{++}_2 =[2 \lambda],\
  M^{+}_2 =[2 \lambda],\
  M^{+}_0 =[2 \lambda],\
  M^{0}_2 =[2 \lambda],\
  M^{0}_0 =
  \begin{bmatrix}
    4\lambda&\quad 2\lambda\\
    2\lambda& \quad 4\lambda
  \end{bmatrix}\ .
\end{equation}
There are therefore two independent eigenvalues,
\begin{equation}
  \label{eq:SM12}
  \Lambda_1= 2\lambda,\quad \Lambda_2=6\lambda\ .
\end{equation}
Applying partial wave unitarity to the eigenvalues of the $s$-wave amplitude matrix, which is a consequence of the optical theorem,
\be \label{pwu}
|a_0|^2\leq |{\rm Im}~a_0|\leq 1\,.
\ee
\Eq{pwu} implies that
\be \label{pwu2}
({\rm Re}~a_0)^2\leq |{\rm Im}~a_0|(1-|{\rm Im}~a_0|)\,.
\ee
Since the right hand side of \eq{pwu2} is bounded by $\tfrac14$, it follows that~\cite{Luscher:1988gc},
\begin{equation}
  \label{eq:SM13}
|{\rm Re}~a_0| \leq \tfrac{1}{2}\ ,
\end{equation}
which translates into
\begin{equation}
  \label{eq:SM14}
  \Lambda_2 \leq 8\pi \ .
\end{equation}
This in turn implies
\begin{equation}
  \label{eq:SM15}
  \lambda \leq \frac{4\pi}{3}, \quad \text{or}\quad m_H \leq
  \sqrt{\frac{8\pi}{3}}\, v \simeq 712\, \text{GeV}\ ,
\end{equation}
a well known result~\cite{Lee:1977yc,Lee:1977eg}.

\subsection{Other known cases}

We applied the method to the complex 2HDM with $\mathbb{Z}_2$ symmetry and we
recover the results of
ref.~\cite{Ginzburg:2003fe,Ginzburg:2005dt}. Next we studied the
general complex 2HDM and we also agree with the results of
ref.~\cite{Ginzburg:2005dt,Kanemura:2015ska}. Both these cases are for
$N=2$. A more 
complicated case is the 3HDM with $S_3$ symmetry. The matrices are
larger but we were able to recover all the results of ref.~\cite{Das:2014fea}.
This makes us confident that the procedure can be generalized to cases
where the unitarity constraints have not been studied.
We consider two new cases.

\subsection{The 3HDM with order-4 CP symmetry}

Consider the potential of the 3HDM with
order-4 CP symmetry given in ref.~\cite{Ivanov:2015mwl},
\begin{align}
  \label{equnit:1}
V =& - m_{11}^2 (\Phi_1^\dagger \Phi_1) - m_{22}^2 (\Phi_2^\dagger
       \Phi_2 + \Phi_3^\dagger \Phi_3)
+ \lambda_1 (\Phi_1^\dagger \Phi_1)^2 + \lambda_2 \left[(\Phi_2^\dagger \Phi_2)^2 + (\Phi_3^\dagger \Phi_3)^2\right]
\nonumber\\
&+ \lambda_3 (\Phi_1^\dagger \Phi_1) (\Phi_2^\dagger \Phi_2 + \Phi_3^\dagger \Phi_3)
+ \lambda'_3 (\Phi_2^\dagger \Phi_2) (\Phi_3^\dagger \Phi_3)
+ \lambda_4 \left[(\Phi_1^\dagger \Phi_2)(\Phi_2^\dagger \Phi_1) +
  (\Phi_1^\dagger \Phi_3)(\Phi_3^\dagger \Phi_1)\right]\nonumber\\
&+ \lambda'_4 (\Phi_2^\dagger \Phi_3)(\Phi_3^\dagger \Phi_2)
+ \left[\lambda_5 (\Phi_3^\dagger\Phi_1)(\Phi_2^\dagger\Phi_1)
+ {\frac{\lambda_6}{2}}\left[(\Phi_2^\dagger\Phi_1)^2 -
   (\Phi_1^\dagger\Phi_3)^2\right]\right. \nonumber\\
&\left.\vphantom{\frac{\lambda_6}{2}}
+ \lambda_8(\Phi_2^\dagger \Phi_3)^2 +
   \lambda_9(\Phi_2^\dagger\Phi_3)(\Phi_2^\dagger\Phi_2-\Phi_3^\dagger\Phi_3)
   + {\rm h.c.}\right],
\end{align}
with all parameters real except for $\lambda_8$, $\lambda_9$ that are
complex. Applying the method we get the following distinct eigenvalues
for the $M$ matrices,
\begin{align}
\Lambda_{1-2}=&\lambda_3\pm\lambda_4\, ,
\\
\Lambda_{3}=&\lambda'_3-\lambda'_4\, ,
\\
\Lambda_{4-5}=&\lambda_3 \pm \sqrt{\lambda_5^2+\lambda_6^2}\, ,
\\
\Lambda_{6-7}=&\lambda_3+2 \lambda_4 \pm 3 \sqrt{\lambda_5^2+\lambda_6^2}\, ,
\\
\Lambda_{8-9}=&\frac{1}{2} \left(2 \lambda_1+2 \lambda_2+\lambda_4'
\pm \sqrt{4 \lambda_1^2-8 \lambda_1 \lambda_2-4 \lambda_1 \lambda_4'+4
 \lambda_2^2+4 \lambda_2  \lambda_4'+8 \lambda_4^2+\lambda_4'^2}\right),
 \\
\Lambda_{10-11}=& 3 \lambda_1+3 \lambda_2+ \lambda_3'
+\frac{\lambda_4'}{2}
\\
&
\pm \sqrt{(-3 \lambda_1-3
  \lambda_2- \lambda_3'-\frac{\lambda_4'}{2})^2- \left(36
  \lambda_1 \lambda_2+12
   \lambda_1 \lambda_3'+6 \lambda_1 \lambda_4'-8 \lambda_3^2-8
   \lambda_3 \lambda_4-2 \lambda_4^2\right)} \nonumber\\
\Lambda_{12-14}=&\text{Roots of:}\nonumber\\
&x^3+x^2 (-2 \lambda_2-2 \lambda_3'+\lambda_4')+x \left(-4
   \lambda_8 \lambda_8^*-4 \lambda_9 \lambda_9^*+4 \lambda_2
  \lambda_3'+\lambda_3'^2-2 \lambda_3'   \lambda_4'\right)\nonumber\\
&
+8 \lambda_2
  \lambda_8 \lambda_8^*+4 \lambda_3' \lambda_9 \lambda_9^*-4
  \lambda_4'
   \lambda_8 \lambda_8^*-4 \lambda_9^2 \lambda_8^*-4 \lambda_8
  \left(\lambda_9^*\right)^2-2 \lambda_2
   \lambda_3'^2+\lambda_3'^2 \lambda_4'=0\, ,
\\
\Lambda_{15-17}=&\text{Roots of:}\nonumber\\
&x^3+x^2 (-6 \lambda_2-3 \lambda_4')+x \left(-36 \lambda_8
   \lambda_8^*-36 \lambda_9 \lambda_9^*+12 \lambda_2 \lambda_3'+24
  \lambda_2 \lambda_4'-3 \lambda_3'^2-6
   \lambda_3' \lambda_4'\right)\nonumber\\
+&216 \lambda_2 |\lambda_8|^2 \!-\! 72 \lambda_3' |\lambda_8|^2+36
   \lambda_3' |\lambda_9|^2-36 \lambda_4' |\lambda_8|^2
+72 \lambda_4' |\lambda_9|^2\!-\!108 \lambda_9^2 \lambda_8^*
\!-\!108 \lambda_8
  \left(\lambda_9^*\right)^2\nonumber\\
-&6 \lambda_2
   \lambda_3'^2-24 \lambda_2 \lambda_3' \lambda_4'-24 \lambda_2
  \lambda_4'^2+2 \lambda_3'^3+9 \lambda_3'^2
   \lambda_4'+12 \lambda_3' \lambda_4'^2+4 \lambda_4'^3=0\, ,
\\
\Lambda_{18-21}=& \text{Roots of:}\nonumber\\
&x^4+ x^3 (-2 \lambda_1-4 \lambda_2-\lambda_3'-\lambda_4')+x^2
   \left(-4 \lambda_8 \lambda_8^*-4 \lambda_9 \lambda_9^*+8 \lambda_1
  \lambda_2+2 \lambda_1 \lambda_3'\right.\nonumber\\
&\left.
+2 \lambda_1 \lambda_4'+4 \lambda_2^2+4 \lambda_2 \lambda_3'+4
  \lambda_2 \lambda_4'-2 \lambda_5^2-2
\lambda_6^2\right)
+x \left(8 \lambda_1 \lambda_8 \lambda_8^*+8
  \lambda_1 \lambda_9 \lambda_9^*\right.\nonumber\\
&
+8 \lambda_2 \lambda_9 \lambda_9^*+4 \lambda_3' \lambda_8
  \lambda_8^*+4 \lambda_4' \lambda_8 \lambda_8^*
-4 \lambda_5 \lambda_6 \lambda_9^*+2
\lambda_6^2 \lambda_8^*+4
  \lambda_9^2 \lambda_8^*+4 \lambda_8
   \left(\lambda_9^*\right)^2\nonumber\\
&
-8 \lambda_1 \lambda_2^2-8 \lambda_1
  \lambda_2 \lambda_3'-8 \lambda_1 \lambda_2
   \lambda_4'-4 \lambda_2^2 \lambda_3'-4 \lambda_2^2 \lambda_4'+8
  \lambda_2 \lambda_5^2
+4 \lambda_2 \lambda_6^2+2 \lambda_3' \lambda_6^2\nonumber\\
&\left.
+2 \lambda_4' \lambda_6^2-4   \lambda_5 \lambda_6 \lambda_9
+2 \lambda_6^2 \lambda_8\right)\nonumber\\
&-16 \lambda_1 \lambda_2 \lambda_9 \lambda_9^*-8
  \lambda_1 \lambda_3' \lambda_8
   \lambda_8^*-8 \lambda_1 \lambda_4' \lambda_8 \lambda_8^*-8
  \lambda_1 \lambda_9^2 \lambda_8^*-8 \lambda_1
   \lambda_8 \left(\lambda_9^*\right)^2 \nonumber \\
&+8 \lambda_2 \lambda_5 \lambda_6 \lambda_9^*
-2 \lambda_3' \lambda_6^2  \lambda_8^*
-2 \lambda_4' \lambda_6^2 \lambda_8^*
+8 \lambda_5^2 \lambda_8 \lambda_8^*
+8  \lambda_5 \lambda_6   \lambda_9 \lambda_8^*+8 \lambda_5
\lambda_6 \lambda_8 \lambda_9^*\nonumber \\
&-2 \lambda_6^2  \left(\lambda_9^*\right)^2
+4 \lambda_6^2 \lambda_9 \lambda_9^*
+8  \lambda_1 \lambda_2^2 \lambda_3'
+8  \lambda_1 \lambda_2^2 \lambda_4'
-8 \lambda_2^2 \lambda_5^2
-4  \lambda_2 \lambda_3'
\lambda_6^2
-4 \lambda_2  \lambda_4' \lambda_6^2\nonumber \\
&
+8 \lambda_2 \lambda_5 \lambda_6 \lambda_9
-2  \lambda_3' \lambda_6^2 \lambda_8
-2  \lambda_4' \lambda_6^2 \lambda_8
-2 \lambda_6^2 \lambda_9^2=0\, .
\end{align}

\subsection{The 3HDM with $\mathbb{Z}_3$ symmetry}

As a second example,
we consider the potential for the 3HDM with a $\mathbb{Z}_3$
symmetry~\cite{Ferreira:2008zy}, that we write in the form
\begin{align}
  \label{equnit:2}
V=&a_1 (\Phi_1^\dagger \Phi_1) + a_2 (\Phi_2^\dagger \Phi_2)
+ a_3 (\Phi_3^\dagger \Phi_3)\nonumber\\
&+r_{1} \left(\Phi_1^\dagger \Phi_1\right)^2
  +r_{2} \left(\Phi_2^\dagger \Phi_2\right)^2
  +r_{3} \left(\Phi_3^\dagger \Phi_3\right)^2
  +2\, r_{4} \left(\Phi_1^\dagger \Phi_1\right)\left( \Phi_2^\dagger \Phi_2\right)
\nonumber\\
  &+2\, r_{5} \left(\Phi_1^\dagger \Phi_1\right)\left( \Phi_3^\dagger
    \Phi_3\right)
+2\, r_{6}\left(\Phi_2^\dagger \Phi_2\right)\left( \Phi_3^\dagger \Phi_3\right)
+2\, r_{7} \left(\Phi_1^\dagger \Phi_2\right)\left( \Phi_2^\dagger \Phi_1\right)
\nonumber\\
&+2\, r_{8} \left(\Phi_1^\dagger \Phi_3\right)\left( \Phi_3^\dagger \Phi_1\right)
+2\, r_{9} \left(\Phi_2^\dagger \Phi_3\right)\left( \Phi_3^\dagger \Phi_2\right)
\nonumber\\
&+ \left[\vb{14}
2\, c_{4}
\left(\Phi_1^\dagger \Phi_2\right)\left( \Phi_1^\dagger \Phi_3\right)
+2\, c_{12}
\left(\Phi_1^\dagger \Phi_2\right)\left( \Phi_3^\dagger \Phi_2\right)
+2\, c_{11}
\left(\Phi_1^\dagger \Phi_3\right)\left( \Phi_2^\dagger \Phi_3\right)
+ {\rm h.c.}
\right]\, ,
\end{align}
where the parameters $a_i$ and $r_i$ are real,
while $c_i$ are complex.
We get the following eigenvalues of the $M$ matrices:
\begin{align}
\label{equnit:7}
\Lambda_{1}=&2 (r_4 - r_7)\, ,
\\
\Lambda_{2}=&2 (r_5 - r_8)\, ,
\\
\Lambda_{3}=&2 (r_6 - r_9)\, ,
\\
\Lambda_{4-5}=&r_1+r_6+r_9
\pm \sqrt{8 |c_4|^2+r_1^2-2 r_1 r_6-2
 r_1 r_9+r_6^2+2 r_6 r_9+r_9^2}\, ,
 \\
\Lambda_{6-7}=&r_2+r_5+r_8
\pm \sqrt{8 |c_{12}|^2+r_2^2-2 r_2
r_5-2 r_2 r_8+r_5^2+2  r_5
   r_8+r_8^2}\, ,
\\
\Lambda_{8-9}=&r_3+r_4+r_7 \pm
\sqrt{8 |c_{11}|^2+r_3^2-2 r_3 r_4-2
r_3 r_7+r_4^2+2 r_4 r_7+r_7^2}\, ,
\\
\Lambda_{10-12}=&
\text{Roots of:}\nonumber\\
&x^3+x^2 (-2 r_1-2 r_2-2 r_3)+x \left(4
   r_1 r_2+4 r_1 r_3+4 r_2 r_3-4 r_7^2-4 r_8^2-4
   r_9^2\right)\nonumber\\
&-8 r_1 r_2 r_3+8 r_1 r_9^2+8 r_2 r_8^2+8 r_3
   r_7^2-16 r_7 r_8 r_9=0\, ,
   \\
\Lambda_{13-15}=&\text{Roots of:}\nonumber\\
&x^3+x^2 (-2 r_4-2 r_5-2 r_6)
\nonumber\\
&+x \left(
-4 |c_{11}|^2 - 4 |c_{12}|^2 -4 |c_4|^2 +4 r_4 r_5+4 r_4
   r_6+4 r_5 r_6\right)\nonumber\\
&
-8 c_4 c_{11}^* c_{12}^*-8 c_{11} c_{12}  c_4^*
+8  r_4 |c_{11}|^2+8  r_5 |c_{12}|^2
+8 r_6   |c_4|^2-8 r_4 r_5 r_6=0\, ,
\\
\Lambda_{16-18}=&\text{Roots of:}\nonumber\\
&x^3+x^2 (-6 r_1-6 r_2-6 r_3)+x \left(36
   r_1 r_2+36 r_1 r_3+36 r_2 r_3-16 r_4^2-16 r_4 r_7\right.\nonumber\\
&\left.
-16 r_5^2
-16 r_5 r_8-16 r_6^2-16 r_6 r_9-4 r_7^2-4 r_8^2-4
   r_9^2\right)-216 r_1 r_2 r_3\nonumber\\
&+96 r_1 r_6^2+96 r_1 r_6
   r_9
+24 r_1 r_9^2+96 r_2 r_5^2+96 r_2 r_5 r_8+24 r_2
   r_8^2+96 r_3 r_4^2\nonumber\\
&
+96 r_3 r_4 r_7+24 r_3 r_7^2-128 r_4 r_5 r_6
-64 r_4 r_5 r_9-64 r_4 r_6 r_8-32 r_4 r_8
   r_9\nonumber\\
&
-64 r_5 r_6 r_7-32 r_5 r_7 r_9-32 r_6 r_7 r_8
-16
   r_7 r_8 r_9=0\, ,
\\
\Lambda_{19-21}=&\text{Roots of:}\nonumber\\
&x^3+x^2 (-2 r_4-2 r_5-2 r_6-4 r_7-4 r_8-4
   r_9)+x \left(-36 c_{11} c_{11}^*-36 c_{12} c_{12}^*\right.
\nonumber\\
&
-36 c_4 c_4^*
+4 r_4 r_5+4 r_4 r_6+8 r_4 r_8+8 r_4 r_9+4 r_5 r_6+8
   r_5 r_7+8 r_5 r_9+8 r_6 r_7\nonumber\\
&\left.
+8 r_6 r_8+16 r_7 r_8
+16 r_7 r_9+16 r_8 r_9\right)
\nonumber\\
&-216 c_4 c_{11}^* c_{12}^*-216 c_{11}
   c_{12} c_4^*+72 c_{11} r_4 c_{11}^*+144 c_{11} r_7 c_{11}^*
\nonumber\\
&+72 c_{12}
   r_5 c_{12}^*+144 c_{12} r_8 c_{12}^*+72 c_4 r_6 c_4^*+144 c_4
   r_9 c_4^*-8 r_4 r_5 r_6-16 r_4 r_5 r_9\nonumber\\
&
-16 r_4 r_6 r_8-32 r_4 r_8 r_9-16 r_5 r_6 r_7-32 r_5 r_7 r_9-32
   r_6 r_7 r_8-64 r_7 r_8 r_9=0\, .
\end{align}

\subsection{Unitarity bounds for specific processes in the NHDM}

In the previous sections, we have presented an algorithm which
computes tree-level unitarity bounds on a given chosen model
of scalar doublets. The method of Lee, Quigg
and Thacker~\cite{Lee:1977yc,Lee:1977eg}, which we have generalized and
optimized, yields
necessary and sufficient conditions for tree unitarity in the scalar
and gauge sectors of any NHDM. A possible shortcoming of this
method is the computational time needed to find all the eigenvalues
for each point in parameter space in the case of  a general scalar doublet
theory, i.e. without symmetries, or with a large $N$.

A complementary approach of finding necessary, although
not sufficient conditions for every
NHDM can be attained by using unitarity bounds on specific processes.
Therefore, we can compute the partial-wave coefficient $a_0$ for
gauge-gauge and gauge-scalar scatterings, using that from the
optical theorem $| \Re(a_0) | < 1/2 $, and that we can use the
Equivalence Theorem to simplify calculations.
The Equivalence Theorem allows us to perform all calculations
from the scalar potential, and therefore, we can always
define
\begin{equation}
| \Re(a_0) | \leq\tfrac12 \Rightarrow | \Re(\mathcal{M}) | \leq 8 \pi \, ,
\end{equation}
where $\mathcal{M}$ stands for the amplitude of the process.
Another important property of this method, also used in the
previous section, is that we will
only consider the quartic couplings contribution at high energy.
We use this simple approach to work out some examples in the NHDM for
an arbitrary $N$.

\subsubsection{The $W^+ W^- \rightarrow W^+ W^-$ process}

As a first example, we consider the process with amplitude
$\mathcal{M}\left( W^+ W^- \rightarrow W^+ W^- \right)$,
which we approximate at high energies to
$\mathcal{M}\left( G^+ G^- \rightarrow G^+ G^- \right)$.
The leading
order contribution for the process will be the quartic coupling
with amplitude
\begin{equation}
\mathcal{M}_Q \left( G^+ G^- \rightarrow G^+ G^- \right) = - \frac{2}{v^2}
\sum_{\beta=2}^{2N} m_\beta^2
\left[ A_{1 \beta} \right]^2 = - \frac{2}{v^2}
\sum_{\beta=2}^{2N} m_\beta^2
\left[S^0_\beta V V \right]^2 \, ,
\end{equation}
where we have used eq.~(\ref{eq:qfeyn41}) and the notation from
section~\ref{sec:sumrules}.
It is straightforward to see that this process enforces the bound
\begin{equation}
\left| \sum_{\beta=2}^{2N} m_\beta^2
\left[S^0_\beta V V \right]^2 \right|
\leq\left( 872 \, \mathrm{GeV} \right)^2 \, .
\end{equation}

It is interesting to note that in the alignment limit of the coupling
$[hVV]_{\mathrm{NHDM}}=[hVV]_{\mathrm{SM}}$ we cannot further constrain any
masses of new neutral scalars. This observation is due to the bound in
the $A$ matrix itself, which is orthogonal, implying that
$\sum_\beta [A_{1 \beta}]^2 = 1$.
This bound is given as a special case of
eq.~(2.10c) of ref.~\cite{Weldon:1984wt},
and it is valid in any NHDM.

\subsubsection{The $Z S^+_a \rightarrow Z S^+_a$ process}

We now turn to the unitarity bounds arising from
the $Z S^+_a \rightarrow Z S^+_a$ scattering.
Using the same reasoning as before, we use the quartic coupling
of eq.~(\ref{eq:qfeyn44}).
We then have
\begin{equation}
\mathcal{M}_Q\left( G^0 S^+_a \rightarrow G^0 S^+_a \right)
= - \frac{2}{v^2} \left[ (D^2_\pm)_{aa} -
(U^\dagger \mu U)_{aa} \right] = -
\frac{2}{v^2} \left[ (D^2_\pm)_{aa} -
Y_{aa} \right] \, ,
\end{equation}
where $Y$ is the quadratic parameter of the lagrangian in the
charged Higgs basis.
Using the optical theorem we find that
\begin{equation}
\left|m^2_{\pm a} - Y_{aa}\right|\leq 4 \pi v^2 \, ,
\end{equation}
Although one cannot predict mass bounds for the charged scalars in this
process, it is possible to study numerically the dependence on
a given choice of $Y$. This parameter in the charged Higgs basis
is a physical one and can, in principle, be measured.

\subsubsection{The $W^+ S^-_a \rightarrow W^+ S^-_a$ process}

As a final example we compute the unitarity bounds for the process
$W^+ S^-_a \rightarrow W^+ S^-_a$.
We follow the same arguments as before and write the
quartic coupling in eq.~(\ref{eq:qfeyn42}) as
\begin{equation}
\mathcal{M}_Q \left( G^+ S^-_a \rightarrow G^+ S^-_a \right)
= - \frac{1}{v^2}
\left[ (B D^2_0 B^\dagger)_{aa} - 2 Y_{aa} \right] \, ,
\end{equation}
where $B_{a \beta}$ is the coupling to $[S^0_\beta W^+ S^-_a]$ and
$(B^\dagger)_{\beta a} $ is the coupling to $[S^0_\beta W^- S^+_a]$.
We can therefore use the optical theorem to obtain,
\begin{equation} \label{BDBD}
\left| (B D^2_0 B^\dagger)_{aa} - 2 Y_{aa} \right|\leq8 \pi v^2 \,.
\end{equation}
It is convenient to rewrite \eq{BDBD} as,
\begin{equation}
\left| \sum_{\beta=2}^{2N} m_\beta^2 [S^0_\beta W^+ S^-_a][S^0_\beta W^- S^+_a]
- 2 Y_{aa} \right| \leq 8 \pi v^2 \, ,
\end{equation}
so that the dependence on cubic couplings is explicit.

\section{\label{sec:conclusions}Conclusions}

We have studied the most general $SU(2)_L \times U(1)_Y$ theory with
$N$ Higgs doublets.
We have stressed the importance of the charged Higgs basis,
where the magnitudes and basis-invariant combinations of phases of
its scalar couplings $Y$ and $Z$ are observables.
The kinetic Lagrangian depends exclusively on a $N \times 2N$
matrix $B$, or equivalently the $2N\times 2N$ real orthogonal matrix
$\tilde{B}$ as defined in eq.~\eqref{tildeB}, which
governs the relation between the neutral scalar components of the scalar doublets in the charged
Higgs basis and the neutral Higgs mass eigenstates.
The matrix $B$ (or $\tilde{B}$) depends on $N-1$ unphysical phases (corresponding
to the non-uniqueness of the charged Higgs basis)
and on $2(N-1)^2$ physical parameters.
Of these,
$N(N-1)$ appear in the special combination
$A = \textrm{Im}(B^\dagger B)$.

Although new parameters beyond $B$ appear in the scalar potential,
many couplings involving the Goldstone bosons $G^0$ and $G^\pm$
can be related to couplings involving $Z^0$ and $W^\pm$,
as expected from consistency with gauge fixing.
We use the crucial eqs.~\eqref{Kprime_2}-\eqref{K_2}
to show that such relations indeed hold.
This is also consistent with bounds from unitarity,
which are discussed in great detail.
In particular,
we develop an efficient algorithm for the inclusion of such bounds
in NHDM and employ it in two new 3HDM models
with a $\mathbb{Z}_3$ and with a order-4 CP symmetry, respectively.
Some model independent necessary constraints are shown,
by applying the optical theorem to selected processes.

In models where the scalar potential exhibits additional symmetries,
some new parameters may appear to arise,
relating the original basis to the charged Higgs basis
(thereby acquiring physical significance).
For example, such a case arises in the the 2HDM with a $\mathbb{Z}_2$-symmetric scalar potential,
where $\beta$ is the angle that rotates the basis in which the $\mathbb{Z}_2$ symmetry
is manifest into the charged Higgs basis.
But, such parameters can always be re-expressed in terms of those discussed here [e.g.,
as shown for $\beta$ in eqs.~(51) and (52) of ref.~\cite{Zs}].
New parameters do arise when fermions are included,
which will be addressed elsewhere \cite{postpone}.

\vspace{2ex}

\acknowledgments
We are very grateful to A. Barroso for collaboration in the
initial stages of this work and to M.J. Gouveia for discussions
on matrices.
J.P.S. is grateful to ``Taberna do Avillez'',
where part of this work was made.
The work of M.P.B, J.C.R. and J.P.S. is supported in part
by the Portuguese \textit{Funda\c{c}\~{a}o para a Ci\^{e}ncia e Tecnologia}
(FCT) under contracts CERN/FIS-NUC/0010/2015 and UID/FIS/00777/2013.
H.E.H. is supported in part by the U.S. Department of Energy grant
number DE-SC0010107.
H.E.H. acknowledges the hospitality and the inspiring working
atmosphere of the Aspen Center for Physics, supported by the
National Science Foundation Grant No.\ PHY-1066293, where some
of the research reported in this work was carried out.

\bigskip\bigskip\bigskip
\centerline{\bf\Large Appendices}

\appendix
\section{\label{app:HiggsBasis}The Higgs basis and the charged Higgs basis}

In section~\ref{subsec:pot},
we have discussed how to determine the physical charged and neutral scalar mass eigenstates starting from a generic basis of scalar fields $\{\Phi_k\}$.   The charged and neutral components of the $k$th scalar doublet can be expressed as linear combinations of the physical charged and neutral scalar mass eigenstates [cf.~\eqs{S+}{S0}, respectively], where the corresponding coefficients define the matrices $U$ and $V$.  To find the physical couplings, we first determine the
scalar potential minimum conditions in the original (generic) basis (these stationarity conditions given in \eq{stationarity}    are complicated coupled cubic equations in the vevs, $v_k$).  Substituting these conditions back into \eqst{V2}{V4} and re-expressing the generic basis scalar fields in terms of the physical charged and neutral scalar mass eigenstates yields the desired expressions for the physical couplings.

Note that in determining the charged and neutral scalar mass eigenstates via  \eqref{S+} and \eqref{S0}, one must decompose
the scalar doublets into their charged and neutral components,
treating each component separately.
This hides an important characteristic of the Higgs potential---namely, physical observables must be invariant under
a unitary transformation among the $N$ scalar \emph{doublets}.
This is known as basis invariance, which is discussed at length in refs.~\cite{BS,BLS,DH}.

It is often more convenient to perform the analysis of the NHDM by first transforming from the generic basis to the so-called Higgs basis~\cite{Georgi:1978ri,LS,BLS,DH}, in which the neutral vev resides \textit{entirely} in the first scalar doublet.
This is achieved through a unitary transformation $X$,
\be \label{A1}
\Phi_j = \sum_{k=1}^N X_{jk} \Phi^H_k,
\ee
such that
\be
X_{j1} = \frac{v_j}{v} \equiv \hat{v}_j\,.
\label{X_k1}
\ee
Since $X$ is unitary, we can invert \eq{A1} to obtain
\be \label{invert}
\Phi_\ell^H=\sum_{j=1}^N X_{j\ell}^*\Phi_j\,.
\ee
We employ the normalization of the vevs such that $\langle{\Phi^0_j}\rangle\equiv v_j/\sqrt{2}$.
Hence, taking the vacuum expectation value of both sides of \eq{invert} and making use of \eq{X_k1} yields
\be
\langle \Phi^{H0}_1 \rangle = v/\sqrt{2}\,,\quad \text{where $v^2\equiv\sum_{j=1}^N |v_j|^2=(246~{\rm GeV})^2$.}
\ee
Moreover, $\half v^2=\sum_j \Phi^*_j \Phi_j=\sum_\ell \Phi^{H*}_\ell\Phi^H_\ell$, since $X$ is unitary.  Thus, one can immediately conclude that
\be
\langle \Phi^{H0}_1 \rangle = v/\sqrt{2}, \ \ \ \ \ \
\langle \Phi^{H0}_k \rangle = 0\ \ (\text{for}~k=2,\ldots,N).
\label{def_Higgs_Basis}
\ee
In light of \eqs{S1+_is_G+}{S10_is_G0},
it follows that the basis where the first scalar field contains
the would-be Goldstone bosons coincides with the basis
where all scalar doublets except the first have zero vev.
Since only $X_{j1}$ is determined,
``the Higgs basis'' is not uniquely determined.   More precisely, the Higgs basis constitutes a
class of bases, since starting in any given Higgs basis, one
can still perform an
$(N-1) \times (N-1)$ unitary transformation on
$\Phi^H_k (k=2,\ldots,N)$ without altering eq.~\eqref{def_Higgs_Basis}.
For example, if $N=2$,
one can still rephase the second Higgs doublet via
$\Phi^H_2 \rightarrow \exp(i \chi) \Phi^H_2$ without leaving the class of
Higgs bases.\footnote{The requirement that no physical observable can depend on the choice
of $\chi$ provided the original motivation for the basis invariance considerations
in the Higgs sector \cite{LS}.}
Comparing eq.~\eqref{X_k1} with eqs.~\eqref{U_k1} and \eqref{V_k1},
we conclude immediately that the charged component of $\Phi^H_1$
is $G^+$,
while the imaginary part of the neutral component is $G^0$.
The difference between
eqs.~\eqref{S+}--\eqref{S0}
and eq.~\eqref{def_Higgs_Basis} is that,
in the latter,
we have also transformed another combination of the
real components of $\Phi_k$, which yields
a real field $H^0$ that is \emph{not} a mass eigenstate.
Thus, we may parameterize
\be
\Phi^H_1 =
\left(
\begin{array}{c}
G^+\\
\tfrac{1}{\sqrt{2}}
\left( v + H^0 + i G^0 \right)
\end{array}
\right).
\ee

One particularly nice feature of the Higgs basis is the simplification obtained in the
stationarity conditions of eq.~\eqref{stationarity},
\be
\mu_{i1} + \lambda_{i1,11} v^2 = (M^2_\pm)_{i1} =0,
\label{stationarity_H}
\ee
and in the masses of eq.~\eqref{Mpm} and eqs.~\eqref{MR}--\eqref{MRI} \cite{NHiggs}:
\ba
(M^2_\pm)_{ij} &=&
\mu_{ij} + v^2 \lambda_{ij,11},
\label{Mpm_H}\\
(M_R^2)_{ij} &=&
\textrm{Re}\left[
\mu_{ij} + v^2
\left( \lambda_{ij,11} + \lambda_{i1,1j} + \lambda_{i1,j1} \right)
\right],
\label{MR_H}\\
\left(M^2_{I}\right)_{ij} &=&
\textrm{Re}\left[
\mu_{ij} + v^2
\left( \lambda_{ij,11} + \lambda_{i1,1j} - \lambda_{i1,j1} \right)
\right],
\label{MI_H}\\
\left(M^2_{RI}\right)_{ij} &=&
- \textrm{Im}\left[
\mu_{ij} + v^2
\left( \lambda_{ij,11} + \lambda_{i1,1j} - \lambda_{i1,j1} \right)
\right].
\label{MRI_H}
\ea
Note that the couplings
in eqs.~\eqref{stationarity_H}--\eqref{MRI_H} are
calculated \emph{in the Higgs basis};
they are \emph{not} the couplings $\mu$ and $\lambda$
of the original generic basis of eq.~\eqref{VH}.
To derive the charged and neutral scalar mass eigenstates, one must now perform an
$(N-1) \times (N-1)$ unitary transformation on
the charged components of
$\Phi^H_k (k=2,\ldots,N)$,
and a $(2N-1) \times (2N-1)$ unitary transformation on
the $2N-1$ neutral scalar components consisting of $H^0$ and the $\Phi^H_k (k=2,\ldots,N)$.

Any matrix $X$ obeying eq.~\eqref{X_k1} will yield a Higgs basis where,
by definition,
only the first scalar doublet has a non-zero vev.
Of the infinitely many choices for $X$ that satisfy \eq{X_k1},
it is particularly interesting to consider the transformation
$X=U$,
where $U$ is the unitary matrix defined in \eq{S+} that yields the physical charged Higgs mass eigenstates~\cite{Nishi:2007nh}.
Notice that we have left the scalar doublet structure intact;
in this procedure, the neutral components transform as the charged components.
Thus, in general, the corresponding neutral partner of the physical charged scalar will not be a mass eigenstate.
We define such a  basis as the \emph{charged Higgs basis},
\be
\Phi_j = \sum_{k=1}^N U_{jk} \Phi^C_k.
\ee
which is a subclass of the class of Higgs bases defined above.
Now,
eqs.~\eqref{U_k1} and \eqref{def_Higgs_Basis} remain valid and we may parameterize
\be \label{chhiggsbasis}
\Phi^H_1 =
\left(
\begin{array}{c}
G^+\\*[2mm]
\tfrac{1}{\sqrt{2}}
\left( v + H^0 + i G^0 \right)
\end{array}
\right),
\ \ \
\Phi^C_2 =
\left(
\begin{array}{c}
H^+_2\\*[2mm]
\tfrac{1}{\sqrt{2}} \varphi^{C0}_2
\end{array}
\right)\,,
\ldots\,,
\ \ \
\Phi^C_N =
\left(
\begin{array}{c}
H^+_N\\*[2mm]
\tfrac{1}{\sqrt{2}} \varphi^{C0}_N
\end{array}
\right),
\ee
where $H_2^+, \ldots, H^+_N$ are the physical charged Higgs mass eigenstate fields,
with corresponding masses $m^2_{\pm, i}$.
In this first step,
the neutral components are transformed by
\be
\left(
\begin{array}{c}
\textrm{Re}(\varphi_1^0)\\
\vdots\\
\textrm{Re}(\varphi_N^0)\\*[2mm]
\textrm{Im}(\varphi_1^0)\\
\vdots\\
\textrm{Im}(\varphi_N^0)
\end{array}
\right)
= \tilde{U}_R\,
\left(
\begin{array}{c}
\textrm{Re}(\varphi_1^{C 0})\\
\vdots\\
\textrm{Re}(\varphi_N^{C 0})\\*[2mm]
\textrm{Im}(\varphi_1^{C 0})\\
\vdots\\
\textrm{Im}(\varphi_N^{C 0})
\end{array}
\right),
\ee
where $\varphi^{C0}_1\equiv H^0+iG^0$, with
\be \label{vbaru}
\tilde{U}_R \equiv
\left(
\begin{array}{cc}
\hspace{1ex}\\
\textrm{Re}\, U &\quad - \textrm{Im}\, U\\
\hspace{2ex}
\\
\textrm{Im}\, U &\quad \phm \textrm{Re}\, U\\
\hspace{1ex}
\end{array}
\right).
\ee
After this first step,
eqs.~\eqref{Mpm_H}--\eqref{MRI_H} become
\ba
(M^2_\pm)_{ij} &=&
Y_{ij} + v^2 Z_{ij,11}
= \delta_{ij}\, m^2_{\pm, i},
\label{Mpm_CH}\\
(M_R^2)_{ij} &=&
\delta_{ij}\, m^2_{\pm, i} + v^2\,
\textrm{Re}\left[Z_{i1,1j} + Z_{i1,j1} \right],
\label{MR_CH}\\
\left(M^2_{I}\right)_{ij} &=&
\delta_{ij}\, m^2_{\pm, i} + v^2\,
\textrm{Re}\left[Z_{i1,1j} - Z_{i1,j1} \right],
\label{MI_CH}\\
\left(M^2_{RI}\right)_{ij} &=&
- v^2\, \textrm{Im}\left[Z_{i1,1j} -Z_{i1,j1} \right],
\label{MRI_CH}
\\
\left(M^2_{RI}\right)^T_{ij} &=&
v^2\, \textrm{Im}\left[Z_{i1,1j} +Z_{i1,j1} \right].
\label{MRIT_CH}
\ea
Thus,
\ba
2 v^2 Z_{i1,1j}
&=&
\left[
(M_R^2)_{ij} + (M_I^2)_{ij} - 2 \delta_{ij}\, m^2_{\pm, i}
\right]
+ i
\left[
\left(M^2_{RI}\right)^T_{ij} - \left(M^2_{RI}\right)_{ij}
\right],
\label{Li11j}\\
2 v^2 Z_{i1,j1}
&=&
\left[
(M_R^2)_{ij} - (M_I^2)_{ij}
\right]
+ i
\left[
\left(M^2_{RI}\right)^T_{ij} + \left(M^2_{RI}\right)_{ij}
\right].
\label{Li1j1}
\ea
In addition,
\ba
\left( M^2_I \right)_{i1}
& = 0 = &
\left( M^2_I \right)_{1j},
\nonumber\\
\left( M^2_{RI} \right)_{i1}
& = 0 = &
\left( M^2_{RI} \right)^T_{1j},
\ea
which implies that the $N+\nth{1}$ row and column of the matrix,
\be
\left(
\begin{array}{cc}
\hspace{1ex}\\
M_R^2 & \quad M_{RI}^2\\
\hspace{2ex}
\\
(M_{RI}^2)^T & \quad M_I^2\\
\hspace{1ex}
\end{array}
\right),
\ee
vanish, corresponding to the massless $\textrm{Im}~\varphi_1^{C0} = G^0$.
As a result,
\ba
2 v^2 Z^\ast_{11,1j}
=
2 v^2 Z_{11,j1}
&=&
(M_R^2)_{1j}
+ i \left(M^2_{RI}\right)_{1j},\quad \text{for $j\neq 1$},
\nonumber\\
2 v^2 Z_{11,11}
&=&
(M_R^2)_{11}.
\ea
The first equality above is a consequence of eq.~\eqref{hermiticity},
and it is consistent with eq.~\eqref{Li11j} because $m^2_{\pm, 1}$
is the mass of $G^+$,
which vanishes.
Notice that the couplings in $Y$ and $Z$ eqs.~\eqref{Mpm_CH}--\eqref{Li1j1} are
calculated \emph{in the charged Higgs basis};
they are \emph{not} the couplings $\mu$ and $\lambda$
of the original generic basis in eq.~\eqref{VH}.

To determine the neutral scalar mass eigenstates, one must diagonalize the
real orthogonal $(2N-1)\times(2N-1)$ squared-mass matrix that mixes
the $(2N-1)$ neutral scalar fields $H^0, \textrm{Re}~\varphi^{C0}_2, \dots,
\textrm{Re}~\varphi^{C0}_N, \textrm{Im}~\varphi^{C0}_2, \dots,
\textrm{Im}~\varphi^{C0}_N$ (which are defined in the charged Higgs basis).
This is achieved thorough the $N\times 2N$ matrix $V^C$
\be
\varphi_k^{C 0} =
\sum_{\beta=1}^{2N} V^C_{k \beta} S_\beta^0,
\ee
where
\ba
V^C_{k1} &=& i \delta_{k1}\ \ \ \ \ (k=1 \dots N),
\nonumber\\
\textrm{Im} (V^C_{1 \beta}) &=& \delta_{1 \beta} \ \ \ \ \ (\beta=1 \dots 2N).
\ea
One can define the real orthogonal $2N\times 2N$ matrix $\tilde{V}^C$ in analogy with \eq{tildeV}, which satisfy equations analogous to eqs.~\eqref{VTV}-\eqref{VVT}.  Deleting the first row and column of $\tilde{V}^C$ yields the matrix that diagonalizes the squared-mass matrix of the neutral scalars fields in the charged Higgs basis.

Performing the neutral scalar squared-mass diagonalization in the charged Higgs basis can be especially useful in some circumstances.
\Eqs{Kprime_2}{K_2}
assume very simple forms in the charged Higgs basis,
\ba
2v^2 Z_{i1,1j}
&=&
-2 (D_{\pm}^2)_{i j} + \left(B D_0^2 B^\dagger \right)_{i j},
\nonumber\\
2 v^2 Z_{i 1,j 1}
&=&
\left(
B D_0^2 B^T \right)_{i j}.
\ea
As a further example,
we notice that the cubic terms of the scalar potential
in eq.~\eqref{V3} may be written in the charged Higgs basis as
\be
V_3
=
v\, \left( S_i^- S_j^+ + \tfrac{1}{2} \varphi_i^{C 0 \ast} \varphi_j^{C 0} \right)
\left[
Z_{ij,k1} \varphi_k^{C 0 \ast}
+
Z_{ij,1k} \varphi_k^{C 0}
\right].
\ee
All couplings with two indices equal to $1$ may be related with
eqs.~\eqref{Li11j}--\eqref{Li1j1}, and hence can be related to the scalar masses.
We find
\ba
v V_3 &=&
\left( S_1^- S_1^+ + \tfrac{1}{2} |\varphi_1^{C 0}|^2 \right)
\left\{
\left( M_R^2\right)_{11} \textrm{Re} \varphi_1^{C 0}
+
\sum_{k = 2}^N
\left[
\left( M_R^2\right)_{1k} \textrm{Re} \varphi_k^{C 0}
+
\left( M_{RI}^2\right)_{1k} \textrm{Im} \varphi_k^{C 0}
\right]
\right\}
\nonumber\\*[2mm]
&+&
\left( S_i^- S_1^+ + \tfrac{1}{2} \varphi_i^{C 0 \ast} \varphi_1^{C 0}  \right)
\left[
\left(M_R^2\right)_{1i} + i \left(M_{RI}^2\right)_{1i}
\right]
\textrm{Re} \varphi_1^{C 0}\ \  +\ \  h.c.
\nonumber\\*[2mm]
&+&
\left( S_i^- S_1^+ + \tfrac{1}{2} \varphi_i^{C 0 \ast} \varphi_1^{C 0}  \right)
\sum_{k = 2}^N
\left\{
\left[
\left[
\left( M_R^2\right)_{ik} - \delta_{ik} m_{\pm, k}^2 \right]
+ i \left( M_{RI}^2\right)^T_{ik}
\right] \textrm{Re} \varphi_k^{C 0}
\right.
\nonumber\\*[1mm]
&&
\hspace{35mm}
\left.
+
\left[
\left( M_{RI}^2\right)_{ik}
+ i  \left[ \left( M_I^2\right)_{ik} - \delta_{ik} m_{\pm, k}^2 \right]
\right] \textrm{Im} \varphi_k^{C 0}
\right\}\ \ +\ \ h.c.
\nonumber\\*[2mm]
&+&
\left( S_i^- S_j^+ + \tfrac{1}{2} \varphi_i^{C 0 \ast} \varphi_j^{C 0}  \right)
2 v^2 Z_{ij,11} \textrm{Re} \varphi_1^{C 0}
\nonumber\\*[2mm]
&+&
\left( S_i^- S_j^+ + \tfrac{1}{2} \varphi_i^{C 0 \ast} \varphi_j^{C 0}  \right)
v^2\,
\sum_{k = 2}^N
\left[
Z_{ij,k1} \varphi_k^{C 0 \ast}
+
Za_{ij,1k} \varphi_k^{C 0}
\right],
\ea
where there are implicit sums over repeated indices, $i=2,\ldots,N$ and $j=2,\ldots,N$.

As in the case of the Higgs basis, the charged Higgs basis is also not uniquely determined.
Indeed, the charged Higgs basis is a
class of bases, since starting in any given charged Higgs basis, one
can separately rephase the $N-1$ scalar doublet fields, $\Phi_2^C, \Phi^C_3,\ldots,\Phi_N^C$,
while preserving the corresponding charged components as mass-eigenstate fields.
It is convenient to keep track of this rephasing degree of freedom.   Thus,  in \eq{S+}, we will
choose the unitary matrix $U$ such that $U_{k1}=\hat{v}_k$ and make some conventional
choice for the overall phases of the column vectors $U_{kj}$, for $j=2,3,\ldots,N$.  Such a choice picks out one of the possible charged Higgs bases.   In this basis, the corresponding physical charged Higgs fields are denoted by $S_a^+$ (for $a=2,3,\ldots,N$).   One can of course transform to any other charged Higgs basis by an appropriate rephasing of $\Phi_2^C, \Phi^C_3,\ldots,\Phi_N^C$, in which case the corresponding physical charged Higgs fields are also rephased, $S_a^+\to e^{i\chi_a}S_a^+$.

\section{\label{app:C2HDM}Connections with the 2HDM}

\subsection{2HDM with a softly-broken $\mathbb{Z}_2$ discrete symmetry}

The Lagrangian shown in sections \ref{subsec:Gauge-higgs} and
\ref{subsec:scalar-scalar} depends on the matrices
$U$ and $V$ introduced in \eqs{S+}{S0}.
This is not the notation commonly used in two Higgs doublet models
(2HDM) \cite{HHG, 2HDM}.
Here we make the connection to the notation used in
the complex two Higgs doublet model (C2HDM), where the $\mathbb{Z}_2$ symmetry
($\Phi_1 \rightarrow \Phi_1$, $\Phi_2 \rightarrow - \Phi_2$)
is softly broken by a complex squared-mass parameter
\cite{Ginzburg:2002wt, Arhrib:2010ju, Barroso:2012wz, Fontes:2014xva}.
In this case,
one transforms to the Higgs basis \cite{LS, BS} through
\be
\left(
\begin{array}{c}
H_1\\
H_2
\end{array}
\right)
=
\left(
\begin{array}{cc}
\phm c_\beta &\quad s_\beta \\
- s_\beta & \quad c_\beta
\end{array}
\right)\,
\left(
\begin{array}{c}
\Phi_1\\
\Phi_2
\end{array}
\right),
\label{changeTOHiggs}
\ee
where
\be \label{cbsb}
\sqrt{2} \langle \Phi^0_1 \rangle\equiv
v_1 = v\, c_\beta\,,\qquad\quad
\sqrt{2} \langle \Phi^0_2 \rangle
\equiv
v_2 = v\, s_\beta\,.
\ee
In \eq{cbsb},
$c_\beta = \cos{\beta}$,
$s_\beta = \sin{\beta}$,
and $v = \sqrt{v_1^2 + v_2^2} = (\sqrt{2} G_F)^{-1/2}$.
Without loss of generality,
we have taken the vevs $v_1$ and $v_2$ real.\footnote{This just
redefines the phase of terms in the scalar potential
sensitive to the relative phase between $\Phi_1$ and $\Phi_2$.}
The doublets in the Higgs basis may be written
\begin{equation}
H_1 =
\begin{pmatrix}
G^+ \\
\frac{1}{\sqrt{2}} (v + h + i G^0)
\end{pmatrix}
\, ,
\ \ \ \ \
H_2 =
\begin{pmatrix}
H^+ \\
\frac{1}{\sqrt{2}} (R + i I)
\end{pmatrix}\, ,
\label{H1H2}
\end{equation}
where $G^\pm$ and $G^0$ are the Goldstone bosons and
$H^\pm$ are the physical charged scalars.
Thus, the matrix $U$ of eqs.~\eqref{S+} is simply
\be
U =
\left(
\begin{array}{cc}
c_\beta &\quad - s_\beta \\
s_\beta & \quad \phm c_\beta
\end{array}
\right).
\ee

Let us parameterize the scalars $\Phi_1$ and
$\Phi_2$ in the original generic basis as
\begin{equation}
\Phi_1 =
\begin{pmatrix}
\varphi_1^+ \\
\frac{1}{\sqrt{2}} (v_1 + \eta_1 + i \chi_1)
\end{pmatrix}
\, ,
\ \ \ \ \
\Phi_2 =
\begin{pmatrix}
\varphi_2^+  \\
\frac{1}{\sqrt{2}} (v_2 + \eta_2 + i \chi_2)
\end{pmatrix}\, .
\label{Phi1Phi2}
\end{equation}
Eqs.~\eqref{changeTOHiggs} and \eqref{H1H2} yield
for the massless would-be Goldstone boson
$G^0 = c_\beta \chi_1 + s_\beta \chi_2$.
We define the orthogonal state
\be
\eta_3 = - s_\beta \chi_1 + c_\beta \chi_2.
\ee
The fields $\eta_1$, $\eta_2$, and $\eta_3$
combine into the mass eigenstates $h_1$, $h_2$, and $h_3$
as
\be
\left(
\begin{array}{c}
h_1\\
h_2\\
h_3
\end{array}
\right)
= R
\left(
\begin{array}{c}
\eta_1\\
\eta_2\\
\eta_3
\end{array}
\right),
\label{h_as_eta}
\ee
where the orthogonal matrix may be parameterized as
\be
R =
\left(
\begin{array}{ccc}
c_1 c_2 & s_1 c_2 & s_2\\
-(c_1 s_2 s_3 + s_1 c_3) & c_1 c_3 - s_1 s_2 s_3  & c_2 s_3\\
- c_1 s_2 c_3 + s_1 s_3 & -(c_1 s_3 + s_1 s_2 c_3) & c_2 c_3
\end{array}
\right)\, .
\label{matrixR}
\ee
Here, $s_i = \sin{\alpha_i}$,
$c_i = \cos{\alpha_i}$ ($i = 1, 2, 3$),
and, without loss of generality,
the angles may be restricted to \cite{ElKaffas:2007rq}
\be
- \pi/2 < \alpha_1 \leq \pi/2,
\hspace{5ex}
- \pi/2 < \alpha_2 \leq \pi/2,
\hspace{5ex}
0 \leq \alpha_3 \leq \pi/2.
\label{range_alpha}
\ee
By definition, we take the masses of the neutral scalars
in increasing order: $m_1 < m_2 < m_3$.

We would like to recombine these expressions into the form of
eqs.~\eqref{Phi-1} and \eqref{tildeV}:
\be
\left(
\begin{array}{c}
\textrm{Re}\, \varphi_1^0\\*[1mm]
\textrm{Re}\, \varphi_2^0\\*[1mm]
\textrm{Im}\, \varphi_1^0\\*[1mm]
\textrm{Im}\, \varphi_2^0
\end{array}
\right)
=
\tilde{V}
\,
\left(
\begin{array}{c}
G^0\\*[1mm]
h_1\\*[1mm]
h_2\\*[1mm]
h_3
\end{array}
\right)
=
\left(
\begin{array}{c}
\textrm{Re}\ V\\
\textrm{Im}\ V
\end{array}
\right)
\,
\left(
\begin{array}{c}
G^0\\*[1mm]
h_1\\*[1mm]
h_2\\*[1mm]
h_3
\end{array}
\right).
\ee
We find
\be
\tilde{V}
=
\left(
\begin{array}{cccc}
0 & c_1 c_2 & -c_3 s_1 - c_1 s_2 s_3 & -c_1 c_3 s_2 + s_1 s_3\\
0 & c_2 s_1 & c_1 c_3 - s_1 s_2 s_3 & -c_3 s_1 s_2 - c_1 s_3\\
c_\beta & - s_2 s_\beta & -c_2 s_3 s_\beta & -c_2 c_3 s_\beta\\
s_\beta & s_2 c_\beta & c_2 s_3 c_\beta & c_2 c_3 c_\beta
\end{array}
\right)
\ee
and
\be
\tilde{B}
=
\left(
\begin{array}{cccc}
0 & \tilde{c}_1 c_2 & -c_3 \tilde{s}_1 - \tilde{c}_1 s_2 s_3
    & -\tilde{c}_1 c_3 s_2 + \tilde{s}_1 s_3\\
0 & c_2 \tilde{s}_1 & \tilde{c}_1 c_3 - \tilde{s}_1 s_2 s_3
    & -c_3 \tilde{s}_1 s_2 - \tilde{c}_1 s_3\\
1 & 0 & 0 & 0\\
0 & s_2 & c_2 s_3  & c_2 c_3
\end{array}
\right),
\label{tildeB_C2HDM}
\ee
which satisfies eqs.~\eqref{properties_B}, as expected.
As we have seen in sections~\ref{subsec:Gauge-higgs} and \ref{subsec:scalar-scalar},
several important couplings,
including the couplings of each neutral scalar with two vector bosons,
involve the special antisymmetric combination
\be
A = \Im (B^\dagger B) =
\left(
\begin{array}{cccc}
0 & \,\,\, -c_2 \tilde{c}_1 &\,\,\, s_2 s_3 \tilde{c}_1 + c_3 \tilde{s}_1 &\,\,\,
    c_3 s_2 \tilde{c}_1 - s_3 \tilde{s}_1\\
c_2 \tilde{c}_1 &\,\,\, 0 &\,\,\, -c_3 s_2 \tilde{c}_1 + s_3 \tilde{s}_1 &\,\,\,
    s_2 s_3 \tilde{c}_1 + c_3 \tilde{s}_1\\
-s_2 s_3 \tilde{c}_1 - c_3 \tilde{s}_1 &\,\,\, c_3 s_2 \tilde{c}_1 - s_3 \tilde{s}_1 &\,\,\,
    0 & \,\,\, c_2 \tilde{c}_1\\
-c_3 s_2 \tilde{c}_1 + s_3 \tilde{s}_1
    & \,\,\,-s_2 s_3 \tilde{c}_1 - c_3 \tilde{s}_1 & \,\,\,-c_2 \tilde{c}_1 & \,\,\,0
\end{array}
\right),
\label{tildeA_C2HDM_1}
\ee
where $\tilde{s}_1 = \sin{(\alpha_1 - \beta)}$
and $\tilde{c}_1 = \cos{(\alpha_1 - \beta)}$.
Notice that, although three angles appear, there are in fact only
two independent parameters in the most general orthogonal and antisymmetric
$4 \times 4$ matrix. Indeed, such a matrix can always be parameterized
as
\be
\left(
\begin{array}{cccc}
0 & \,\,\,- \bar{c}_1 &\,\,\, \phm\bar{s}_1 \bar{c}_2 &\,\,\, \bar{s}_1 \bar{s}_2\\
\bar{c}_1 &\,\,\, 0 &\,\,\, - \bar{s}_1 \bar{s}_2 &\,\,\, \bar{s}_1 \bar{c}_2\\
- \bar{s}_1 \bar{c}_2 &\,\,\,\phm \bar{s}_1 \bar{s}_2 &\,\,\, 0 &\,\,\, \bar{c}_1\\
- \bar{s}_1 \bar{s}_2 &\,\,\, - \bar{s}_1 \bar{c}_2 &\,\,\, - \bar{c}_1 & \,\,\,0
\end{array}
\right),
\label{tildeA_C2HDM_2}
\ee
where $\bar{c}_k = \cos{\theta_k}$
and $\bar{s}_k = \sin{\theta_k}$,
for $k=1,2$.
Thus,
of the three angles in the matrix $R$,
only two combinations can be determined by measurements involving solely
the neutral scalars.

The complete set of Feynman rules for the C2HDM is presented on a webpage
\cite{WebPageC2HDM}.
We have checked explicitly that the couplings in sections
\ref{subsec:Gauge-higgs} and \ref{subsec:scalar-scalar}
reproduce the C2HDM Feynman rules in \cite{WebPageC2HDM}.
This is a highly non-trivial cross-check since the expressions
are very complicated when written in terms of the angles
$\alpha_i$ and $\beta$.
Moreover,
for those couplings involving masses
the equality is only obtained when using the relation
\be
m_3^2 = \frac{m_1^2\, R_{13} (R_{12} \tan{\beta} - R_{11})
+ m_2^2\ R_{23} (R_{22} \tan{\beta} - R_{21})}{R_{33} (R_{31} - R_{32} \tan{\beta})},
\label{m3_derived}
\ee
which holds in the C2HDM \cite{Fontes:2014xva}.
The following relations are also useful:
\ba
A_{1,\, i+1}
&=&
- \cos{\beta}\,  R_{i 1} - \sin{\beta}\,  R_{i 2},
\nonumber\\
A_{i+1,\, j+1}
&=&
- \epsilon_{i j k}
\left[ \cos{\beta}\, R_{k 1} + \sin{\beta}\,  R_{k 2}
\right].
\ea

In the limit of the real 2HDM ($s_2 \rightarrow 0$ and $s_3 \rightarrow 0$),
one finds
\be
A
=
\left(
\begin{array}{cccc}
\phm 0 &\,\,\, - \tilde{c}_1 &\,\,\, \phm\tilde{s}_1 &\,\,\, \phm 0\\
\phm\tilde{c}_1 &\,\,\, \phm 0 & \,\,\,\phm 0 &\,\,\, \phm\tilde{s}_1\\
-\tilde{s}_1 &\,\,\, \phm 0 & \,\,\,\phm 0 &\,\,\, \phm\tilde{c}_1\\
\phm 0 & \,\,\,- \tilde{s}_1 &\,\,\, - \tilde{c}_1 &\,\,\,\phm 0
\end{array}
\right).
\ee

It is also interesting to consider what happens in the SM,
where
\be
\phi^0 = h + i G^0
=
\left(
\begin{array}{cc}
i& \ \ 1
\end{array}
\right)
\left(
\begin{array}{c}
G^0\\
h
\end{array}
\right).
\ee
Thus, $V= (i \ \ 1)$,
and
\be
V^\dagger V
=
\left(
\begin{array}{cc}
1 &\,\,\, -i\\
i & \,\,\,\phm 1
\end{array}
\right),
\ee
confirming that $\textrm{Re} (V^\dagger V) = \bm{1}_{2 \times 2}$
and $\textrm{Im} (V^\dagger V) $ is antisymmetric.

\subsection{Basis-independent treatment of the most general 2HDM}

By using the basis-independent techniques introduced in
refs.~\cite{DH,Haber:2006ue}, one can analyze the most general
CP-violating 2HDM (with no additional symmetries imposed on the scalar
potential) in terms of quantities that are independent of the choice of
basis for the two scalar doublet fields.
All physical observables of the theory can be expressed in terms
of such basis-independent quantities.
It is instructive to see how this formalism is related to the treatment
of the NHDM given in section~\ref{sec:model}.

Since the notation of ref.~\cite{Haber:2006ue} differs somewhat from the notation
used in this paper, we provide here a brief introduction to the basis-independent
treatment of the 2HDM.  We begin with the scalar potential in a generic basis
given in \eq{VH}.  The vevs of the scalar doublet fields are given by
\be \label{vhat}
\langle\Phi_i\rangle=
\frac{v}{\sqrt{2}}\begin{pmatrix} 0\\ \hat{v}_i\end{pmatrix}\, ,
\ee
where $v=2m_W/g\simeq 246$~GeV and $\hat{v}=(\hat{v}_1,\hat{v}_2)$ is a complex vector of unit norm.
A second unit vector $\hat{w}$ can be defined that is orthogonal to $\hat{v}$,
\be \label{what}
\hat{w}_j=\hat{v}_i^*\epsilon_{ij}\,,
\ee
where $\epsilon_{12}=-\epsilon_{21}=1$ and $\epsilon_{11}=\epsilon_{22}=0$.
Indeed, the complex dot product,
$\hat{v}_j^*\hat{w}_j=0$, where the sum over the repeated index $j$ is implicit.

It is convenient to define two hermitian projection operators,
\be
V_{ij}\equiv \hat v_i \,\hat v_j^\ast\,,\qquad\quad W_{ij}\equiv \hat w_i \,\hat w_j^\ast=\delta_{ij}-V_{ij}\,.
\ee
Note that $\hat{v}$ and $\hat{w}$ are eigenvectors of the matrix $V$.
The matrices $V$ and $W$ can be used to define the following manifestly
basis-invariant real quantities that depend on the scalar potential parameters defined in the generic basis of scalar fields [cf.~\eq{VH}],
\ba
Y_1 &\equiv & \Tr(\mu V)\,,\label{syvv1}\\
Y_2&\equiv& \Tr(\mu W)\,,\label{syvv2}\\
Z_1&\equiv& 2\lambda_{ij,k\ell}\,V_{ji}V_{\ell k}\,,\label{szvv1}\\
Z_2&\equiv&  2\lambda_{ij,k\ell}\,W_{ji}W_{\ell k}\,,\label{szvv2}\\
Z_3 &\equiv&  2\lambda_{ij,k\ell}\,V_{ji}W_{\ell k}\,,\label{szvv3}\\
Z_4 &\equiv &  2\lambda_{ij,k\ell}\,V_{jk}W_{\ell i}\,.\label{szvv4}
\ea
In addition, we shall define the following pseudo-invariant
(potentially complex) quantities,
\ba
Y_3&\equiv & \mu_{ij}\,v_i^\ast w_j\,, \label{syvv3} \\
Z_5&\equiv & 2\lambda_{ij,k\ell} \,v_i^\ast w_j v_k^\ast w_\ell\,.\label{szvv5}\\
Z_6&\equiv & 2\lambda_{ij,k\ell}\, v_i^\ast v_j v_k^\ast w_\ell\,.\label{szvv6}\\
Z_7&\equiv & 2\lambda_{ij,k\ell}\, v_i^\ast w_j w_k^\ast w_\ell\,.\label{szvv7}
\ea

The significance of the quantities defined by \eqst{syvv1}{szvv7}
become clearer after rewriting the scalar potential in the Higgs basis.  Using the notation of \eqs{vhat}{what}, the Higgs basis fields can be defined as,\footnote{\Eq{h1h2} defines the Higgs basis scalar fields for one particular choice in the class of Higgs bases.}
\be \label{h1h2}
H_1\equiv \hat{v}^*_i\Phi_i\,,\qquad H_2\equiv \hat w_i^*\Phi_i\,.
\ee
In particular, note that the vevs of the Higgs basis fields are
\be \label{hbasis}
\langle H_1^0\rangle=\frac{v}{\sqrt{2}}\,,\qquad\quad \langle H_2^0\rangle=0\,,
\ee
as required by \eq{def_Higgs_Basis}.
That is, starting from the scalar potential defined in the generic
basis [cf.~\eq{VH}], we simply set $\hat v=(1,0)$ and $\hat{w}=(0,1)$.
Applying these results to \eqst{syvv1}{szvv7}, we see that $Y_{1,2,3}$
are the coefficients of the squared mass terms and $Z_{1,2,\ldots,7}$
are the coefficients of the quartic terms
of the scalar potential when expressed in terms of the Higgs basis fields.
In particular,
\ba
V_H&=& Y_1 H_1^\dagger H_1+ Y_2 H_2^\dagger H_2 +[Y_3
H_1^\dagger H_2+{\rm h.c.}]
+\half Z_1(H_1^\dagger H_1)^2+\half Z_2(H_2^\dagger H_2)^2
\nonumber\\
&&\quad
+Z_3(H_1^\dagger H_1)(H_2^\dagger H_2)
+Z_4( H_1^\dagger H_2)(H_2^\dagger H_1)\nonumber\\
&&\quad
+\left\{\half Z_5 (H_1^\dagger H_2)^2 +\big[Z_6 (H_1^\dagger
H_1) +Z_7 (H_2^\dagger H_2)\big] H_1^\dagger H_2+{\rm
h.c.}\right\}\,.
\ea
In the Higgs basis, the minimization of the scalar potential yields
\be \label{pmincond}
Y_1=-\half Z_1 v^2\,,\qquad\quad Y_3=-\half Z_6 v^2\,.
\ee

One key observation is that the Higgs basis as defined by \eq{hbasis} is unique only up to an overall phase redefinition of the Higgs basis field
$H_2\to e^{i\chi} H_2$.    Indeed, in light of \eq{h1h2}, the phase freedom in defining the Higgs basis simply corresponds to $\hat{w}\to e^{-i\chi}\hat{w}$.  It then follows that $Y_3$, $Z_5$, $Z_6$ and $Z_7$ also acquire a phase under $H_2\to e^{i\chi} H_2$,
\be
 [Y_3, Z_6, Z_7]\to e^{-i\chi}[Y_3, Z_6, Z_7] \quad{\rm and}\quad
Z_5\to  e^{-2i\chi} Z_5\,,
\ee
which is why these quantities were called pseudo-invariants above.  In contrast, $Y_1$, $Y_2$ and $Z_{1,2,3,4}$ are invariant under $H_2\to e^{i\chi} H_2$.

Our goal now is to evaluate the matrices $B$ and $A$ defined in \eqs{def_B}{A_definition}, respectively.
This requires the diagonalization of the charged Higgs and neutral Higgs squared-mass matrices.  First, we consider
the charged Higgs squared-mass matrix, which is given by \eq{Mpm}.  It is convenient to rewrite this matrix as follows.
Following ref.~\cite{Haber:2006ue}, we note that we can expand an
hermitian second-ranked tensor in terms of the eigenvectors of
$V$,
\be \label{aab}
A_{ij}=
\Tr(VA)V_{ij}+\Tr(WA)W_{ij}+(\hat v^*_{k}
\hat w_\ell A_{k\ell})\hat v_i \hat w^*_j+
(\hat w^*_{k}\hat v_\ell
 A_{k\ell})\hat w_i \hat v^*_j\,.
\ee
Applying \eq{aab} to the hermitian  charged Higgs squared-mass matrix, it follows that
\be
(M^2_\pm)_{ij} = (Y_1+\half Z_1 v^2)V_{ij}+(Y_2+\half Z_3 v^2)W_{ij}+(Y_3+\half Z_6 v^2)\hat{v}_i\hat{w}^*_j+
(Y^*_3+\half Z^*_6 v^2)\hat{w}_i\hat{v}^*_j\,,
\ee
after making use of \eqst{syvv1}{szvv6}.
After imposing the scalar potential minimum conditions
given in \eq{pmincond},
we end up with
\be
(M^2_\pm)_{ij} =(Y_2+\half Z_3 v^2)W_{ij}\,.
\ee
The diagonalization of $M^2_\pm$ is straightforward.
Defining the diagonalization matrix $U$ as in \eq{S+}, it follows that
\be \label{u2hdm}
U=\begin{pmatrix} \hat{v}_1 & \,\,\,\hat{w}_1 \\
\hat{v}_2 &\,\,\, \hat{w}_2\end{pmatrix}\,,
\ee
which satisfies,
\be
U^\dagger W\,U=\begin{pmatrix} 0 & \,\,\, 0\\ 0 & \,\,\, 1\end{pmatrix}\,.
\ee
That is,
of the two eigenvalue of $M^2_\pm$,
one is zero,
corresponding to the charged Goldstone boson,
and the other is $m_{H^\pm}^2=Y_2+\half Z_3 v^2$.

Next, we obtain the matrix that diagonalizes the neutral Higgs
squared-mass matrix.
In this analysis, it will prove useful to first perform the
diagonalization in the Higgs basis, since this allows us to
easily identify the relevant basis-independent quantities.
This has been carried out in ref.~\cite{Haber:2006ue}.
Here we summarize the main results that we need for our present analysis.
The three physical neutral Higgs boson mass-eigenstates
are determined by diagonalizing the $3\times 3$ real symmetric squared-mass
matrix that is defined in the Higgs basis~\cite{BLS,Haber:2006ue},
\be  \label{mtwo}
\mathcal{M}^2=v^2\left( \begin{array}{ccc}
Z_1&\,\, \Re Z_6 &\,\, -\Im Z_6\\
\Re Z_6 &\,\, \half (Z_{345}+Y_2/v^2) & \,\,
- \half \Im Z_5\\ -\Im Z_6 &\,\, - \half \Im Z_5 &\,\,
 \half (Z_{345}+Y_2/v^2)-\Re Z_5\end{array}\right),
\ee
where $Z_{345}\equiv Z_3+Z_4+\Re Z_5$.
To identify the neutral Higgs mass-eigenstates,
we diagonalize the squared-mass matrix $\mathcal{M}^2$.
The diagonalization matrix is a $3\times 3$
real orthogonal matrix that depends on three angles:
$\theta_{12}$, $\theta_{13}$ and~$\theta_{23}$,
\be \label{mixingmatrix}
\begin{pmatrix} h_1\\ h_2 \\ h_3\end{pmatrix}
=
\begin{pmatrix} c_{12} c_{13} & \quad -s_{12}c_{23}-c_{12}s_{13}s_{23} &
\quad -c_{12}s_{13}c_{23}+s_{12}s_{23} \\
s_{12} c_{13} & \quad c_{12}c_{23}-s_{12}s_{13}s_{23} &
\quad -s_{12}s_{13}c_{23}-c_{12}s_{23} \\
s_{13} & \quad c_{13}s_{23} & c_{13}c_{23}\end{pmatrix}\begin{pmatrix}
\sqrt{2}\Re H_1^0 -v \\ \sqrt{2}\Re H_2^0\\ \sqrt{2}\Im H_2^0 \end{pmatrix}\,,
\ee
where the $h_i$ are the mass-eigenstate neutral Higgs fields,
$c_{ij}\equiv\cos\theta_{ij}$ and $s_{ij}\equiv\sin\theta_{ij}$.
Under the rephasing $H_2\to e^{i\chi}H_2$,
\be \label{rephasing}
\theta_{12}\,,\, \theta_{13}~{\hbox{\text{are invariant, and}}}\quad
\theta_{23}\to  \theta_{23}-\chi\,.
\ee
As shown in ref.~\cite{Haber:2006ue}, the invariant angles
$\theta_{12}$ and $\theta_{13}$ are in fact basis-independent
quantities---that is, they can be expressed explicitly in
terms of basis-independent combinations of quantities defined
in any scalar field basis.\footnote{More generally,
one can show that any quantity defined in the Higgs basis
that is invariant under the rephasing of $H_2\to e^{i\chi}H_2$
can be rewritten explicitly in a basis-independent form.}

The neutral Goldstone boson and the physical neutral Higgs states ($h_0\equiv G^0$ and $h_{1,2,3}$, respectively) are then given by:
\be \label{hsubk}
h_j=\frac{1}{\sqrt{2}}\biggl\{q_{j1}^*\left(H_1^0-\frac{v}{\sqrt{2}}\right)
+q_{j2}^*H_2^0 e^{i\theta_{23}}+{\rm h.c.}\biggr\}\,,
\ee
where
the $q_{j1}$ and $q_{j2}$ are invariant
combinations of  $\theta_{12}$ and $\theta_{13}$,
which are exhibited in Table~\ref{tabinv}.
In particular, note that the quantities $q_{j1}$ and $q_{j2}$
are basis-invariants and the neutral fields $h_k$ are also
invariant with respect to a rephasing of the Higgs basis field $H_2$.
%
\begin{table}[t!]
\centering
\begin{tabular}{|c||c|c|}\hline
$\phaa j\phaa $ &\phaa $q_{j1}\phaa $ & \phaa $q_{j2} \phaa $ \\
\hline
$0$ &  $i$ & $0$ \\
$1$ & $c_{12} c_{13}$ & $-s_{12}-ic_{12}s_{13}$ \\
$2$ & $s_{12} c_{13}$ & $c_{12}-is_{12}s_{13}$ \\
$3$ & $s_{13}$ & $ic_{13}$ \\ \hline
\end{tabular}
\caption{Invariant combinations of the neutral Higgs
boson mixing angles $\theta_{12}$ and $\theta_{13}$,
where $c_{ij}\equiv\cos\theta_{ij}$ and
$s_{ij}\equiv\sin\theta_{ij}$.
\label{tabinv}}
\end{table}

To identify the diagonalizing matrix $V$ defined in
\eq{S0},\footnote{Note that in the notation used here,
$h_{\beta-1}\equiv S^0_{\beta}$, where $\beta=1,2,3,4$.}
we make use of \eq{h1h2} to rewrite \eq{hsubk} as follows,
\be \label{Phisubi}
\Phi_i=\left(\begin{array}{c}G^+\hat v_i+H^+ \hat w_i\\[6pt]
\displaystyle
\frac{v}{\sqrt{2}}\hat v_i+\frac{1}{\sqrt{2}}\sum_{j=0}^3
\left(q_{j1}\hat v_i+q_{j2}e^{-i\theta_{23}}\hat w_i\right)h_j
\end{array}\right)\,,
\ee
for $i=1,2$.  Hence, it immediately follows that
\be \label{v2hdm}
V_{ij}=q_{j1}\hat{v}_i+q_{j2}e^{-i\theta_{23}}\hat{w}_i\,.
\ee

Using \eqs{u2hdm}{v2hdm}, we can now evaluate the matrices $B$ and $A$.
First,
\be \label{beemat}
B=U^\dagger V=\begin{pmatrix} i & \quad q_{11} & \quad q_{21} & \quad q_{31} \\
0 &\quad  q_{12}e^{-i\theta_{23}} &
\quad q_{22}e^{-i\theta_{23}} &\quad q_{32}e^{-i\theta_{23}} \end{pmatrix}\,.
\ee
It is straightforward to check that
\be
BB^\dagger=2\begin{pmatrix} 1 & \,\,\, 0 \\ 0 &
\,\,\, 1\end{pmatrix}\,,\qquad \quad \Re(B^\dagger B)=\bm{1}_{4 \times 4}
\ee
as noted in \eq{bbid}.

We immediately see that $B$ is not an invariant matrix in
light of \eq{rephasing}.
Nevertheless, in \eq{kinetic} we note that the matrix $B$
always appears along with the charged Higgs or Goldstone fields,
namely $B_{a\beta}S_a^-$ (and its hermitian conjugate).
Recall that under the rephasing of the Higgs basis field
$H_2\to e^{i\chi} H_2$, we have $\hat{w}\to e^{-i\chi}\hat{w}$,
whereas $\hat{v}$ is invariant.
\Eq{Phisubi} implies that $G^\pm$ is invariant and
\be
H^{\pm}\to e^{\pm i\chi}H^{\pm}\,.
\ee
From \eq{beemat} we see that $B_{1j}$ is invariant and
$B_{2j}\to  e^{i\chi}B_{2j}$ in light of \eq{rephasing},
whereas $S_1^-=G^-$ is invariant and $S_2^-=H^-\to e^{-i\chi}H^-$
as noted above.
Hence, the combination $B_{a\beta}S_a^-$ is invariant as expected.

Next, we construct the orthogonal matrix $\tilde B$ defined in \eq{tildeB},
\be
\tilde{B}
=
\left(
\begin{array}{c}
\Re B\\*[1mm]
\Im B
\end{array}
\right)
=
\begin{pmatrix}
0 & \quad q_{11} & \quad q_{21}
    & \quad q_{31} \\
0 & \quad \Re(q_{12} e^{-i\theta_{23}})
    & \quad \Re(q_{22} e^{-i\theta_{23}})
    & \quad \Re(q_{32} e^{-i\theta_{23}})  \\
1 & \quad 0 & \quad 0 & \quad 0 \\
0 & \quad \Im(q_{12} e^{-i\theta_{23}})
    & \quad \Im(q_{22} e^{-i\theta_{23}})
    & \quad \Im(q_{32} e^{-i\theta_{23}})
\end{pmatrix}\,.
\ee
Using the results of Table~\ref{tabinv},
\be \label{btil}
\tilde B=\begin{pmatrix} 0 & \quad c_{12} c_{13}
    &\quad  s_{12} c_{13} &\quad s_{13}\\
0 & \quad -s_{12} c_{23}-c_{12}s_{13} s_{23}
    & \quad c_{12} c_{23} -s_{12}s_{13} s_{23} & \quad c_{13} s_{23} \\
1 & \quad 0 & \quad 0 & \quad 0 \\
0 & \quad s_{12} s_{23}-c_{12}s_{13} c_{23}
    & \quad -s_{12}s_{13}c_{23}-c_{12}s_{23} & \quad c_{13} c_{23}
\end{pmatrix}\,.
\ee
Indeed, one easily checks that $\tilde B^T \tilde B= \bm{1}_{4 \times 4}$.

In the 2HDM, the charged Higgs basis and the Higgs basis coincide.
Thus, the matrix $\tilde B$ rotates the Higgs basis fields into
the neutral Higgs mass eigenstate fields (which includes
the massless Goldstone field, $G^0=\sqrt{2}\Im H_1^0$).
More precisely, comparison with \eq{mixingmatrix} yields
\be
\begin{pmatrix} G^0\\ h_1\\ h_2 \\ h_3\end{pmatrix}
=
\tilde B^T
\begin{pmatrix}
\sqrt{2}\Re H_1^0 -v \\ \sqrt{2}\Re H_2^0\\
\sqrt{2}\Im H_1^0 \\ \sqrt{2}\Im H_2^0
\end{pmatrix}\,.
\ee

Finally, we evaluate the matrix $A$,
\be \label{afirst}
A=\Im(B^\dagger B)
=
\begin{pmatrix} 0
    & \quad -q_{11} & \quad -q_{21} & \quad -q_{31} \\
q_{11} & \quad 0 & \quad \Im(q_{12}^* q_{22})
    & \quad \Im(q_{12}^* q_{32}) \\
q_{21} &  \quad -\Im(q_{12}^* q_{22}) &\quad 0
    &  \quad \Im(q_{22}^* q_{32}) \\
q_{31} &  \quad -\Im(q_{12}^* q_{32})
    & \quad -\Im(q_{22}^* q_{32}) & \quad 0
\end{pmatrix}\,,
\ee
after using \eq{beemat}.
Once again, we can use the results of Table~\ref{tabinv} to obtain,
\be \label{amat}
A=
\begin{pmatrix}
0 & \quad -c_{12} c_{13}
    & \quad -s_{12} c_{13} & \quad -s_{13} \\
c_{12} c_{13} & \quad 0 &  \quad s_{13}
    & \quad -s_{12} c_{13} \\
s_{12} c_{13} & \quad -s_{13} & \quad 0
    & \quad c_{12} c_{13} \\
s_{13} & \quad s_{12} c_{13} & \quad -c_{12} c_{13}
    & \quad 0
\end{pmatrix}\,.
\ee
As expected, the matrix $A$ is invariant, as it depends only on the
invariant angles $\theta_{12}$ and $\theta_{13}$ [cf.~\eq{rephasing}].
It is now straightforward to check that the interaction Lagrangian involving
the coupling of the gauge bosons to the scalars given in
\eqst{VVS_lag}{GG_CHCH} reproduce the corresponding 2HDM results
given in ref.~\cite{Haber:2006ue}.

The power of the notation introduced in ref.~\cite{Haber:2006ue}
is clear in \eq{amat}, which depends on the only two invariant angles.
In contrast, the notation of \eqs{h_as_eta}{matrixR} conventionally
used in the C2HDM community, leads  to the matrix $A$ given
in \eq{tildeA_C2HDM_1} which seems to depend on three angles.
The true physical content of  \eq{tildeA_C2HDM_1}
becomes apparent only after rewriting it as in \eq{tildeA_C2HDM_2},
which as in the case of  \eq{amat} depends
only on two independent parameters.

Note that $A$ is a real orthogonal antisymmetric matrix,
as required by \eq{roa}.
Indeed, the most general $4\times 4$ real orthogonal antisymmetric
matrix depends on two parameters,
which we have identified with the two invariant angles
of the neutral Higgs squared-mass matrix diagonalization.

In the 2HDM, the special form of the $q_{j1}$ and $q_{j2}$
allow us to rewrite \eq{afirst} as
\be
A_{00}=0\,,\qquad\quad A_{j0}
=-A_{0j}=q_{j1}\,,\qquad\quad A_{ij}
=\epsilon_{ijk}q_{k1}\,,
\ee
where 0 labels the first row and column of $A$,
and the indices $i,j,k=1,2,3$ (with an implicit sum over $k$)
correspond to the second, third and fourth rows and columns of $A$.
This allows one to simplify the expression for the $Zh_i h_j$ vertex.
This is special to the case of $N=2$ and does not generalize
to the NHDM with $N>2$.

Given the explicit forms for the matrices $A$ and $B$ given
by \eqs{amat}{beemat} respectively, one can immediately check
that \eq{properties_B} is satisfied.

\section{\label{app:count}Counting parameters that govern the $A$ and $B$ matrices}

The matrices $A$ and $B$ enter the expression for the interaction
Lagrangian that couples the Higgs mass eigenstates to the gauge
bosons and Goldstone bosons.
The matrix $A$ is manifestly invariant under a change of the
scalar basis used in expressing the NHDM Lagrangian in terms
of interaction-eigenstate scalar fields.
The matrix $B$ is a pseudo-invariant quantity that depends
on $N-1$ unphysical phases.
However, these phases can be completely absorbed into
the definition of the $N-1$ physical charged Higgs fields.
In this appendix, we discuss the number of parameters that
govern the $A$ and $B$ matrix in the NHDM.

\subsection{Independent parameters of the matrix $A$}
\label{app:Acount}

The key properties of the matrix $A$ are given in \eq{roa}.
Namely, $A$ is an arbitrary real orthogonal antisymmetric $2N\times 2N$
matrix.
First, we note that $A$ is a $2N\times 2N$ nonsingular matrix
such that $\det A=1$.
Since $A^{T}A=\idtN$, it follows that $\det A=\pm 1$,
which implies that $A$ is nonsingular.
Moreover,
for any even-dimensional $2N\times 2N$ antisymmetric matrix $A$,
the \textit{pfaffian} of $A$, denoted by
pf~\!$A$, is defined by
\be
{\rm pf}~\!A=\frac{1}{2^{n} n!}\,\epsilon\lsub{i_1 j_1 i_2 j_2 \cdots i_n
  j_n} A_{i_1 j_1}A_{i_2 j_2}\cdots A_{i_n j_n}\,,
\ee
where $\epsilon$ is the rank-$2N$ Levi-Civita tensor, and
the sum over repeated indices is implied.
A well-known result states that for any antisymmetric
matrix $A$,\footnote{For a discussion of
the properties of the pfaffian, see, e.g., ref.~\cite{pfaff}.}
\be \label{pfm}
\det A=[{\rm pf}~A]^2.
\ee
In particular, if $A$ is also orthogonal then $\det A=1$,
in which case ${\rm pf}~A=\pm 1$.

Next, we note that the eigenvalues of any real antisymmetric
matrix $A$ are purely imaginary.
Moreover if $\lambda$ is an eigenvalue of $A$ then $\lambda^*$
is also an eigenvalue (see, e.g., ref.~\cite{Gallier}).
Thus, the eigenvalues of a $2N\times 2N$ antisymmetric matrix $A$
can be denoted by $\pm ia_i$, ($i=1,2,\ldots,n$)
where the $a_i$ are real and positive.
We now exploit the real normal form of a nonsingular $2N\times 2N$
real antisymmetric matrix $A$ (see, e.g., appendix D.4
of ref.~\cite{Dreiner:2008tw}).
In particular, there exists a real orthogonal matrix $Q$ such that
\be \label{takagiA}
Q^{T}AQ=  {\rm diag}\left\{\begin{pmatrix} \phm 0 & a_1 \\ -a_1 & 0
\end{pmatrix}\,,\,\begin{pmatrix} \phm 0 & a_2 \\ -a_2 & 0
\end{pmatrix}\,,\,\cdots \,,\begin{pmatrix} \phm 0 & a_N \\ -a_N & 0
\end{pmatrix}\right\}\,,
\ee
is written in block diagonal form with
$2\times 2$ matrices appearing along the diagonal
and the $a_i$ are real and positive.
Note that the $a_i$ are the positive square roots of
the eigenvalues of $A^{T}A$.

If in addition, $A$ is a real orthogonal matrix,
then we may use the fact that the eigenvalues of a
real orthogonal matrix are complex numbers of unit modulus.
In light of the above results, it follows that
$a_i=1$ for all $i=1,2,\ldots, n$.
Thus,
\be \label{QAQ}
Q^{T}AQ= J\equiv {\rm diag} \underbrace{\left\{\begin{pmatrix} \phm 0 & 1 \\ -1 & 0
\end{pmatrix}\,,\,\begin{pmatrix} \phm 0 & 1\\ -1 & 0
\end{pmatrix}\,,\,\cdots \,,\begin{pmatrix} \phm 0 & 1 \\ -1 & 0
\end{pmatrix}\right\}}_{N}\,.
\ee
Hence, we conclude that any real orthogonal antisymmetric
$2N\times 2N$ matrix $A$ can be parameterized by
\be \label{AQ}
A=QJQ^{T}\,,
\ee
where $J$ is defined in \eq{QAQ} and $Q$ is a real orthogonal matrix.
We now employ the well-known property of the pfaffian that
${\rm pf}(QJQ^{T})={\rm pf}~J\,\det Q$.
In light of ${\rm pf}~J=1$, it follows that
\be \label{detqm}
\det Q={\rm pf}~A\,,
\ee
which determines the sign of $\det Q$.

As discussed in appendix D.4 of ref.~\cite{Dreiner:2008tw},
the orthogonal matrix $Q$ in \eq{AQ} is unique up to multiplication
on the right by a $2N\times 2N$ real orthogonal matrix $S$ that
satisfies $SJS^{T}=J$.  Such a matrix $S$ is an element of Sp($N,\mathbb{R}$)$~\!\cap\!~$O(2$N$)~$\iso$~U($N$),
where a proof of this isomorphism is given in
ref.~\cite{Baker}.\footnote{See problem 1.12 on p.~41 and
its solution on p.~306 of ref.~\cite{Baker}.
The proof of this result consists of representing an arbitrary
complex unitary $N\times N$ matrix as a real $2N\times 2N$ matrix.
Following section 1.6 of ref.~\cite{Baker}, the corresponding real
$2N\times 2N$ matrix can be identified by $\tilde{U}_R$ given in \eq{Ureal}.
Indeed, one can check that $\tilde{U}_R$ is a $2N\times 2N$
orthogonal symplectic matrix.}
Since O($2N$) is parameterized by $N(2N-1)$ continuous parameters
and U($N$) is parameterized by $N^2$ parameters, we can use the
freedom to multiply $Q$ on the right by $S$ to remove $N^2$
parameters from $Q$.
This leaves
$N(2N-1)-N^2=N(N-1)$ parameters in $Q$ that cannot be removed.

That is, a real orthogonal antisymmetric $2N\times 2N$
matrix $A$ can be parameterized by $N(N-1)$ continuous parameters.

\subsection{Independent parameters of the matrix $B$}
\label{app:Bcount}

To determine the number of independent parameters that govern the
$N\times 2N$ matrix $B$, it is more convenient to consider the real
orthogonal $2N\times 2N$ matrix $\tilde B$.
The transpose of this matrix rotates the charged Higgs basis
fields into the neutral Higgs mass eigenstate fields $S_\beta^0$,
\be
S_\beta^0=\sum_{k=1}^{2N} \tilde B_{k\beta}\mathcal{H}^0_k\, ,
\ee
where $S_1^0\equiv G^0$ is the neutral Goldstone boson field, and
\be
\mathcal{H}^0_k=(H^0\,,\,\Re\,\varphi^{C0}_2\,,\,
\ldots\,,\,\Re\,\varphi^{C0}_N\,,\,G^0\,,\,\Im\,\varphi^{C0}_2\,,\,
\ldots\,,\,\Im\,\varphi^{C0}_N)\,.
\ee
Note that
$H^0\equiv \Re\,\varphi^{C0}_1$ has the properties of the SM Higgs boson,
$G^0\equiv \Im\,\varphi^{C0}_1$ is the neutral Goldstone boson,
and $\tfrac{1}{\sqrt{2}}\varphi^{C0}_k$ ($k=1,2,\ldots,N$) are the neutral
components of the $N$ scalar fields in the charged Higgs basis [cf.~\eq{chhiggsbasis}].
It then follows that
\be
\tilde B_{j1}=\delta_{j,N+1}\,,\qquad\quad \tilde B_{N+1,k}=\delta_{k1}\,.
\ee
It is convenient to remove the first column and the $N+\nth{1}$
row of $\tilde B$ to eliminate the neutral Goldstone bosons state.
The resulting $(2N-1)\times (2N-1)$ matrix will be called~$\mathcal{R}$.
We label $\mathcal{R}_{k\beta}$ with row and column
indices that take on values from $1,2,\ldots,2N$,
but excluding $k=N+1$ and $\beta=1$.
That is,
\be \label{Sbeta}
S_\beta^0=\sum_{\substack{k=1\\[3pt] k\neq N+1}}^{2N} \mathcal{R}_{k\beta}\mathcal{H}^{0}_k\,,\qquad \text{for $\beta=2,3,\ldots,2N$}\,.
\ee

We can parameterize the matrix $\mathcal{R}$ as follows.
First we define the $2\times 2$ orthogonal matrix,
\be
r_{ij}=\begin{pmatrix} \cos\theta_{ij} & -\sin\theta_{ij} \\ \sin\theta_{ij} &  \phm\cos\theta_{ij}\end{pmatrix}\,.
\ee
We then define the $(2N-1)\times (2N-1)$ matrix $R$
as a matrix whose matrix elements are given by
\be
R_{k\ell}=\begin{cases} \delta_{k\ell} &  \text{for $k, \ell\neq i,j$}
\,,\\ r_{ij}\,, &\text{for $k, \ell=i,j$}\,.\end{cases}
\ee
Then, $\mathcal{R}$ can be written recursively as~\cite{Murnaghan}
\be
\mathcal{R}=R_{12}R_{13}\cdots R_{1,2N-1} \mathscr{R}_{2N-2}\,,
\ee
where $\mathscr{R}_{2N-2}$ can be expressed in block matrix form
in terms of a $(2N-2)\times (2N-2)$ real orthogonal matrix $\mathcal{R}_{2N-2}$,
\be
\mathscr{R}_{2N-2}\equiv \begin{pmatrix} 1 & \quad 0 \\ 0 &\quad   \mathcal{R}_{2N-2}\end{pmatrix}\,.
\ee

It is always possible to express $\mathcal{R}_{2N-2}$ by making use
of the decomposition given in \eq{rcur},
\be
\mathcal{R}_{2N-2}=R_c U_R\,,
\ee
where $R_c$ is a $(2N-2)\times (2N-2)$ matrix given in block form by
\be
R_c=\exp\begin{pmatrix} C & \quad \phm D \\ D & \quad -C\end{pmatrix}\,,
\ee
where $C$ and $D$ are $(N-1)\times (N-1)$ real antisymmetric matrices,
and the matrix $U_R$ is a $(2N-2)\times (2N-2)$
real representation of the U($N-1$) subgroup of SO($2N-2$).
Thus, our final expression for the matrix $\mathcal{R}$ is
\be \label{Rdecomp}
\mathcal{R}=R_{12}R_{13}\cdots R_{1,2N-1} \begin{pmatrix} 1 & \quad 0 \\ 0 &\quad  R_c U_R\end{pmatrix}\,.
\ee

We have already noted in section~\ref{subsec:matrix_B} that under
the rephasing of the physical charged Higgs fields,
$S^+_a\to e^{i\chi_a}S_a^+$, we must also transform
$B_{a\beta}\to e^{i\chi_a}B_{a\beta}$ (in both cases for
$a=2,3,\ldots,N$), so that the combinations $B_{a\beta}S_a^-$
and its charged conjugate appearing in the interaction Lagrangian are invariant.
Since $U_R$ in \eq{Rdecomp} is a real representation of U($N-1$),
which depends on $(N-1)^2$ parameters, it is convenient to
decompose this matrix into the product,
\be
U_R=U_{Rc}U_{Rd}\,,
\ee
where
\be
 U_{Rd}\in \underbrace{{\rm U}(1)\times{\rm U}(1)\times\cdots\times{\rm U}(1)}_{N-1}\,,
\ee
is a real representation of an element of the diagonal
${\rm U}(1)\times{\rm U}(1)\times\cdots\times{\rm U}(1)$
subgroup of U($N-1$), which depends on $N-1$ parameters,
and $U_{Rc}$ incorporates the remaining
$(N-1)(N-2)$ degrees of freedom of $U_R$.
Then, we see that
$U_{Rd}$ represents the freedom
to perform separate rephasings of the $N-1$ scalar doublets with
zero vevs.
That is, the degrees of freedom in the
matrix $U_{Rd}$ represent the freedom to redefine the charged Higgs basis.
Hence, the $N-1$ parameters that govern the matrix $U_{Rd}$ are unphysical.

The remaining parameters that describe the matrix
$\mathcal{R}$ are physical.  We can count these parameters as follows.
First, the product $R_{12}R_{13}\cdots R_{1,2N-1}$ consists of
$2N-2$ angles $\theta_{12}$, $\theta_{13},\ldots,\theta_{1,2N-1}$.
Second, the number of parameters that govern the $(2N-2)\times (2N-2)$
matrix $R_c$ is equal to the number of parameters needed to express the
two $(N-1)\times (N-1)$ real antisymmetric matrices $C$ and $D$.
This provides $(N-1)(N-2)$ additional parameters.
Finally, we must include the $(N-1)(N-2)$ parameters that govern the matrix $U_{Rc}$.
Thus, we have $2N-2+2(N-1)(N-2)=2(N-1)^2$ parameters.
These are basis-invariant parameters, since they do not depend
on the choice of the charged Higgs basis.

The end result is that the matrix $\widetilde B$ can be expressed in terms of $2(N-1)^2$ physical parameters.  Indeed, this number can be obtained starting with the $(N-1)(2N-1)$ parameters that describe the $(2N-1)\times (2N-1)$ real orthogonal matrix $\mathcal{R}$ [cf.~\eq{Sbeta}], and then subtracting the $N-1$ unphysical phases by absorbing them into the definition of the physical charged Higgs fields as described above.

\subsection{\label{embed}The embedding of the U($N$) subgroup inside SO($2N$)}

We begin with the following theorem,
which is useful in the analysis of spontaneous symmetry breaking of
an SO($2N$) symmetric potential in a theory with a second-rank
antisymmetric tensor multiplet of scalars~\cite{symmbreak1,symmbreak2,symmbreak3}.
\vskip 0.1in

{\bf Theorem}: Suppose that $\Sigma_0$
is a $2N\times 2N$ real antisymmetric matrix that satisfies
$\Sigma_0^{T}\Sigma_0=\Sigma_0\Sigma_0^{T}=c^2 \bm{1}_{2N \times 2N}$
for some real number $c$.
Then, if the generators of the Lie algebra of SO($2N$),
henceforth denoted by $\mathfrak{so}(2N)$,
in the defining ($2N$-dimensional)
representation are given by $\{T_a,X_b\}$, where the $iT_a$ and
$iX_b$ are real antisymmetric $2N\times 2N$ matrices that satisfy:
\ba
T_a\Sigma_0+\Sigma_0 T_a^{T}&=&0\,,\\
X_b\Sigma_0-\Sigma_0 X_b^{T}&=&0\,,
\ea
then the $T_a$ span a $\mathfrak{u}(N)$ Lie subalgebra of
$\mathfrak{so}(2N)$, while the remaining generators,
$X_b$, span elements of $\mathfrak{so}(2N)$
whose exponentials comprise
the SO($2N$)$/$U($N$) homogeneous space.
Moreover, $\Tr(T_a X_b)=0$.
\vskip 0.1in

{\bf Proof:} First, we show that if
$\Sigma_0^{T}\Sigma_0=\Sigma_0\Sigma_0^{T}=c^2 \bm{1}_{2N \times 2N}$
and $T_a\Sigma_0+\Sigma_0 T_a^{T}=0$, then the $T_a$ span an U($N$) Lie
subalgebra.  Note that these two conditions imply:
\be \label{tat3}
c^2\, T_a^{T}=-\Sigma_0^{T} T_a\Sigma_0\,.
\ee
In light of \eq{takagiA}, there exists a real orthogonal matrix $W$ such that
$WMW^{T}={\rm diag}(\mathcal{J}_1\,,\mathcal{J}_2\,,\ldots\,,\mathcal{J}_n$)
is block diagonal, where each block is a $2\times 2$ matrix of
the form
$\mathcal{J}_n\equiv{\phantom{-}0\,\,\,\, z_n\choose -z_n
\,\,\,\,0\phantom{_n}}$,
where $z_n\in\mathbb{R}$ and the $z_n^2$ are the eigenvalues
of $MM^{T}$ (or $M^{T} M$).
Applying this result to $\Sigma_0$, note that the eigenvalues of
$\Sigma_0\Sigma_0^{T}$ are all degenerate and equal to $c^2$.
Moreover, since the matrix
\be \label{jdef}
\tilde{J}\equiv \begin{pmatrix} 0 &\quad \bm{1}_{N \times N} \\ -\bm{1}_{N \times N} & \quad 0\end{pmatrix}\,.
\ee
satisfies $\tilde{J}\tilde{J}^{T}
=\bm{1}_{2N \times 2N}$, it follows that one can find real
orthogonal matrices $W_1$ and $W_2$
such that $W_1\Sigma_0 W_1^{T}=cW_2 \tilde{J} W_2^{T}={\rm diag}(c\mathcal{J},
c\mathcal{J},\ldots,c\mathcal{J})$, where $\mathcal{J}$ is the $2\times 2$ matrix,
\be \label{caljdef}
\mathcal{J}\equiv\begin{pmatrix}\phm 0 & \quad 1 \\ -1 &\quad  0\end{pmatrix}\,.
\ee
That is, the factorization of
$\Sigma_0$ and $c\tilde{J}$ both yield the same block diagonal matrix
consisting of $N$ identical $2\times 2$ blocks consisting of $c\mathcal{J}$.
Thus, there exists a real orthogonal matrix $V=W_2^{-1}W_1$ such that
$V\Sigma_0 V^{T}=c\,\tilde{J}$.
The inverse of this result is $V\Sigma_0^{T} V^{T}=-c\,\tilde{J}$
(since $\tilde{J}^{T}=-\tilde{J}$).
We now define $\widetilde T_a\equiv VT_aV^{T}$.
Then \eq{tat3} implies that
\be \label{tildetat3}
\widetilde T_a^{T}
=\frac{-1\phantom{0}}{c^2}V\Sigma_0^{T} V^{T}
\widetilde T_a V\Sigma_0 V^{T}
= \tilde{J}\widetilde T_a \tilde{J}\,.
\ee
Likewise, one
can use the same matrix $V$ to define $\widetilde X_b\equiv VX_b V^{T}$.
By an analogous computation, $c^2 X^{T}=\Sigma_0^{T} X\Sigma_0$, which
implies that $\widetilde X_b^{T}=-\tilde{J}\widetilde X_b \tilde{J}$.

Recall that $T_a$ and $X_b$ are both antisymmetric $2N\times 2N$
matrices.  Then,
$\widetilde T_a\equiv VT_aV^{T}$ and $\widetilde X_a\equiv VX_aV^{T}$
are also antisymmetric.  Hence, it follows that
\be \label{jtxj}
\widetilde T_a= -\tilde{J}\widetilde T_a \tilde{J}\,,\qquad
\widetilde X_a= \tilde{J}\widetilde X_a \tilde{J}\,.
\ee
Using the explicit form for $\tilde{J}$, \eq{jtxj} implies that $\widetilde{T}_a$ and
$\widetilde{X}_b$ take the following block form:
\be \label{blockform}
i\,\widetilde T_a=\begin{pmatrix}\phm A & \quad B \\ -B & \quad A\end{pmatrix}\,,\qquad\qquad
i\widetilde X_b=\begin{pmatrix}C& \quad \phm D \\D & \quad -C\end{pmatrix}\,,
\ee
where $A$, $B$, $C$ and $D$ are $N\times N$ real
matrices such that $A$, $C$ and $D$ are antisymmetric and $B$ is
symmetric.
Thus, we have exhibited a similarity transformation
(note that $V^{T}=V^{-1}$) that transforms
the basis of the Lie algebra spanned by the $T_a$ to one that is
spanned by the $\widetilde T_a$.  Moreover, consider the isomorphism
that maps $i\,\widetilde T_a$
given in \eq{blockform} to the $N\times N$ matrix
$A+iB$.  Since $(A+iB)^\dagger=(A-iB)^{T}=-(A+iB)$, we see that the
$A+iB$ are anti-hermitian generators (which are not generally
traceless) that span a $\mathfrak{u}(N)$ subalgebra of $\mathfrak{so}(2N)$.
We can check the number of $\mathfrak{u}(N)$ generators by counting the
number of degrees of freedom of one real antisymmetric and one real
symmetric matrix: $\half N(N-1)+\half N(N+1)=N^2$, as expected.

Finally, multiplying the two equations $c^2\,T_a^{T}=-\Sigma_0^{T} T_a\Sigma_0$ and $c^2 X_b^{T}=\Sigma_0^{T} X_b\Sigma_0$, it follows that
$c^2\,T_a^{T} X_b^{T}=-\Sigma_0^{T} T_a X_b\Sigma_0$ (after employing $\Sigma_0^{T}\Sigma_0=c^2 I_{2n}$).  Taking the trace yields $\Tr T_a
X_b=-\Tr T_a X_b$, and we conclude that
$\Tr T_a X_b=0$.

To show that the
$\{T_a,X_b\}$ span the full $\mathfrak{so}(2N)$ Lie algebra,
we have already noted above that there
are $N^2$ generators, $\{T_a\}$.
In addition, there are $N(N-1)$  generators, $\{X_a\}$,
corresponding to the number of parameters describing two real
antisymmetric matrices [see \eq{blockform}].
Thus, the total number
of generators is $N(2N-1)$ which matches the total
number of $\mathfrak{so}(2N)$
generators.

Any element of the SO($2N$) group is an exponential of an element
of the corresponding $\mathfrak{so}(2N)$ Lie algebra.
This provides many possible choices for parameterizing an
arbitrary element of the SO($2N$) group.
We shall choose $\widetilde T_a$ and $\widetilde X_b$ as generators
of the $\mathfrak{so}(2N)$ Lie algebra.
Exponentiating the appropriate linear combinations of
generators [cf.~\eq{blockform}] allows us to express any
element $\mathcal{R}_{2N}\in$ SO($2N$) in the following form,
\be \label{rcur}
\mathcal{R}_{2N}=R_c \tilde{U}_R\,,
\ee
where
\be
R_c\equiv\exp \begin{pmatrix} C&\,\,\, \phm D \\ D &\,\,\,  -C\end{pmatrix}\,,\qquad\quad \tilde{U}_R\equiv\exp \begin{pmatrix}\phm A &\quad  B \\ -B &\quad  A\end{pmatrix}\,,
\ee
where $A$, $B$, $C$ and $D$ are $N\times N$ real
matrices such that $A$, $C$ and $D$ are antisymmetric and
$B$ is symmetric.
Based on the discussion below \eq{blockform},
we recognize $\tilde{U}_R$ as the $2N$-dimensional real
representation of the group U($N$).
That is, given an $N\times N$ unitary matrix $U$,
one can identity,
\be
\tilde{U}_R\equiv\left(\begin{array}{cc}\Re U &\ -\Im U \\
\Im U & \ \phantom{-}\Re U \end{array}\right)\,,
\ee
as a $2N\times 2N$ real orthogonal matrix that provides the
explicit form for the embedding of U($N$) inside SO($2N$).

Since the exponential of any element of the Lie algebra
$\mathfrak{so}(2N)$ yields an element of SO($2N$), one can also
choose a different order in the product of exponentials to
parameterize an element of SO($2N$).
For example,
one can also express any element $\mathcal{R}_{2N}\in$ SO($2N$)
in the following form,
\be \label{rcurp}
\mathcal{R}_{2N}=\widetilde{W}_R \tilde{R}_c\,,
\ee
where
\be \label{rpdef}
\tilde{R}_c\equiv\exp \begin{pmatrix} C'& \,\,\, \phm D'\\ D' &\,\,\,  -C'\end{pmatrix}\,,\qquad\quad
\widetilde{W}_R=\left(\begin{array}{cc}\Re W&\ -\Im W \\
\Im W & \ \phantom{-}\Re W \end{array}\right)\,,
\ee
where $C'$ and $D'$ are real $N\times N$ antisymmetric matrices and $W$
is an $N\times N$ unitary matrix.
In general, $C'\neq C$, $D' \neq D$ and $W\neq U$.

\section{\label{app:cubic} Cubic couplings of the Goldsone boson and physical Higgs scalars}

It is quite remarkable that the cubic scalar couplings that involve a single or two Goldstone fields
can be simplified to expressions that involve either squared mass differences or squared masses of
the corresponding physical Higgs scalars, as exhibited in \eqst{eq:feyn1}{eq:feyn4}.  This was first noted in the context of the CP-conserving 2HDM in Ref.~\cite{Gunion:2002zf,Grinstein:2013fia}.

Achieving such simplified forms for these couplings is rather laborious.
It is instructive to provide some of the details of the derivation.
As an example,
we demonstrate below how to derive the coupling between a neutral scalar
(here denoted by $S_p^0$) and two Goldstone bosons $G^0=S_1^0$
obtained in eq.~\eqref{eq:feyn4}.
Starting from eq.~\eqref{V3_mass},
there seem to be three relevant terms:
those with $(\beta, \gamma, \delta) = (p,1,1)$,
$(\beta, \gamma, \delta) = (1,p,1)$,
and $(\beta, \gamma, \delta) = (1,1,p)$.
However, the latter two vanish due to eqs.~\eqref{V_k1}--\eqref{omk}.
Indeed,
when $\beta=1$,
the result involves
\be
(V^\dagger)_{1 k} v_l + v_k^\ast V_{l 1}
=
\left(-i \frac{v^\ast_k}{v}\right) v_l + v_k^\ast \left(i \frac{v_l}{v}\right)
= 0.
\ee
Applying eqs.~\eqref{V_k1}--\eqref{omk} to the remaining term,
we find for $4 v\, V_3[S_p^0 G^0 G^0]$,
\ba
&&
\frac{2}{v} \lambda_{ij,kl}
v_i^\ast v_j
\left[ \left( V^\dagger\right)_{pk} v_l + v^\ast_k V_{lp}\right]
\nonumber\\
&=&
\left[
- 2 \left( U D_{\pm}^2 U^\dagger \right)_{kj} + \left( V D_0^2 V^\dagger\right)_{kj}
\right]
\hat{v}_j \left( V^\dagger\right)_{pk}
+
\left[
- 2 \left( U D_{\pm}^2 U^\dagger \right)_{il} + \left( V D_0^2 V^\dagger\right)_{il}
\right]
\hat{v}_i^\ast \left( V \right)_{lp}
\nonumber\\
&=&
\left( V^\dagger V D_0^2 V^\dagger\right)_{pj} \hat{v}_j
+
\hat{v}_i^\ast
\left( V D_0^2 V^\dagger V \right)_{ip}
\nonumber\\
&=&
\left( V^\dagger V D_0^2\right)_{p\theta}
\textrm{Re}\left[ (V^\dagger)_{\theta j}\, \hat{v}_j\right]
+
\textrm{Re}\left[ \hat{v}_i^\ast V_{i \theta}\right]
\left( D_0^2 V^\dagger V \right)_{\theta p}
\nonumber\\
&=&
\left[
V^\dagger V D_0^2\,
\textrm{Im}\left( V^\dagger V \right)
\right]_{p 1}
-
\left[
\textrm{Im}\left( V^\dagger V \right) D_0^2
V^\dagger V
\right]_{1 p}
\nonumber\\
&=&
\left[
\textrm{Re}\left( V^\dagger V \right) D_0^2\,
\textrm{Im}\left( V^\dagger V \right)
\right]_{p 1}
-
\left[
\textrm{Im}\left( V^\dagger V \right) D_0^2
\textrm{Re}\left( V^\dagger V \right)
\right]_{1 p}
\nonumber\\
&=&
-
2 m_p^2
\left[
\textrm{Im}
\left(
V^\dagger V
\right)
\right]_{1 p},
\ea
thus reproducing eq.~\eqref{eq:feyn4}.
The most crucial step is the first,
where eq.~\eqref{Kprime_2} was used to relate these
couplings with the mass matrices.
The third line above is obtained by
using eqs.~\eqref{omU} and \eqref{Dpm2},
where the (11) entry vanishes,
which shows that the charged boson masses do not contribute,
as expected.
The fourth line is obtained by breaking
$(V^\dagger)_{\theta j} \hat{v}_j$ in its real and imaginary
parts, and then using eqs.~\eqref{ImVDomega} and \eqref{D02}
to show that the imaginary part involves
the vanishing entries of $D_0^2$.
Eq.~\eqref{ReVDomega} yields the fifth line.
To proceed,
we break the remaining $V^\dagger V$ matrix into its real and imaginary parts.
According to eq.~\eqref{BIA} the real part is the unit matrix,
while the imaginary part is antisymmetric.
The two terms involving the symmetric matrix
$\textrm{Im}(V^\dagger V) D_0^2 \textrm{Im}(V^\dagger V)$
cancel each other.
Given eqs.~\eqref{D02} and \eqref{BIA},
we reach the last line.

It is noteworthy that it took such a long calculation
to obtain such a simple result.
In fact, it turns out that such proofs are simpler when performed in the charged Higgs basis.

\section{\label{app:SR}Generalized sum rules}

In this appendix,
we rederive in detail the sum rules obtained by Gunion, Haber and
Wudka in ref.~\cite{Gunion:1990kf}.
Following the conventions of ref.~\cite{Gunion:1990kf},
we indicate vector bosons with indices $a, b, c, \ldots$ and
scalars with indices $i, j, k,\ldots$.
The Feynman rules for the cubic vertices are
\ba
\label{eq:1}
A^\alpha_a A^\beta_b A^\gamma_c: &&
i\, g_{abc}\left[ (p_a -p_b)^\gamma
+ (p_b -p_c)^\alpha + (p_c -p_a)^\beta\right]
\equiv i\, g_{abc}\ \Gamma^{\alpha\beta\gamma}(p_a,p_b,p_c),
\\[+2mm]
A^\alpha_a A^\beta_c \phi_i: &&
i\, g_{abi}\ g^{\alpha\beta},
\\[+2mm]
A_a^\alpha \phi_i \phi_j: &&
i\, g_{aij}\ (p_i-p_j)^\alpha,
\ea
with all momenta incoming, and the Feynman rules for the quarticic vertices are
\be
A_a^\alpha A_b^\beta \phi_i \phi_j:
i\, g_{abij}\ g^{\alpha \beta},
\ee

For the sum rule involving four gauge
bosons we also need the relation between the quartic and cubic term. We
adopt here the conventions of Cornwall, Levin and
Tiktopoulos~\cite{Cornwall:1974km}.
We just need the relevant terms,
\begin{equation}
\label{eq:2}
\mathcal{L} = - D_{abcd}\ W_{a\mu} W^\mu_b W_{c\nu} W^\nu_d
- C_{abc}\ \partial_\nu W_{a\mu} W^\mu_b W^\nu_c + \cdots
\end{equation}
This gives the following Feynman rules,
\begin{eqnarray}
A_a^\mu A_b^\nu A_c^\rho A_d^\sigma
:&&
-8i\, \left(D_{abcd}\ g^{\mu\nu} g^{\rho\sigma}
+D_{acbd}\ g^{\mu\rho} g^{\nu\sigma} + D_{adbc}\ g^{\mu\sigma}
g^{\nu\rho}\right),
\label{eq:3a}\\[+2mm]
A_a^\alpha A_b^\beta A_c^\gamma
:&&
-C_{abc}\  \left[ (p_a -p_b)^\gamma
+ (p_b -p_c)^\alpha + (p_c -p_a)^\beta\right] =
-C_{abc}\ \Gamma^{\alpha\beta\gamma}(p_a,p_b,p_c),
\nonumber\\
&&
\label{eq:3b}
\end{eqnarray}
where we note, for future reference, that comparing eq.~\eqref{eq:1}
and eq.~\eqref{eq:3b} we get,
\begin{equation}
\label{eq:4}
C_{abc}= -i\, g_{abc}  .
\end{equation}

Cornwall, Levin and Tiktopoulos~\cite{Cornwall:1974km},
and independently Llewellyn Smith~\cite{LlewellynSmith:1973yud})
show that, in order for unitarity to hold, the couplings $C_{abc}$ and
$D_{abcd}$ must be those of a gauge theory.
In particular,
\begin{eqnarray}
D_{a b c d} &=& \frac{1}{8}  (C_{ace} C_{bde} - C_{ade} C_{cbe}  ),
\label{eq:5a}
\\[+2mm]
0 &=& C_{abe} C_{cde} -C_{ace} C_{bde} - C_{ade} C_{cbe},
\label{eq:5b}
\end{eqnarray}
where the last relation is the Jacobi identity.
This means that $C_{abc}$ are the structure constants of the gauge group.
As we want to write everything in terms of the structure
constants $g_{abc}$ of ref.~\cite{Gunion:1990kf},
we use eq.~\eqref{eq:4} to obtain
\begin{eqnarray}
D_{a b c d} &=& - \frac{1}{8}  (g_{ace} g_{bde} - g_{ade} g_{cbe}  ),
\label{eq:6a}
\\[+2mm]
0 &=&
g_{abe} g_{cde} -g_{ace} g_{bde} - g_{ade} g_{cbe}.
\label{eq:6b}
\end{eqnarray}
\bigskip

\subsection{$AAAA$ Sum Rules}
\bigskip
\subsubsection{The amplitudes}

\begin{figure}[t!]
	\centering
	\includegraphics[width=0.6\linewidth]{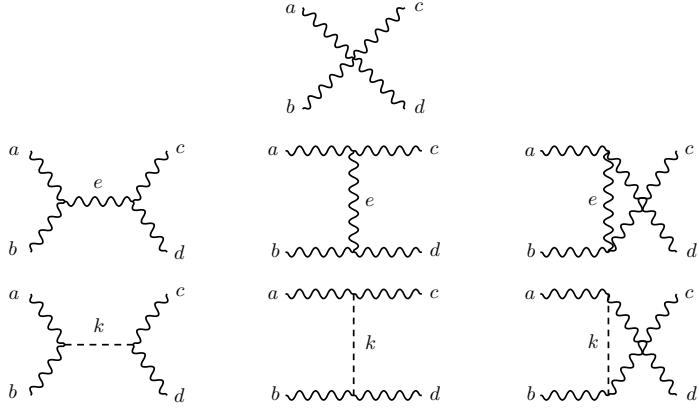}
	\caption{Amplitudes for the scattering $A_a A_b \to A_c A_d$}
	\label{fig:AAAA}
\end{figure}

The diagrams contributing to the scattering
$A_a(p_1) + A_b(p_2) \to A_c(p_3) + A_d(p_4)$
are given in figure~\ref{fig:AAAA}.
In an obvious notation we will name the amplitudes according to the
Mandelstam variables channel ($s, t$ or $u$) and by the particle being
exchanged.
We get
\begin{equation}
  \label{eq:18}
  \mathcal{M}^{\rm 4 Point} =(-i)^2  8
\left(D_{abcd}\, g_{\alpha\beta} g_{\gamma\delta} +D_{acbd}\, g_{\alpha\gamma}
g_{\beta\delta} + D_{adbc}\, g_{\alpha\delta} g_{\beta\gamma}\right)
f^{\alpha \beta \epsilon \delta},
\end{equation}
\clearpage

%
%

\begin{align}
  \label{eq:5}
  \mathcal{M}_s^{A} =&(-i) (i\, g_{abe}) (- i\, g_{cde}) (-i)
\Gamma_{\alpha\beta\mu}(p_1,p_2,-p_1-p_2) \Gamma_{\delta\gamma\nu}(-p_4,-p_3,-p_1-p_2)
\nonumber\\
&\hskip 5mm
\frac{\left[g^{\mu\nu} -
(p_1+p_2)^\mu (p_1+p_2)^\nu/M_e^2 \right]}
{s -M_e^2}
f^{\alpha \beta \epsilon \delta},
\end{align}
\begin{align}
  \label{eq:13}
  \mathcal{M}_t^{A} =&(-i) (-i\, g_{ace}) ( i\, g_{bde}) (-i)
\Gamma_{\alpha\mu\gamma}(p_1,-p_1+p_3,-p_3)
\Gamma_{\beta\delta\nu}(p_2,-p_4,p_4-p_2)
\nonumber\\
&\hskip 5mm
\frac{\left[g^{\mu\nu} -
(p_1-p_3)^\mu (p_1-p_3)^\nu/M_e^2 \right]}
{t -M_e^2}
f^{\alpha \beta \epsilon \delta},
\end{align}
\begin{align}
  \label{eq:14}
  \mathcal{M}_u^{A} =&(-i) (-i\, g_{ade}) ( - i\, g_{cbe}) (-i)
\Gamma_{\alpha\mu\delta}(p_1,-p_1+p_4,-p_4)
\Gamma_{\beta\gamma\nu}(p_2,-p_3,p_3-p_2)
\nonumber\\
&\hskip 5mm
\frac{\left[g^{\mu\nu} -
(p_1-p_4)^\mu (p_1-p_4)^\nu/M_e^2 \right]}
{u -M_e^2}
f^{\alpha \beta \epsilon \delta},
\end{align}
\begin{equation}
  \label{eq:15}
  \mathcal{M}_s^{\phi} = (-i) (i g_{abk}) (i g_{cdk}) i\
     \frac{g_{\alpha\beta}\ g_{\gamma\delta}}{s-M_k^2}
f^{\alpha \beta \epsilon \delta},
\end{equation}
\begin{equation}
  \label{eq:16}
  \mathcal{M}_t^{\phi} = (-i) (i g_{ack}) (i g_{bdk}) i\
     \frac{g_{\alpha\gamma}\ g_{\beta\delta}}{t-M_k^2}
f^{\alpha \beta \epsilon \delta},
\end{equation}
\begin{equation}
  \label{eq:17}
  \mathcal{M}_u^{\phi} = (-i) (i g_{adk}) (i g_{bck}) i\
     \frac{g_{\alpha\delta}\ g_{\beta\gamma}}{u-M_k^2}
f^{\alpha \beta \epsilon \delta},
\end{equation}

\noindent
where
\be
f^{\alpha \beta \epsilon \delta} =
\epsilon^\alpha(p_1) \epsilon^\beta(p_2) \epsilon^\gamma(p_3) \epsilon^\delta(p_4).
\ee

\subsubsection{The high energy limit}

When the gauge bosons are longitudinally polarized,
the diagrams of
figure~\ref{fig:AAAA} grow with energy for large center of mass energy
$\sqrt{s}$.
The most divergent behavior arises from the first four diagrams
that grow like $E^4$. To see this, one has to use the expression for the
polarization vector for the longitudinal case, which is given by
\begin{equation}
\label{eq:20}
\epsilon_L=(\gamma\beta,\gamma \vec \beta/\beta) \simeq
\frac{p^\mu}{M} + \mathcal{O} \left(\frac{1}{\gamma^2} \frac{E}{M}\right).
\end{equation}
To determine the coefficients of the high energy behavior
(see eq.~\eqref{eq:3} below) we cannot use the
approximate expression on the right-hand side of eq.~\eqref{eq:20}
because we would then lose contributions from the $E^4$ terms that modify
the $E^2$ terms.
So we should use consistently the definitions of the left-hand side
and expand the result in powers of $s$, $t$ or $u$.
For instance, for particle $a$ we have,
\begin{eqnarray}
\label{eq:21}
\epsilon_a^L=(\gamma_a\beta_a,\gamma_a \vec \beta_a/\beta_a),
&&
\beta_a=\frac{\sqrt{E_a^2-M_a^2}}{E_a},
\nonumber\\
\gamma_a= \frac{1}{\sqrt{1-\beta_a^2}},
&&
E_a = \frac{s +M_a^2 -M_b^2}{2 \sqrt{s}},
\end{eqnarray}
and similarly for the other particles.
Next we use the kinematics for the process
\begin{equation}
\label{eq:42}
A_a(p_1)+A_b(p_2)\to A_c(p_3)+A_d(p_4),
\end{equation}
in order to write
\begin{align}
\label{eq:41}
&p_1=(E_a,0,0,\beta_a E_a),&&
p_2=(E_b,0,0,-\beta_b E_b),\\
&p_3=(E_c,\beta_c E_c \sin\theta,0,\beta_c E_c \cos\theta),&&
p_4=(E_d,-\beta_d E_d \sin\theta,0,-\beta_d E_d \cos\theta),\\
&\epsilon_a^L=(\gamma_a \beta_a,0,0, \gamma_a),&&
\epsilon_b^L=(\gamma_b \beta_b,0,0, -\gamma_b),\\
&\epsilon_c^L=(\gamma_c \beta_c,\gamma_c \sin\theta,0, \gamma_c \cos\theta),&&
\epsilon_d^L=(\gamma_d \beta_d,-\gamma_d \sin\theta,0, -\gamma_d \cos\theta).
\end{align}
We then use these expressions to evaluate all the amplitudes. In the
end we substitute $\cos\theta$ in terms of the Mandelstam variable
$t$, through the relation
\begin{equation}
\label{eq:43}
\cos\theta=\frac{t-M_a^2-M_c^2+ 2 E_a E_c}{2 E_a E_c \beta_a \beta_c}.
\end{equation}
At this point all the amplitudes are expressed in terms of the
Mandelstam variables and the masses. As the Mandelstam variables are
not independent we can still use the relation
\begin{equation}
\label{eq:44}
s + t + u = M_a^2+M_b^2+M_c^2+M_d^2,
\end{equation}
to express the result in terms of just two independent
variables.
Which ones should be used will depend on the problem.
Next we want to isolate the terms that grow with $E^4$ and $E^2$.
To achieve this we make the scaling
\begin{equation}
\label{eq:45}
s\to s/x, \quad t\to t/x, \quad u\to u/x,
\end{equation}
and perform an expansion for small $x$. The terms in $E^4$ are the
coefficients of $x^{-2}$ and the terms that grow like $E^2$ are the
coefficients of $x^{-1}$.
Therefore,
we can write for each amplitude
\begin{equation}
\label{eq:3}
M_i = A^s_i s^2 + A^t_i t^2 + A^{st}_i s t + B^s_i s + B^t_i t + \text{constant},
\end{equation}
where we assumed that the independent Mandelstam variables are $s$ and $t$.
We did this consistent expansion using \texttt{FeynCalc} and
\texttt{Mathematica} for the Lorentz algebra and series expansion,
respectively. To have an idea of what is involved we just write the
exact amplitude for the s-channel exchange of a gauge boson:
\clearpage
\begin{align}
&  \mathcal{M}_s^A=\frac{-g_{abe} g_{cde}}{4 M_a^2
	M_b^2 M_e^2 \left(s\!-\!M_e^2\right)
	\sqrt{\frac{M_c^2 s}{\left(M_c^2!-\!M_d^2+s\right)^2}}
	\left(M_c^2-M_d^2+s\right) \sqrt{\frac{M_d^2
			s}{\left(M_d^2\!-\!M_c^2+s\right)^2}}
	\left(M_d^2\!-\!M_c^2+s\right)}\nonumber\\[+2mm]
&\hskip 3mm
\left[ \sqrt{\frac{M_a^2
		s}{\left(M_a^2-M_b^2+s\right)^2}}
\left(M_a^2-M_b^2+s\right) \sqrt{\frac{M_b^2
		s}{\left(-M_a^2+M_b^2+s\right)^2}}
\left(-M_a^2+M_b^2+s\right) \right.\nonumber\\[+2mm]
&\hskip 3mm
\left(M_a^4
\left(-M_c^4+M_c^2
\left(s-M_e^2\right)+M_d^4+M_d^2 \left(3
M_e^2-s\right)+M_e^2 s\right)\right.\nonumber\\[+2mm]
&\hskip 3mm
+M_a^2 \left(2
M_b^2 M_e^2
\left(M_c^2+M_d^2+s\right)+M_c^4
\left(s-M_e^2\right)+M_c^2 \left(2 M_d^2
M_e^2+M_e^2 (s-2 t)-s^2\right)\right.\nonumber\\[+2mm]
&\hskip 3mm
\left.
+3 M_d^4
M_e^2-M_d^4 s+M_d^2 M_e^2 s-2 M_d^2
M_e^2 t+M_d^2 s^2-2 M_e^2 s t\right)\nonumber\\[+2mm]
&\hskip 3mm
+M_b^4
\left(M_c^4+M_c^2 \left(3
M_e^2-s\right)-\left(M_d^2+M_e^2\right)
\left(M_d^2-s\right)\right)+M_b^2 \left(M_c^4 \left(3
M_e^2-s\right) \right.\nonumber\\[+2mm]
&\hskip 3mm
\left.
+M_c^2 \left(2 M_d^2
M_e^2+M_e^2 (s-2 t)+s^2\right)+M_d^4
\left(s-M_e^2\right)+M_d^2 \left(M_e^2 (s-2
t)-s^2\right)-2 M_e^2 s t\right) \nonumber\\[+2mm]
&\hskip 3mm
\left.\left.
+M_e^2 s
\left(M_c^2+M_d^2+s\right)
\left(M_c^2+M_d^2-s-2 t\right)\right)\right],
\end{align}
This is a quite complicated expression,
but making the series expansion as described above gives simply
\begin{align}
\label{eq:19}
\mathcal{M}_s^{A} =&\frac{1}{M_a M_b M_c M_d}\left[
\frac{1}{4} g_{abe} g_{cde}\ s^2
+\frac{1}{2} g_{abe} g_{cde} s t \right. \nonumber\\[+2mm]
& \hskip 25mm 
+ \frac{1}{4}g_{abe} g_{cde} \left[M_e^2
+\frac{(M_a^2\!-\!M_b^2)
	(M_c^2\!-\!M_d^2)}{M_e^2}\right]\ s \nonumber\\[+2mm]
&\left. \hskip 25mm 
+\frac{1}{2} g_{abe} g_{cde}
(M_a^2+M_b^2+M_c^2+M_d^2 +M_e^2)\ t + \text{constant} \right]
\end{align}

If we had not taken the exact expression for the polarization vectors
in eq.~\eqref{eq:20} but only the approximate expression, we would have obtained
instead of eq.~\eqref{eq:19} the following expression,
\begin{align}
\label{eq:46}
\mathcal{M}_s^{A} =&\frac{1}{M_a M_b M_c M_d}\left[
\frac{1}{4}\, g_{abe} g_{cde}\ s^2
+\frac{1}{2}\, g_{abe} g_{cde}\ s t \right.\nonumber\\[+2mm]
&\hskip 25mm 
+\frac{1}{4} g_{abe} g_{cde} \left[M_e^2
\!- \!2 M_a^2  \!- \!2 M_b^2  \!- \!M_c^2  \!- \!M_d^2\right] s
\nonumber\\[+2mm]
&\left. \hskip 25mm 
+\frac{1}{2} g_{abe} g_{cde}
\left[- M_a^2-  M_b^2 + M_e^2\right] t + \text{constant} \right],
\end{align}
which shows that the $E^4$ terms are correct but there is a difference
in the $E^2$ terms as previously anticipated. Note that no such problem arises for the exchange
of the scalars, as the most divergent
terms are of order $E^2$. Nevertheless we adopt the same procedure for
all the diagrams.

\subsubsection*{The $E^4$ terms}

The first four diagrams yield terms that grow like $E^4$. To simplify the
expressions we redefine the coefficients $\hat A_i =A_i M_a M_b M_c M_d$.
The corresponding $\hat A_i$ coefficients are given in table~\ref{tab:1}.
\renewcommand{\arraystretch}{1.4}
\begin{table}[htb]
	\centering
	\begin{tabular}{|c|c|c|c|}\hline
		Diagram & $\hat A_i^s$ &$\hat A_i^t$ &$\hat A_i^{st}$ \\[+1mm]
		\hline\hline
		$\mathcal{M}^{\rm 4 Point}$ & $- 2\left( D_{abcd}+D_{adbc}\right)$
		&$-2 \left(D_{acbd}+D_{adbc}\right)$ &
		$-4  D_{a d b c}$ \\[+1mm]  \hline
		$\mathcal{M}_s^{A}$ & $\frac{1}{4}\, g_{abe}\, g_{cde}$
		&$ 0 $ &
		$\frac{1}{2}\, g_{abe}\, g_{cde}$ \\[+1mm]  \hline
		$\mathcal{M}_t^{A}$ & $0 $
		&$\frac{1}{4}\, g_{ace}\, g_{bde} $ &
		$\frac{1}{2}\, g_{ace}\, g_{bde}$ \\[+1mm]  \hline
		$\mathcal{M}_u^{A}$ & $\frac{1}{4}\, g_{ade}\, g_{cbe}  $
		&$- \frac{1}{4}\, g_{ade}\, g_{cbe}$ &
		$0 $ \\[+1mm]  \hline \hline
		$\sum \mathcal{M}_i^{A}$ & $0 $
		&$0 $ &  $0 $ \\[+1mm]  \hline
	\end{tabular}
	\caption{Coefficients $\hat A_i$.}
	\label{tab:1}
\end{table}
The last line in this table is obtained after we use the relations
found in ref.~\cite{Cornwall:1974km} and given in eq.~\eqref{eq:6a}.
Namely,
\begin{align}
D_{a b c d} =& - \frac{1}{8}  (g_{ace}\, g_{bde} - g_{ade}\, g_{cbe}),
\label{eq:23} \\[+2mm]
D_{a c b d} =& - \frac{1}{8} (g_{abe}\, g_{cde} + g_{ade}\, g_{cbe}),
\label{eq:24}
\\[+2mm]
D_{a d b c} =& -\frac{1}{8} (-g_{abe}\, g_{cde} - g_{ace}\, g_{bde}),
\label{eq:25}
\end{align}
where the antisymmetry of the constants $g_{abc}$ was used. So the more
divergent terms cancel only with the gauge part. The constraints that emerge simply imply that
we must have a spontaneously broken gauge
theory~\cite{Cornwall:1974km,LlewellynSmith:1973yud}
as given in
eqs.~\eqref{eq:6a}-\eqref{eq:6b}.

\subsubsection*{The $E^2$ terms}

Having shown that a spontaneously broken gauge theory assures that the
the most divergent high energy behavior cancels, we consider next the terms that diverge
like $E^2$. Here the gauge theory part is not enough to achieve
cancellation and we get constraints on the gauge boson couplings to
scalars.
For convenience we define
\begin{equation}
\label{eq:22}
\hat B_i\equiv 4\, M_a  M_b M_c M_d\ B_i.
\end{equation}
We also note that once we use eq.~\eqref{eq:19} there is no
contribution at this order for $B_i^{st}$.
\begin{table}[htb]
	\centering
	\begin{tabular}{|c|c|c|}\hline
		Diagram & $\hat B_i^s$ &$\hat B_i^t$  \\[+1mm]
		\hline\hline
		$\mathcal{M}^{\rm 4 Point}$ & $8
		\left( D_{abcd}+D_{adbc}\right)  \Sigma $
		&$-8 \left(D_{acbd}+D_{adbc}\right) \Sigma$
		\\[+1mm]  \hline
		$\mathcal{M}_s^{A}$ & $g_{abe} g_{cde} \left[M_e^2 +\frac{(M_a^2-M_b^2)
			(M_c^2-M_d^2)}{M_e^2}\right]$
		&$ 2 g_{abe} g_{cde} (\Sigma +M_e^2) $ \\[+2mm]  \hline
		$\mathcal{M}_t^{A}$ & $-2 g_{ace} g_{bde} (\Sigma-M_e^2) $
		&$g_{ace} g_{bde}\left[M_e^2 -2 \Sigma + \frac{(M_a^2 -M_c^2)
			(M_b^2-M_d^2)}{M_e^2}\right]$ \\[+2mm]  \hline
		$\mathcal{M}_u^{A}$ &$- g_{ade}\, g_{cbe} \left[M_e^2 +
		2 \Sigma + \frac{(M_a^2 -M_d^2)
			(M_c^2-M_b^2)}{M_e^2}\right] $
		&$- g_{ade}\, g_{cbe} \left[- M_e^2 +
		4 \Sigma + \frac{(M_a^2 -M_d^2)
			(M_c^2-M_b^2)}{M_e^2}\right]$ \\[+2mm]  \hline \hline
		$\mathcal{M}_s^{\phi}$ &$ - g_{abk}\, g_{cdk}$
		&$ 0 $  \\[+1mm]  \hline
		$\mathcal{M}_t^{\phi}$ & $0 $
		&$- g_{ack}\, g_{bdk}$  \\[+1mm]  \hline
		$\mathcal{M}_u^{\phi}$ & $g_{adk}\, g_{bck}  $
		&$g_{adk}\, g_{bck}$  \\[+1mm]  \hline \hline
	\end{tabular}
	\caption{Coefficients $\hat B_i$. We have defined $\Sigma=(M_a^2 +M_b^2+M_c^2+ M_d^2)$.}
	\label{tab:2}
\end{table}
The results are summarized in table~\ref{tab:2}.

\subsubsection{The sum rule of ref.~\cite{Gunion:1990kf}}

To obtain the sum rule in eq.~(2.4) of ref.~\cite{Gunion:1990kf} we take
as independent the Mandelstam variables $s$ and $t$. The coefficients of
the terms growing with $s$ and $t$ must vanish. If we take the
coefficient of $s$ we obtain the desired sum rule. To show this, we notice
first that the sum of all contributions to $\hat B_i^s$ is
(sums implied)
\begin{align}
\label{eq:27}
&  8 \left( D_{abcd}+D_{adbc}\right)  \Sigma +
g_{abe} g_{cde} \left[M_e^2 +\frac{(M_a^2-M_b^2)
	(M_c^2-M_d^2)}{M_e^2}\right]\nonumber\\[+1mm]
&-2 g_{ace} g_{bde} (\Sigma-M_e^2)
- g_{ade}\, g_{cbe} \left[M_e^2 +
2 \Sigma + \frac{(M_a^2 -M_d^2)
	(M_c^2-M_b^2)}{M_e^2}\right]
\\[+1mm]
&=g_{abk}\, g_{cdk} -g_{adk}\, g_{bck},
\nonumber
\end{align}
where we have defined $\Sigma=(M_a^2 +M_b^2+M_c^2+ M_d^2)$.
Now we use eqs.~\eqref{eq:23}--\eqref{eq:25} to obtain
\begin{equation}
\label{eq:26}
8 \left( D_{abcd} + D_{adbc} \right) \Sigma = \left(g_{ade}\,
g_{cbe} + g_{abe}\, g_{cde} \right) \Sigma\, .
\end{equation}
Inserting this result into eq.~\eqref{eq:27} we obtain,
\begin{align}
\label{eq:27b}
& g_{abe} g_{cde} \left[\Sigma +M_e^2 +\frac{(M_a^2-M_b^2)
	(M_c^2-M_d^2)}{M_e^2}\right]- g_{ace} g_{bde} (2\Sigma-2M_e^2)\nonumber
\\[+1mm]
&
- g_{ade}\, g_{cbe} \left[M_e^2 +
\Sigma + \frac{(M_a^2 -M_d^2)
	(M_c^2-M_b^2)}{M_e^2}\right]=g_{abk}\, g_{cdk} -g_{adk}\, g_{bck}.
\end{align}
Now we use the Jacobi identity of eq.~\eqref{eq:6b} in the form
\begin{equation}
\label{eq:28}
\left(g_{abe} g_{cde} -g_{ace} g_{bde} - g_{ade} g_{cbe}\right) \Sigma =0,
\end{equation}
and subtract it from eq.~\eqref{eq:27b}. We then obtain the sum rule
of eq.~(2.4) of ref.~\cite{Gunion:1990kf},
\clearpage
\begin{align}
\label{eq:7}
&\sum_e{}' g_{abe}\, g_{cde} \left[ M_e^2 + \frac{(M_a^2 -M_b^2)
	(M_c^2-M_d^2)}{M_e^2}\right]
\nonumber\\[+2mm]
&- \sum_e{}' g_{ade}\, g_{cbe}
\left[ M_e^2 + \frac{(M_a^2 -M_d^2) (M_c^2-M_b^2)}{M_e^2}\right]
\nonumber\\[+2mm]
&
-\sum_e g_{ace}\, g_{bde}\left(M_a^2+M_b^2+M_c^2+M_d^2-2 M_e^2\right) =
\sum_k\left(g_{abk}\, g_{cdk} - g_{adk}\, g_{bck}\right),
\end{align}
where the prime in $\sum'$ indicates that the sum only runs over
massive gauge bosons.

\subsubsection{Another sum rule}

If we take the coefficient of $t$,
that is the sum of the $\hat B_i^t$ we obtain another sum rule:
\begin{align}
\label{eq:8}
&\sum_e{}' g_{ace}\, g_{bde} \left[ M_e^2 + \frac{(M_a^2 -M_c^2)
	(M_b^2-M_d^2)}{M_e^2}\right] \nonumber\\[+2mm]
&+ \sum_e{}'
g_{ade}\, g_{cbe} \left[ M_e^2 + \frac{(M_a^2 -M_d^2) (M_b^2-M_c^2)}{M_e^2}\right]
\nonumber\\[+2mm]
&
-\sum_e g_{abe}\, g_{cde}\left(M_a^2+M_b^2+M_c^2+M_d^2-2 M_e^2\right) =
\sum_k\left(g_{ack}\, g_{bdk} - g_{adk}\, g_{bck}\right).
\end{align}
Notice, however,
that eq.~\eqref{eq:8} is not independent of eq.~\eqref{eq:7}.
It is just the result of crossing symmetry from the $s$-channel
to the $t$-channel.

\subsection{$AAA\phi$ Sum Rules}
\bigskip

\subsubsection{The Amplitudes}

\begin{figure}[h!]
	\centering
	\includegraphics[width=0.6\linewidth]{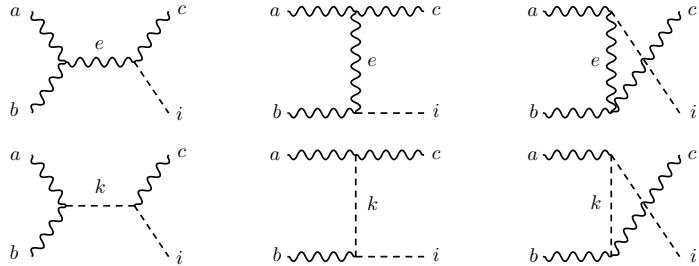}
	\caption{Amplitudes for the scattering $A_a A_b \to A_c \phi_i$}
  \label{fig:AAAS}
\end{figure}

The diagrams contributing to the scattering $A_a + A_b \to A_c + \phi_i$
are given in figure~\ref{fig:AAAS}.
%
%
%

\noindent
The corresponding amplitudes are
\begin{align}
\label{eq:6}
  \mathcal{M}_s^{A} =&(-i) (i\, g_{abe}) ( i\, g_{eci}) (-i)
\Gamma_{\alpha\beta\mu}(p_1,p_2,-p_1-p_2)\ g_{\gamma\nu}
\nonumber\\
&\hskip 5mm
\frac{\left[g^{\mu\nu} -
(p_1+p_2)^\mu (p_1+p_2)^\nu/M_e^2 \right]}
{s -M_e^2}
f^{\alpha \beta \gamma},
\end{align}
\begin{align}
\label{eq:29}
  \mathcal{M}_t^{A} =&(-i) (-i\, g_{ace}) ( i\, g_{ebi}) (-i)
\Gamma_{\alpha\mu\gamma}(p_1,-p_1+p_3,-p_3) \
g_{\beta\nu}
\nonumber\\
&\hskip 5mm
\frac{\left[g^{\mu\nu} -
(p_1-p_3)^\mu (p_1-p_3)^\nu/M_e^2 \right]}
{t -M_e^2}
f^{\alpha \beta \gamma},
\end{align}
\begin{align}
\label{eq:30}
  \mathcal{M}_u^{A} =&(-i) (i\, g_{bce}) (i\, g_{aei}) (-i) \
\Gamma_{\beta\gamma\nu}(p_2,-p_3,p_3-p_2)  \ g_{\alpha\mu}
\nonumber\\
&\hskip 5mm
\frac{\left[g^{\mu\nu} -
(p_1-p_4)^\mu (p_1-p_4)^\nu/M_e^2 \right]}
{u -M_e^2}
f^{\alpha \beta \gamma},
\end{align}
\begin{equation}
\label{eq:31}
  \mathcal{M}_s^{\phi} = (-i) (i g_{abk}) (i g_{cik}) i\
     \frac{g_{\alpha\beta}\ (-2 p_4 - p_3)_\gamma}{s-M_k^2}\
f^{\alpha \beta \gamma},
\end{equation}
\begin{equation}
\label{eq:32}
  \mathcal{M}_t^{\phi} = (-i) (i g_{ack}) (i g_{bik}) i\
     \frac{g_{\alpha\gamma}\ (-2 p_4 + p_2)_\beta}{t-M_k^2}\
f^{\alpha \beta \gamma},
\end{equation}
\begin{equation}
\label{eq:33}
  \mathcal{M}_u^{\phi} = (-i) (i g_{aik}) (i g_{bck}) i\
     \frac{(-2 p_4+p_1)_{\alpha}\ g_{\beta\gamma}}{u-M_k^2}\
f^{\alpha \beta \gamma},
\end{equation}
where
\be
f^{\alpha \beta \gamma}
=
\epsilon^\alpha(p_1) \epsilon^\beta(p_2) \epsilon^\gamma(p_3).
\ee

\subsubsection{The high energy limit}

In this case it is convenient to choose as independent variables $s$
and $u$. The results are summarized in table~\ref{tab:3}. Again we
used a definition similar to eq.~\eqref{eq:22},
\begin{equation}
  \label{eq:34}
  \hat B_i\equiv 2\, M_a  M_b M_c \ B_i.
\end{equation}
\begin{table}[htb]
  \centering
  \begin{tabular}{|c|c|c|}\hline
    Diagram & $\hat B_i^s$ &$\hat B_i^u$  \\[+1mm]
    \hline\hline
    $\mathcal{M}_s^{A}$ & $-g_{abe}
     g_{eci}\left[\frac{M_a^2-M_b^2+M_e^2}{2 M_e^2}\right] $
    &$ -g_{abe}\, g_{eci} $ \\[+2mm]  \hline
    $\mathcal{M}_t^{A}$ & $g_{ace}\, g_{ebi}
     \left[\frac{M_a^2-M_c^2+M_e^2}{2 M_e^2} \right]$
    &$ g_{ace}\, g_{ebi}
\left[\frac{M_a^2-M_c^2-M_e^2}{2 M_e^2}\right]$\\[+2mm]  \hline
    $\mathcal{M}_u^{A}$ &$g_{aei}\, g_{bce}$
    &$g_{aei}\, g_{bce} \left[\frac{-M_b^2+M_c^2+M_e^2}{2 M_e^2}\right]$ \\[+2mm]  \hline \hline
    $\mathcal{M}_s^{\phi}$ &$ g_{abk}\, g_{cik}$
    &$ 0 $  \\[+1mm]  \hline
    $\mathcal{M}_t^{\phi}$ & $-g_{ack}\, g_{bik} $
    &$-g_{ack}\, g_{bik}$  \\[+1mm]  \hline
    $\mathcal{M}_u^{\phi}$ & $0  $
    &$g_{aik}\, g_{bck}$  \\[+1mm]  \hline \hline
  \end{tabular}
  \caption{Coefficients $\hat B_i$. }
  \label{tab:3}
\end{table}

\subsubsection{The sum rule of ref.~\cite{Gunion:1990kf}}

To obtain the sum rule in eq.~(2.5) of ref.~\cite{Gunion:1990kf},
we take
as independent Mandelstam variables $s$ and $u$. The coefficients of
the terms growing with $s$ and $u$ must vanish. If we take the
coefficient of $s$ we obtain the desired sum rule.
\begin{align}
\label{eq:9}
&\sum_e{}' \left[g_{abe}\, g_{eci} \left[ \frac{M_a^2 -M_b^2+M_e^2}{2
  M_e^2}\right] -
g_{ace}\, g_{ebi} \left[ \frac{M_a^2 -M_c^2+M_e^2}{2 M_e^2}\right]
- g_{bce}\, g_{eai}\right]\nonumber\\[+2mm]
&\hskip 20mm
 = \sum_k\left(g_{cik}\, g_{abk} - g_{bik}\, g_{ack}\right).
\end{align}

\subsubsection{Another sum rule}

If we take the coefficient of $u$ we get another sum rule,
\begin{align}
\label{eq:10}
&\sum_e{}' \left[g_{ace}\, g_{ebi} \left[ \frac{M_c^2 -M_a^2+M_e^2}{2
  M_e^2}\right] -
g_{bce}\, g_{aei} \left[ \frac{M_c^2 -M_b^2+M_e^2}{2 M_e^2}\right]
+ g_{abe}\, g_{eci}\right]\nonumber\\[+2mm]
&\hskip 20mm
 =
\sum_k\left(g_{aik}\, g_{bck} - g_{bik}\, g_{ack}\right),
\end{align}
which
is just the crossed version of eq.~\eqref{eq:9}.

\bigskip\bigskip
\subsection{$AA\phi\phi$ Sum Rules}
\bigskip

\subsubsection{The Amplitudes}

\begin{figure}[h!]
	\centering
	\includegraphics[width=0.6\linewidth]{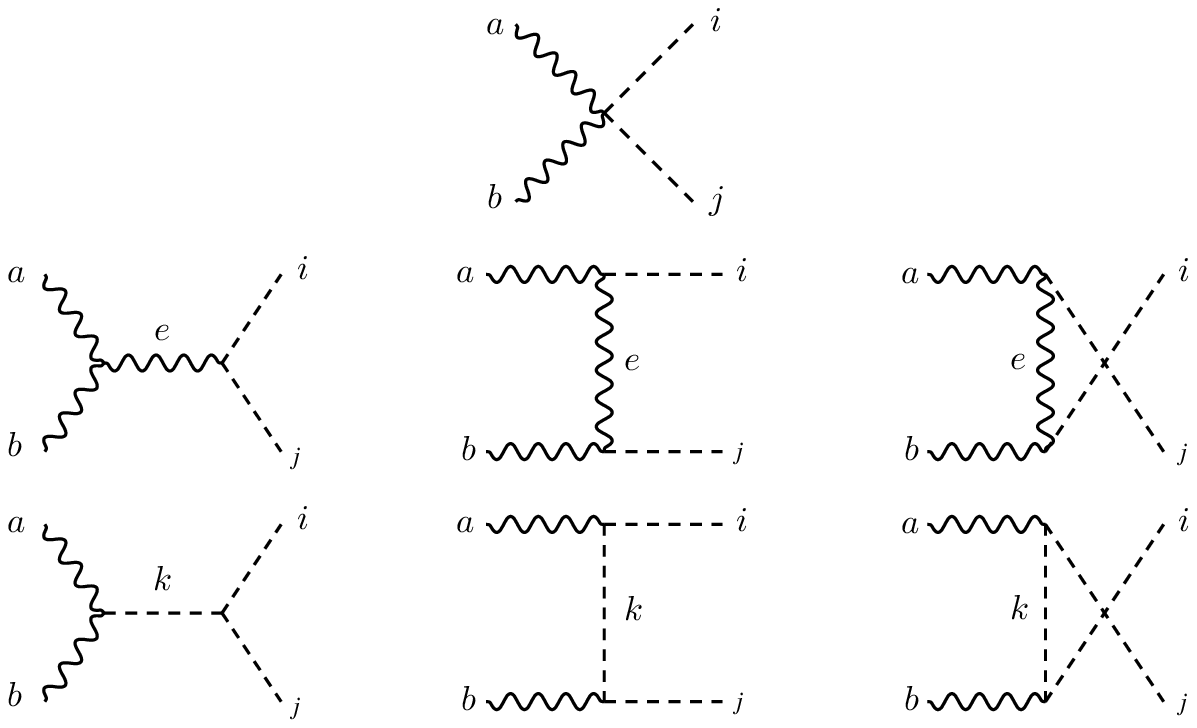}
  \caption{Amplitudes for the scattering $A_a A_b \to \phi_i \phi_j$}
  \label{fig:AASS}
\end{figure}

The diagrams contributing to the scattering $A_a + A_b \to \phi_i +
\phi_j$
are given in figure~\ref{fig:AASS}.

The diagram with the triple Higgs vertex does not exhibit any bad high energy behavior, and
therefore one does not have to consider it here. The other amplitudes are
\begin{equation}
\label{eq:35}
  \mathcal{M}^{\rm 4 Point} =(-i) (i\, g_{abij}\ g_{\alpha\beta}\
\epsilon^\alpha(p_1) \epsilon^\beta(p_2),
\end{equation}
\clearpage

\begin{align}
\label{eq:36}
  \mathcal{M}_s^{A} =&(-i) (i\, g_{abe}) ( i\, g_{eij}) (-i)
\Gamma_{\alpha\beta\mu}(p_1,p_2,-p_1-p_2)\ (- p_3 + p_4)_\nu
\nonumber\\
&\hskip 5mm
\frac{\left[g^{\mu\nu} -
(p_1+p_2)^\mu (p_1+p_2)^\nu/M_e^2 \right]}
{s -M_e^2}
\epsilon^\alpha(p_1) \epsilon^\beta(p_2),
\end{align}
\begin{align}
\label{eq:37}
  \mathcal{M}_t^{A} =&(-i) (i\, g_{aei}) ( i\, g_{bej}) (-i)
\frac{\left[g_{\alpha\beta} -
(p_1-p_3)_\alpha (p_1-p_3)_\beta/M_e^2 \right]}
{t -M_e^2}
\epsilon^\alpha(p_1) \epsilon^\beta(p_2),
\end{align}
\begin{align}
\label{eq:38}
  \mathcal{M}_u^{A} =&(-i) (i\, g_{aej}) (i\, g_{bei}) (-i) \
\frac{\left[g_{\alpha\beta} -
(p_1-p_4)_\alpha (p_1-p_4)_\beta/M_e^2 \right]}
{u -M_e^2}
\epsilon^\alpha(p_1) \epsilon^\beta(p_2),
\end{align}
\begin{equation}
\label{eq:39}
  \mathcal{M}_t^{\phi} = (-i) (i g_{aik}) (i g_{bkj}) i\
     \frac{(-2 p_3 + p_1)_\alpha \ (2 p_4 - p_2)_\beta}{t-M_k^2}\
\epsilon^\alpha(p_1) \epsilon^\beta(p_2),
\end{equation}
\begin{equation}
\label{eq:40}
  \mathcal{M}_u^{\phi} = (-i) (i g_{ajk}) (i g_{bki}) i\
     \frac{(-2 p_4+p_1)_{\alpha}\ (2 p_3 -p_2)_\beta}{u-M_k^2}\
\epsilon^\alpha(p_1) \epsilon^\beta(p_2).
\end{equation}

\subsubsection{The high energy limit}

In this case it is convenient to choose as independent variables $t$
and $u$. The results are summarized in table~\ref{tab:4}. Again we
used a definition similar to eq.~\eqref{eq:22},
\begin{equation}
  \label{eq:34new}
  \hat B_i\equiv  M_a  M_b \ B_i.
\end{equation}
\begin{table}[htb]
  \centering
  \begin{tabular}{|c|c|c|}\hline
    Diagram & $\hat B_i^t$ &$\hat B_i^u$  \\[+1mm]
    \hline\hline
    $\mathcal{M}_s^{\rm 4 Point}$ & $- \frac{1}{2}\, g_{abij}$
    &$ - \frac{1}{2}\, g_{abij} $ \\[+2mm]  \hline
    $\mathcal{M}_s^{A}$ & $-\frac{1}{2}\ g_{abe}\, g_{eij} $
    &$ \frac{1}{2}\, g_{abe} g_{eij}$ \\[+2mm]  \hline
    $\mathcal{M}_t^{A}$ & $\frac{1}{4}\frac{g_{aei}\, g_{bej}}{M_e^2}$
    &$ 0$\\[+2mm]  \hline
    $\mathcal{M}_u^{A}$ &$0$
    &$\frac{1}{4} \frac{g_{aej}\, g_{bei}}{ M_e^2}$
 \\[+2mm]  \hline \hline
    $\mathcal{M}_t^{\phi}$ & $g_{aik}\, g_{bkj}$
    &$0$  \\[+1mm]  \hline
    $\mathcal{M}_u^{\phi}$ & $0  $
    &$g_{ajk}\, g_{bki}$  \\[+1mm]  \hline \hline
  \end{tabular}
  \caption{Coefficients $\hat B_i$. }
  \label{tab:4}
\end{table}

\subsubsection{The sum rule of ref.~\cite{Gunion:1990kf}}

To obtain the sum rule in eq.~(2.6) of ref.~\cite{Gunion:1990kf},
we take
as independent Mandelstam variables $t$ and $u$. The coefficients of
the terms growing with $t$ and $u$ must vanish. If we take the
coefficient of $t$ we obtain the desired sum rule.
\begin{align}
\label{eq:11}
&\sum_k g_{aik} g_{bkj} - \frac{1}{2} g_{abij}
+ \frac{1}{4} \sum_e{}' \frac{g_{aei}\, g_{ebj}}{M_e^2}
- \sum_e \frac{1}{2} g_{abe}\, g_{eij}=0.
\end{align}

\subsubsection{Another sum rule}

If we take the coefficient of $u$ we get another sum rule,
\begin{align}
\label{eq:12}
&\sum_k g_{ajk} g_{bki} - \frac{1}{2} g_{abij}
+ \frac{1}{4} \sum_e{}' \frac{g_{aej}\, g_{bei}}{M_e^2}
+\sum_e \frac{1}{2} g_{abe}\, g_{eij}=0,
\end{align}
which is just the crossed version of eq.~\eqref{eq:11}.

\section{\label{app:ZZS=WWS}Proof that $[Z Z S_\beta^0] = [W^+ W^- S_\beta^0]$}

Theories based on $SU(2)_L \times U(1)_Y$ involve the operators
\ba
\hat{T}_+ \hat{T}_- + \hat{T}_- \hat{T}_+
=
\hat{T}^2 - \hat{T}_3^2,\qquad\quad
\hat{Q} = \hat{T}_3 + \hat{\mathcal{Y}}.
\ea
When acting on some neutral field,
\be
\Phi_k^0 = \frac{1}{\sqrt{2}} \left( v_k + \varphi_k^0 \right),
\ee
The operators $\hat{\mathcal{Y}}$, $\hat{T}$ and $ \hat{T}_3$, when acting on a neutral ($Q=0$) field $\Phi^0_k$, yield the corresponding eigenvalues $y_k$,
$T_k(T_k + 1)$ and $(T_3)_k = - y_k$, respectively.
Thus,
\ba
{\cal L}
&=&
\sum_{k=1}^N \left(D^\mu \Phi_k \right)^\dagger \left( D_\mu \Phi_k\right)
\nonumber\\
&\supset&
\frac{g^2}{2}
\left(|v_k|^2 + v_k^\ast \varphi_k^0 + v_k \varphi_k^{0 \ast}\right)
\left(
\left[ T_k(T_k + 1) - y_k^2  \right]
W_\mu^- W^{+\, \mu}
+
\frac{1}{c_W^2}
\left[ y_k^2  \right]
Z_\mu Z^\mu
\right),
\label{cubic_VVS}
\ea
where there is an implicit sum running over all neutral fields
in the theory,
from $1$ to $N_0$.\footnote{Notice that nothing was assumed
about the exact representations used, nor the value of $N_0$.}
Thus,
\ba
M_W^2 &=& \frac{g^2}{2}
\sum_k \left[ T_k(T_k + 1) - y_k^2  \right] |v_k|^2,
\nonumber\\
M_Z^2 &=& \frac{g^2}{2} \frac{1}{c_W^2}
\sum_k  \left[ 2 y_k^2  \right] |v_k|^2,
\ea
and
\be
\rho = \frac{M_W^2}{c_W^2 M_Z^2}
=
\frac{\sum_k \left[ T_k(T_k + 1) - y_k^2  \right] |v_k|^2}{
\sum_k  \left[ 2 y_k^2  \right] |v_k|^2}.
\label{rho_geral}
\ee

We now turn to the cubic couplings.
Using
\be
\varphi_k^0 =
\sum_{\beta=1}^{2N_0} V_{k \beta} S_\beta^0,
\label{general_V}
\ee
we find
\be
\left( v_k^\ast \varphi_k^0 + v_k \varphi_k^{0 \ast}\right)
=
2 v \sum_\alpha \Re\left[ (V^\dagger)_{\alpha k} \frac{v_k}{v}\right]
 S_\alpha^0,
\ee
where $v^2 = \sum_k |v_k|^2$.
Substituting in eq.~\eqref{cubic_VVS},
we find the corresponding Feynman rules as
\ba
[Z_\mu Z_\nu S_\alpha^0]^F
&=&
i \frac{g^2 v}{c_W^2} \sum_k
\left[ 2 y_k^2  \right]
 \Re\left[ (V^\dagger)_{\alpha k} \frac{v_k}{v}\right]
 g_{\mu \nu},
\nonumber\\
{}
[W^+_\mu W^-_\nu S_\alpha^0]^F
&=&
i g^2 v \sum_k
\left[ T_k(T_k + 1) - y_k^2  \right]
 \Re\left[ (V^\dagger)_{\alpha k} \frac{v_k}{v}\right]
 g_{\mu \nu}.
\ea
Defining the ratio of Feynman rules as
\ba
[Z_\mu Z_\nu S_\alpha^0]
&=&
\frac{[Z_\mu Z_\nu S_\alpha^0]^F}{[Z_\mu Z_\nu S_\alpha^0]^F_\textrm{SM}},
\nonumber\\
{}
[W_\mu^+ W_\nu^- S_\alpha^0]
&=&
\frac{[W_\mu^+ W_\nu^- S_\alpha^0]^F}{[W_\mu^+ W_\nu^- S_\alpha^0]^F_\textrm{SM}},
\ea
we get
\be
\rho_3 \equiv
\frac{[W_\mu^+ W_\nu^- S_\alpha^0]}{[Z_\mu Z_\nu S_\alpha^0]}
=
\frac{\sum_k
\left[ T_k(T_k + 1) - y_k^2  \right]
 \Re\left[ (V^\dagger)_{\alpha k}\, v_k \right]}{
\sum_k
\left[ 2 y_k^2  \right]
 \Re\left[ (V^\dagger)_{\alpha k}\, v_k \right]}.
\label{rho3_geral}
\ee
Notice the similarity between eqs.~\eqref{rho_geral} and
\eqref{rho3_geral}.

One knows from experiment that $\rho=1$ to high precision~\cite{rhoparm}.
Barring a fine tuning of the various vevs,
that can only occur if \emph{all} representations
of the theory with scalar fields with non-vanishing vevs
satisfy
\be \label{TTY}
T_k(T_k + 1) =  3 y_k^2.
\ee
Moreover, if \eq{TTY} is satisfied, then
eq.~\eqref{rho3_geral} implies that
$[W_\mu^+ W_\nu^- S_\alpha^0]$ is necessarily
equal to $[Z_\mu Z_\nu S_\alpha^0]$.
Note that this conclusion does not depend on whether the scalar potential is (or is not) CP-conserving.

We now apply eq.~\eqref{rho3_geral} in the case of the NHDM where
$T_k (T_k + 1) = 2 y_k^2 = 1/2$ for all $k$,
and the matrix $V$ in eq.~\eqref{general_V} coincides
with the matrix $V$ in eq.~\eqref{S0}.
Then,
eqs.~\eqref{ReVDomega} and \eqref{A_definition}
imply that
\be
\sum_k
 \Re\left[ (V^\dagger)_{\alpha k} \frac{v_k}{v} \right] =
- A_{1 \alpha}.
\ee
This agrees with the couplings obtained in eq.~\eqref{VVS_lag}.

\end{document}